\documentclass[twocolumn]{aastex631}
\usepackage{graphicx}
\usepackage{amssymb,amsfonts,amsmath,amstext,amsgen,amsopn,amsxtra,indentfirst,times}

\usepackage{savesym}
\savesymbol{tablenum}
\usepackage{siunitx}
\DeclareSIUnit \parsec {pc}
\DeclareSIUnit \mag {mag}
\DeclareSIUnit \Rjup {R_J}
\restoresymbol{SIX}{tablenum}
\usepackage{chemformula}
\usepackage{booktabs}
\usepackage{xspace}
\usepackage{mathptmx}
\usepackage{moresize}

\DeclareUnicodeCharacter{2212}{-}

\shorttitle{Intercomparison of Brown Dwarf Model Grids and Atmospheric Retrieval Using Machine Learning}
\shortauthors{Lueber et al.}

\begin{document}

\title{Intercomparison of Brown Dwarf Model Grids and Atmospheric Retrieval Using Machine Learning}

\correspondingauthor{Anna Lueber}
\email{anna.lueber@unibe.ch}

\author[0000-0001-6960-0256]{Anna Lueber}
\affiliation{Ludwig Maximilian University, University Observatory Munich, Scheinerstrasse 1, Munich D-81679, Germany}
\affiliation{University of Bern, Center for Space and Habitability, Gesellschaftsstrasse 6, CH-3012, Bern, Switzerland}

\author[0000-0003-4269-3311]{Daniel Kitzmann}
\affiliation{University of Bern, Center for Space and Habitability, Gesellschaftsstrasse 6, CH-3012, Bern, Switzerland}

\author[0000-0003-0652-2902]{Chloe E.\ Fisher}
\affil{University of Oxford, Department of Physics, Denys Wilkinson Building, Oxford OX1 3RH, UK}

\author[0000-0003-2649-2288]{Brendan P.\ Bowler}
\affil{The University of Texas at Austin, Department of Astronomy, 2515 Speedway, Stop C1400, Austin, TX 78712, U.S.A.}

\author[0000-0002-6523-9536]{Adam J.\ Burgasser}
\affil{Department of Astronomy \& Astrophysics, University of California San Diego, La Jolla, CA 92093, U.S.A.}

\author[0000-0002-5251-2943]{Mark Marley}
\affil{Department of Planetary Sciences and Lunar and Planetary Laboratory, University of Arizona, Tuscon, AZ}

\author[0000-0003-1907-5910]{Kevin Heng}
\affiliation{Ludwig Maximilian University, University Observatory Munich, Scheinerstrasse 1, Munich D-81679, Germany}
\affiliation{University of Warwick, Department of Physics, Astronomy \& Astrophysics Group, Coventry CV4 7AL, United Kingdom}
\affiliation{University of Bern, ARTORG Center for Biomedical Engineering Research, Murtenstrasse 50, CH-3008, Bern, Switzerland}

\begin{abstract}
Understanding differences between sub-stellar spectral data and models has proven to be a major challenge, especially for self-consistent model grids that are necessary for a thorough investigation of brown dwarf atmospheres.  Using the supervised machine learning method of the random forest, we study the information content of 14 previously published model grids of brown dwarfs (from 1997 to 2021).  The random forest method allows us to analyze the predictive power of these model grids, as well as interpret data within the framework of Approximate Bayesian Computation (ABC). Our curated dataset includes 3 benchmark brown dwarfs (Gl 570D, $\epsilon$ Indi Ba and Bb) as well as a sample of 19 L and T dwarfs; this sample was previously analyzed in \cite{Lueber2022ApJ...930..136L} using traditional Bayesian methods (nested sampling). We find that the effective temperature of a brown dwarf can be robustly predicted independent of the model grid chosen for the interpretation.  However, inference of the surface gravity is model-dependent.  Specifically, the \texttt{BT-Settl}, \texttt{Sonora Bobcat} and \texttt{Sonora Cholla} model grids tend to predict $\log{g} \sim 3$--4 (cgs units) even after data blueward of 1.2 $\mu$m have been disregarded to mitigate for our incomplete knowledge of the shapes of alkali lines. Two major, longstanding challenges associated with understanding the influence of clouds in brown dwarf atmospheres remain: our inability to model them from first principles and also to robustly validate these models.
\end{abstract}


\section{Introduction}
\label{sect:intro}

Brown dwarfs are objects that are intermediate in mass between giant exoplanets and stars.  Hundreds of brown dwarfs have been detected and spectrally characterized, making them an interesting training set for confronting models with data.  It has become standard practice to match measured brown dwarf spectra with pre-computed, self-consistent model grids, typically using a simple interpolation method. More recently, Bayesian retrieval methods, which were originally adapted from the Earth and planetary sciences (e.g., \citealt{Rodgers2000imas.book.....R, Irwin2008JQSRT.109.1136I}) to analyze spectra of exoplanetary atmospheres (e.g., \citealt{Madhusudhan2009ApJ}), may be used to infer posterior distributions of parameters from the analysis of spectra \citep{Line2017ApJ, Burningham2017MNRAS, Madhusudhan2018haex.bookE.104M, Molliere2019A&A...627A..67M, Kitzmann2020ApJ}.

Despite decades of model development efforts by the community studying brown dwarfs (e.g., \citealt{Burrows1997ApJ...491..856B, Burrows2006ApJ...640.1063B, Hubeny2007ApJ...669.1248H, Allard2012EAS....57....3A, Morley2012ApJ...756..172M, Saumon2012ApJ...750...74S, Phillips2020AA...637A..38P, Marley2021ApJ...920...85M, Karalidi2021ApJ...923..269K}; for reviews, see \citealt{Helling2014A&ARv..22...80H, Marley2015ARA&A}), including detailed calculations of the atomic and molecular opacities, the physical and chemical processes in brown dwarf atmospheres remain incompletely understood.  Different research groups tend to employ different approaches for implementing radiative transfer, chemistry and clouds.  The computer codes as well as important input data, such as thermochemical data or opacities, for calculating model grids often tend to be proprietary, making intercomparison of these grids challenging (e.g., \citealt{2008MNRAS.391.1854H}).  It is not always straightforward to identify discrepancies in outcomes between model grids employing different assumptions and methods.  Furthermore, model grids are typically constructed by uniforming sampling the parameters (either linearly or log-linearly), which is sub-optimal if one wishes to use these grids for Bayesian inference \citep{Fisher2022ApJ...934...31F}.

In the current study, we adopt the following approach: even without access to the computer codes used to generate them, we treat the model grids as training sets for a supervised machine learning method.  Such an approach was previously used to study the spectra of exoplanets at various spectral resolutions \citep{Marquez2018NatAs...2..719M, Fisher2020AJ....159..192F, Guzman2020AJ....160...15G}.  It enables three different lines of inquiry:
\begin{itemize}
    
    \item To analyze the information content of each model grid in isolation.  Specifically, the predictability of each parameter and its influence across wavelength may be quantified.

    \item The information content of a pair of model grids may be compared to quantify differences in parameter sensitivity across wavelength.  Unsurprisingly, one of our main findings is that inferring surface gravity from spectra is a model-dependent exercise.

    \item Having trained the machine learning procedure on a specific model grid, we may then use it to interpret measured spectra within the framework of Approximate Bayesian Computation \citep{Sisson2018handbook}.
    
\end{itemize}

This multi-faceted approach was first implemented by \cite{Oreshenko2020AJ}, but involving only three brown dwarfs and three model grids. While one may be skeptical about using machine learning to perform inference, our approach is \textit{no worse than the standard approach of fitting pre-computed model grids to spectra of brown dwarfs using an interpolation method}.  Such an approach not only offers an order-of-magnitude speed-up compared to traditional Bayesian inference methods (since the models no longer need to be computed on-the-fly), the information content analysis is not easily performed using classical methods (e.g. computing Jacobians), especially for model grids where the computer code is proprietary.  We are thus introducing a novel approach for both analyzing the information content of a large suite of models and interpreting a sample of spectra of brown dwarfs, which to our knowledge has not been done in the peer-reviewed literature. We note that various approaches for using machine learning to perform atmospheric retrieval already exist in the astrophysical literature (e.g., \citealt{Waldmann2016ApJ...820..107W, Marquez2018NatAs...2..719M, Zingales2018AJ....156..268Z, Cobb2019AJ....158...33C, Fisher2020AJ....159..192F, Matchev2022ApJ...930...33M, ArdevolMartinez2022A&A...662A.108A, Vasist2023A&A...672A.147V}).

This study is structured as follows: In Section~\ref{sect:Grids} we describe the publicly available atmospheric model grids used in this study. In Section~\ref{sect:random_forest_grids}, we briefly describe the machine learning algorithm deployed, and discuss effects of grid sizes and grid interpolation. In Sections~\ref{sect:Grid_comparison} we evaluate consistency between model grids, including the influence of grid parameter limits. In Section~\ref{sect:random_forest_retrievals} we apply our random forest retrieval infrastructure to three benchmark brown dwarfs and a sequence of L and T dwarf spectral standards to assess parameter inference and the quality of model fits on  measured spectra. In Section~\ref{sect:Discusion} we summarize the outcomes of our investigation and implications for future work.

\section{Model Grids of Brown Dwarf Atmospheres}
\label{sect:Grids}

In this study we use eleven different legacy model grids spanning 25 years of development in the literature. Some grids have separate versions that account for clouds, yielding a total of 14 different grids that are used in this study. All model grids are self-consistent, meaning that energy is conserved through the atmosphere based on radiative transfer and a temperature-pressure profile that accounts for chemical composition. The only free parameters are the physical parameters that describe the boundary conditions of the atmosphere, the effective temperature $T_\mathrm{eff}$ and surface gravity $\log{g}$, as well as elemental abundances and (in some models) cloud properties and equilibrium versus non-equilibrium chemistry. Most of the grids assume solar composition \citep{AndersGrevesse1989GeCoA..53..197A}, which we adopt throughout this study. Parameter ranges for effective temperature and surface gravity, as well as other properties, for each model grid are listed in Table~\ref{tab:properties}. 

\begin{figure*}
\begin{center}
\includegraphics[width=0.98\columnwidth]{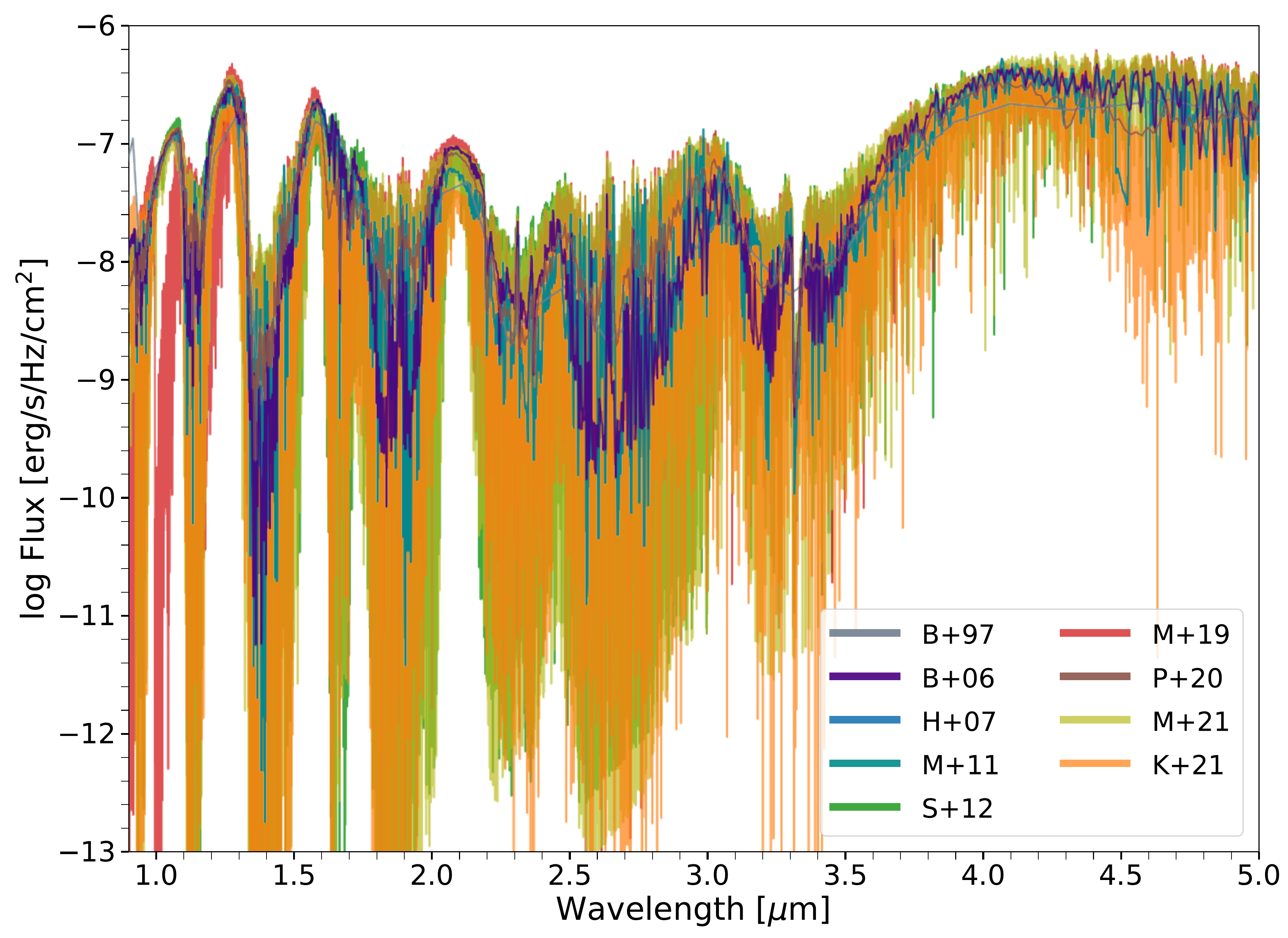}
\includegraphics[width=0.99\columnwidth]{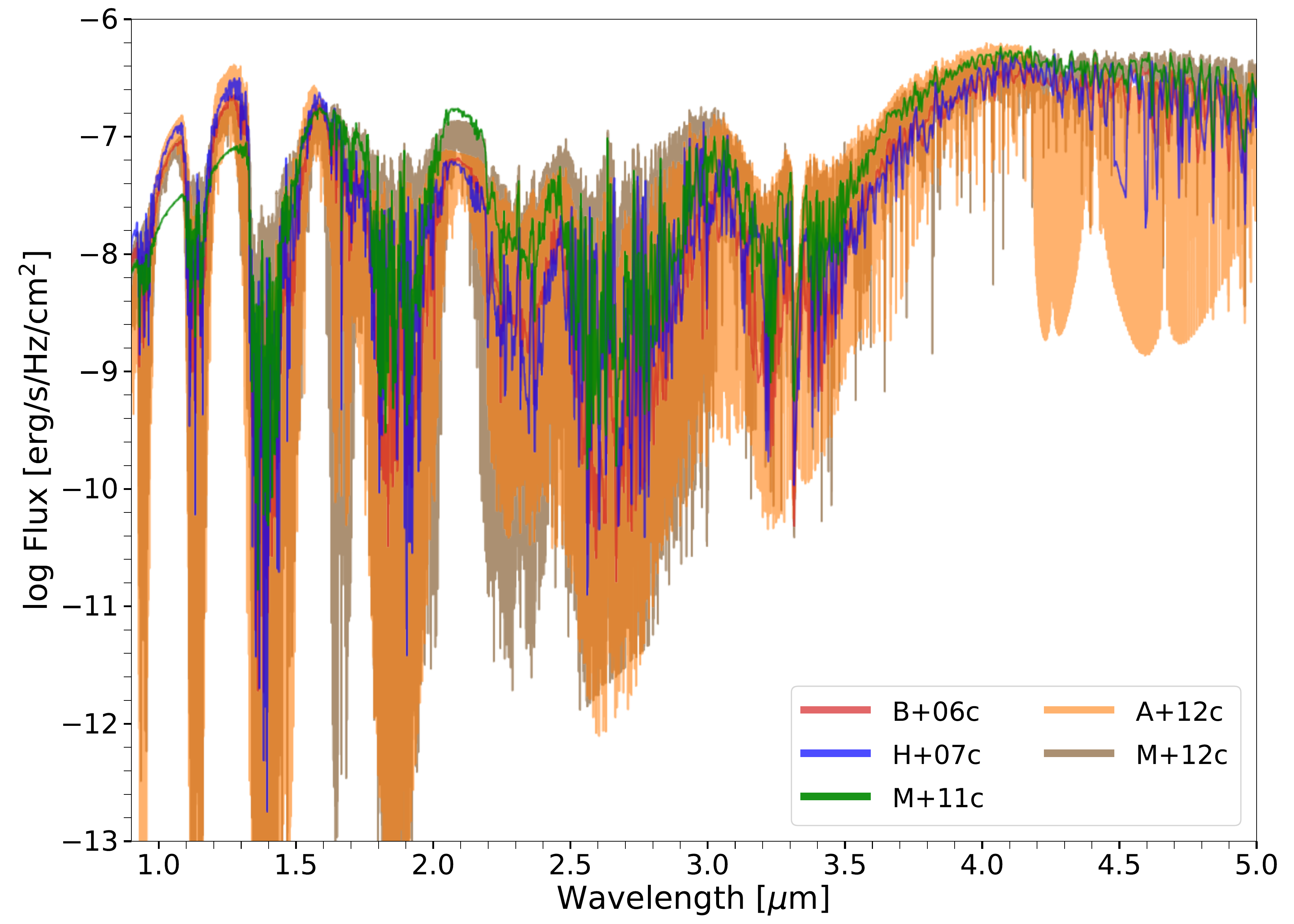}
\end{center}
\vspace{-0.1in}
\caption{Examples of spectra from all of the model grids for a typical late T dwarf with $T_\mathrm{eff}= 800$ K and $\log{g}$ = 5. Left panel: cloud-free models.  Right panel: cloudy models.}
\label{fig:Spectra}
\end{figure*}

To highlight differences between the various grids, we show synthetic spectra for a typical late T dwarf with $T_\mathrm{eff}$~=~800~K and $\log{g}$ = 5 in Figure~\ref{fig:Spectra}. As the figure illustrates, there are large variations among both cloudy and non-cloudy model spectra in both continuum levels and depths of absorptions. This variation is not surprising given the different modeling approaches involved and the development of opacity data over the 25 years these various grids have been produced. 

Here, we provide a brief summary of each model:

\begin{itemize}
    \item \citet{Burrows1997ApJ...491..856B} — This atmosphere grid (hereafter B+97) utilizes the two-stream source function approach \citep{Toon1989JGR....9416287T} with convective adjustment to obtain the atmospheric $T$-$P$ profile in radiative–convective equilibrium. The chemical composition is modeled with an equilibrium chemistry model, assuming that the elements \ch{Ti}, \ch{Al}, \ch{Ca}, and \ch{V} have been depleted by the formation of dust grains and are neglected during the chemical equilibrium calculations. Other potential cloud-forming gas phase species are considered to be limited by their saturation vapor pressure curves. Even though the impact of condensation is approximated in the chemical equilibrium calculations, the radiative impact of clouds is not taken into account, so this is effectively a cloud-free model. These models are calculated for effective temperatures between 125~K and 1200~K and are thus primarily applicable to T dwarfs and colder brown dwarfs. 
    
    \item \citet{Burrows2006ApJ...640.1063B} — This cloud-free and cloudy atmosphere grid (hereafter B+06 and B+06c) uses a variant of the stellar atmosphere code \texttt{TLUSTY} \citep{Hubeny1988CoPhC..52..103H, Hubeny1992LNP...401..375H, HubenyLanz1995ApJ...439..875H}, called \texttt{COOLTLUSTY}, with modifications as described in \cite{Burrows2002ApJ}, \cite{Sudarsky2003ApJ...588.1121S}, and \cite{Hubeny2003ApJ...594.1011H}. \texttt{COOLTLUSTY} self-consistently solves a set of radiative transfer equations for selected frequency points by the Rybicki radiative transfer scheme \citep{Rybicki1971JQSRT..11..589R, Mihalas1978stat.book}, and calculates the atmospheric temperature profile based on radiative and convective equilibrium using a Newton-Raphson method. The chemical composition is obtained using the chemical equilibrium code \texttt{SOLGASMIX} \citep{BurrowsSharp1999ApJ...512..843B, Burrows2003ApJ}, with updated thermochemical data for \ch{Ti} and \ch{V} compounds, silicates and calcium aluminates, \ch{H-}, as well as metal hydrides. The chemistry model incorporates prescriptions for rainout and depletion due to condensate formation of refractory silicates, aluminates, titanates, iron, water, and ammonia (see \citealt{Hubeny2007ApJ...669.1248H} for a detailed description). Clouds are assumed to be uniformly distributed without holes in the cloud deck, and are parameterized by geometric and optical thicknesses, and log-normal particle size distributions with pre-defined modal particle sizes. Convection is treated in the mixing-length formalism, with a mixing length of one pressure scale height. The calculation of the temperature profile in radiative-convective equilibrium follows the approach described in \citet{Saumon1995ApJS...99..713S}. These models extend over effective temperatures between 700~K and 2000~K and are thus applicable to L and T dwarfs. 

    \item \citet{Hubeny2007ApJ...669.1248H} — This cloud-free and cloudy atmosphere grid (hereafter H+07 and H+07c) is an adaptation of the B+06 models, using \texttt{COOLTLUSTY}.
    This grid uses an updated opacity database as described in \citet{SharpBurrows2007ApJS..168..140S}, and deploys prescriptions for equilibrium and non-equilibrium chemistry; for this analysis, we use the non-equilibrium set. These models extend over effective temperatures between 700~K and 1800~K and are thus applicable from mid L to T dwarfs. 
    
    \item \citet{Madhusudhan2011ApJ...737...34M} — This cloud-free and cloudy atmosphere grid (hereafter M+11 and M+11c) is an adaptation of the H+07 models, using \texttt{COOLTLUSTY}. Thick-clouds (AE type; see \citealt{Madhusudhan2011ApJ...737...34M} for details) are modeled following the approach of \citet{Burrows2006ApJ...640.1063B}; gas-phase equilibrium chemical abundances are computed using \citet{BurrowsSharp1999ApJ...512..843B} and \citet{SharpBurrows2007ApJS..168..140S}; and opacities are drawn from \citet{Burrows2001RvMP...73..719B} and \citet{SharpBurrows2007ApJS..168..140S}. These models also extend over effective temperatures between 700~K and 1800~K and are thus applicable from mid L to T dwarfs.

    \item \citet{Allard2012EAS....57....3A} — This cloudy atmosphere grid, also known as \texttt{BT-Settl} (hereafter A+12c), is based on the stellar atmosphere modeling code \texttt{PHOENIX} \citep{Hauschildt1992JQSRT..47..433H, Hauschildt1997ApJ...488..428H}, which solves the radiative transfer equation using an accelerated lambda iteration approach (ALI) in combination with opacity sampling. This grid is an updated version of the \texttt{AMES-Cond/Dusty} atmosphere model grid \citep{Allard2001ApJ...556..357A} incorporating the BT2 water line list \citep{Barber2006MNRAS.368.1087B} and updated solar element abundances from \citet{Asplund2009ARA&A..47..481A}. The cloud model of \texttt{BT-Settl} allows condensates to settle below the photosphere by accounting for vertical mixing by convection and overshooting \citep{Allard2011ASPC..448...91A}. These models have the broadest effective temperature range, between 500~K and 2400~K, and are thus applicable for all L and T dwarfs.
    
    \item \citet{Morley2012ApJ...756..172M} — This cloudy atmosphere grid (hereafter M+12c) is appropriate for T and Y dwarfs (400~K to 1300~K), and include ``cold'' condensates of \ch{Na2S}, \ch{KCl}, \ch{ZnS}, \ch{MnS}, and \ch{Cr}. The cloud structures and opacities are calculated using a modification of the cloud model developed by  \citet{Ackerman2001ApJ} that accounts for the location and vertical extent of cloud layers by balancing the upward transport of vapor and condensate by turbulent mixing in the atmosphere with the downward transport of condensate by sedimentation. The latter is described by a free parameter that quantifies the efficiency of particle sedimentation ($f_{cld}$), with higher sedimentation efficiencies producing thinner clouds with larger particle sizes. The cloud model is coupled to a self-consistent, one-dimensional atmosphere model that computes atmospheric $T$-$P$ profiles in radiative–convective equilibrium  \citep{McKay1989Icar...80...23M, Marley1996Sci, Burrows1997ApJ...491..856B, Marley1999Icar..138..268M, Marley2002ApJ, Saumon2008ApJ...689.1327S, Fortney2008ApJ...683.1104F}. Solar element abundances and equilibrium chemistry are assumed throughout the grid.

    \item \citet{Saumon2012ApJ...750...74S} — This cloud-free atmosphere grid (hereafter S+12) is based on the model sets of \citet{Saumon2008ApJ...689.1327S} and M+12c, but uses updated ab-initio calculations of collision-induced absorption (CIA) of molecular hydrogen (\ch{H2}) and new opacities of ammonia (`BYTe', \citealt{Yurchenko2011MNRAS}). These models extend from 300~K and 1500~K, and are thus applicable to T and Y dwarfs.

    \item \cite{Malik2019AJ....157..170M} — To get the convergent cloud-free atmospheric solution in radiative-convective equilibrium, the open-source radiative transfer code \texttt{HELIOS} (\citealt{Malik2017AJ....153...56M, Malik2019AJ....157..170M}; hereafter M+19) employs an improved hemispheric two-stream method \citep{HengKitzmann2017ApJS..232...20H, Heng2018ApJS..237...29H} with non-isotropic scattering and convective adjustment. Table 1 of \cite{Malik2019AJ....157..170M} lists the considered opacity sources and the corresponding line lists. The majority of opacities are computed with \texttt{HELIOS-K} \citep{Grimm2015ApJ} at a resolution of 10$^{−2}$ cm$^{−1}$ using a Voigt profile with a wing cutoff of 100 cm$^{-1}$. The \texttt{ExoMol} and \texttt{Hitran} online databases contain pressure broadening as a standard feature. The \ch{Na} and \ch{K} treatment, which is thoroughly explained in Appendix A of \cite{Malik2019AJ....157..170M}, is based on \cite{Burrows2000ApJ...531..438B} and \cite{Burrows2003ApJ...583..985B}. Based on the elemental abundances listed in Table 1 of \cite{Asplund2009ARA&A..47..481A}, \texttt{FastChem} \citep{Stock2018MNRAS.479..865S, Stock2022MNRAS.517.4070S} is used to compute the equilibrium gas-phase chemistry. The removal of species owing to condensation is included for \ch{H2O}, \ch{Na}, \ch{K}, \ch{TiO}, \ch{VO}, and \ch{SiO} and is described in Appendix B of \cite{Malik2019AJ....157..170M}. \texttt{HELIOS} uses the k-distribution method with the correlated-k approximation. A total of 300 wavelength bins and 20 Gaussian quadrature points throughout the range of 0.33~$\mu$m to 10~cm are used for integration over wavelength. A post-processing resolution of 3000 is used for the spectra. These models are computed from 200~K and 3000~K.

    \item \cite{Phillips2020AA...637A..38P} — These atmosphere models are generated using the 1D radiative-convective equilibrium code \texttt{ATMO} \citep{Amundsen2014A&A...564A..59A, Drummond2016A&A...594A..69D, Drummond2019MNRAS.486.1123D, Tremblin2015ApJ, Tremblin2016ApJ, Tremblin2017ApJ...841...30T, Tremblin2017ApJ...850...46T, Goyal2018MNRAS.474.5158G, Goyal2019MNRAS.482.4503G}, developed to study hot Jupiters and brown dwarf atmospheres. \texttt{ATMO 2020} now incorporates several significant upgrades, including a new \ch{H}-\ch{He} equation of state, ab initio quantum molecular dynamics calculations, and updated molecular opacities. The plane-parallel 1D model uses a Newton-Raphson solver to calculate the $T$-$P$ structure of an atmosphere self-consistent with the hydrostatic equilibrium for a given surface gravity. The treatment of collisionally broadened potassium resonance doublet was improved (see \citealt{Phillips2020AA...637A..38P} for details). Three distinct grids of model simulations were created, each computed self-consistently with the temperature-pressure structure of the atmosphere. One grid employed equilibrium chemistry, and the other used non-equilibrium chemistry caused by vertical mixing with varying mixing intensities. This study focuses on the solar metallicity, non-equilibrium, and weak vertical mixing grid (hereafter P+20). These models are computed from 200~K and 1800~K.

    \item \citet{Marley2021ApJ...920...85M} — This atmosphere grid, also known as \texttt{Sonora Bobcat} (hereafter M+21), are cloud-free models computed for radiative–convective equilibrium atmospheres utilizing a layer-by-layer convective adjustment approach, allowing for solutions with detached convective zones. Condensation from the gas phase is included for rainout chemistry, as described by \cite{Lodders2006asup.book....1L}, but no explicit cloud opacity is included. These models use the k-distribution method to calculate wavelength-dependent gas absorption \citep{Goody1989JQSRT..42..539G}, with updated opacities and equilibrium constants for the gas phase chemistry (see \citealt{Marley2021ApJ...920...85M} for details). These models extend from 200~K to 2400~K.
    
    \item \citet{Karalidi2021ApJ...923..269K}  — This cloud-free atmosphere grid called \texttt{Sonora Cholla} is an extension of the M+21 grid by including non-equilibrium chemistry of key molecular species such as \ch{CH4}, \ch{H2O}, \ch{CO}, and \ch{NH3}. These models extend from 500~K and 1500~K, and are thus applicable to T dwarfs.
    
\end{itemize}

Our selection of model grids is by no means complete, especially since some of the grids were omitted as they were superceded by more recent versions using the same code heritage (e.g., \citealt{Chabrier2000ApJ, Allard2001ApJ...556..357A, Saumon2008ApJ...689.1327S, Allard2011ASPC..448...91A, Witte2011A&A...529A..44W}).

\begin{deluxetable*}{lccccccccccc}
\tablecaption{Properties of the atmosphere model grids used in this study.}
\label{tab:properties}
\tablehead{
\colhead{}  & \colhead{} & \colhead{} & \colhead{} & \colhead{} & \colhead{} & \multicolumn{3}{c}{Original grid} & \colhead{} & \multicolumn{2}{c}{Enlarged grid}\\
\cline{7-9} \cline{11-12}
\colhead{Model \hspace{2cm}}  & \colhead{$\lambda$ ($\,\mu$m)} & \colhead{Average $\lambda/\Delta\lambda$} & \colhead{$T_\text{eff}$ (K)} & \colhead{$\log{g}$ (cm s$^{-2}$)} & \colhead{[M/H]} & \colhead{\# models} & \colhead{$\Delta T_\text{eff}$} & \colhead{$\Delta \log{g}$} & & \colhead{\# models} & \colhead{$\Delta \log{g}$}}
\startdata
B+97  & 0.87 -- 2,500 & 22 & 128  -- 1,196  & 3.3 -- 5.5  &  0 &  44  & 100 & 0.5 &  & 290 &  0.05\\
B+06  & 0.43 -- 300         & 764 & 700  -- 2,000  & 4.5 -- 5.5  & -0.5, 0, +0.5 & 44  & 100 & 0.5 & & 560 &  0.025\\
B+06c & 0.43 -- 300 & 764 & 700  -- 2,000  & 4.5 -- 5.5  & 0 & 298 & 50 & 0.1 & & 1,080 &  0.025\\
H+07  \tablenotemark{*}  & 0.43 -- 300         & 764 & 700  -- 1,800  & 4.5 -- 5.5  & 0 & 318 & 100 & 0.5 & &  480 &  0.025\\
H+07c \tablenotemark{*} & 0.43 -- 300         & 764 & 700  -- 1,800  & 4.5 -- 5.5  & 0 & 208 & 100 & 0.5 & &  480 &  0.025 \\
M+11  \tablenotemark{*}  & 0.43 -- 300         & 764 & 700  -- 1,800  & 4.5 -- 5.5  & 0 & 108 & 100 & 0.5 & & 240 &  0.05\\
M+11c \tablenotemark{*} & 0.43 -- 300         & 764 & 600  -- 1,700  & 3.5 -- 5.5  & 0 & 149 & 50-100 & 0.1-0.5 &  & 641 &  0.05\\
A+12c  & 10$^{-4}$ -- 10$^4$ & 353,673 & 500  -- 2,400  & 2.0 -- 5.5  & 0 & 216 & 50-100 & 0.5 &  & 1,237 &  0.05\\
M+12c  & 0.6 -- 30    & 87,530 & 400  -- 1,300  & 4.0 -- 5.5  &0 & 182 & 50-100 & 0.5 &  & 396 &  0.05\\
S+12  & 0.6 -- 30    & 87,530 & 300  -- 1,500  & 4.0 -- 5.5  & 0& 158 & 50-100 & 0.5 &  & 1,259 &  0.025\\
M+19  &  0.4 -- 10$^4$   & 2,630 & 200  -- 3,000  & 1.4 -- 6.0 & 0 & 696 & 100 & 0.2 &  & 6,699 & 0.020\\
P+20 &  0.2 -- 1,980   & 2,500	& 200 -- 1800	& 2.5 -- 5.5 & 0 & 139  & 50-100 & 0.5 &	& 2,441	& 0.025\\
M+21  & 0.4 -- 50    & 155,936 & 200  -- 2,400  & 3.25 -- 5.5 & -0.5, 0, +0.5 & 429 & 25-100 & 0.25& & 3,700 & 0.025\\
K+21  \tablenotemark{*}  & 1.0 -- 250  & 55,622 & 500  -- 1,300  & 3.25 -- 5.5 & 0 & 459 & 50 & 0.25 & & 1,360 & 0.025\\
\hline
\enddata
\tablenotetext{*}{Disequilibrium chemistry model grid.}
\tablerefs{B+97: \citet{Burrows1997ApJ...491..856B}, B+06/c: \citet{Burrows2006ApJ...640.1063B}, H+07/c: \citet{Hubeny2007ApJ...669.1248H}, M+11/c: \citet{Madhusudhan2011ApJ...737...34M}, A+12c: \citet{Allard2012EAS....57....3A}, M+12c: \citet{Morley2012ApJ...756..172M}, S+12: \citet{Saumon2012ApJ...750...74S}, M+19: \citet{Malik2019AJ....157..170M}, P+20: \citet{Phillips2020AA...637A..38P}, M+21: \citet{Marley2021ApJ...920...85M}, K+21: \citet{Karalidi2021ApJ...923..269K}.}
\end{deluxetable*}

\section{Analyzing the Information Content of a Model Grid}
\label{sect:random_forest_grids}

\begin{figure}
\centering
\includegraphics[width=\columnwidth]{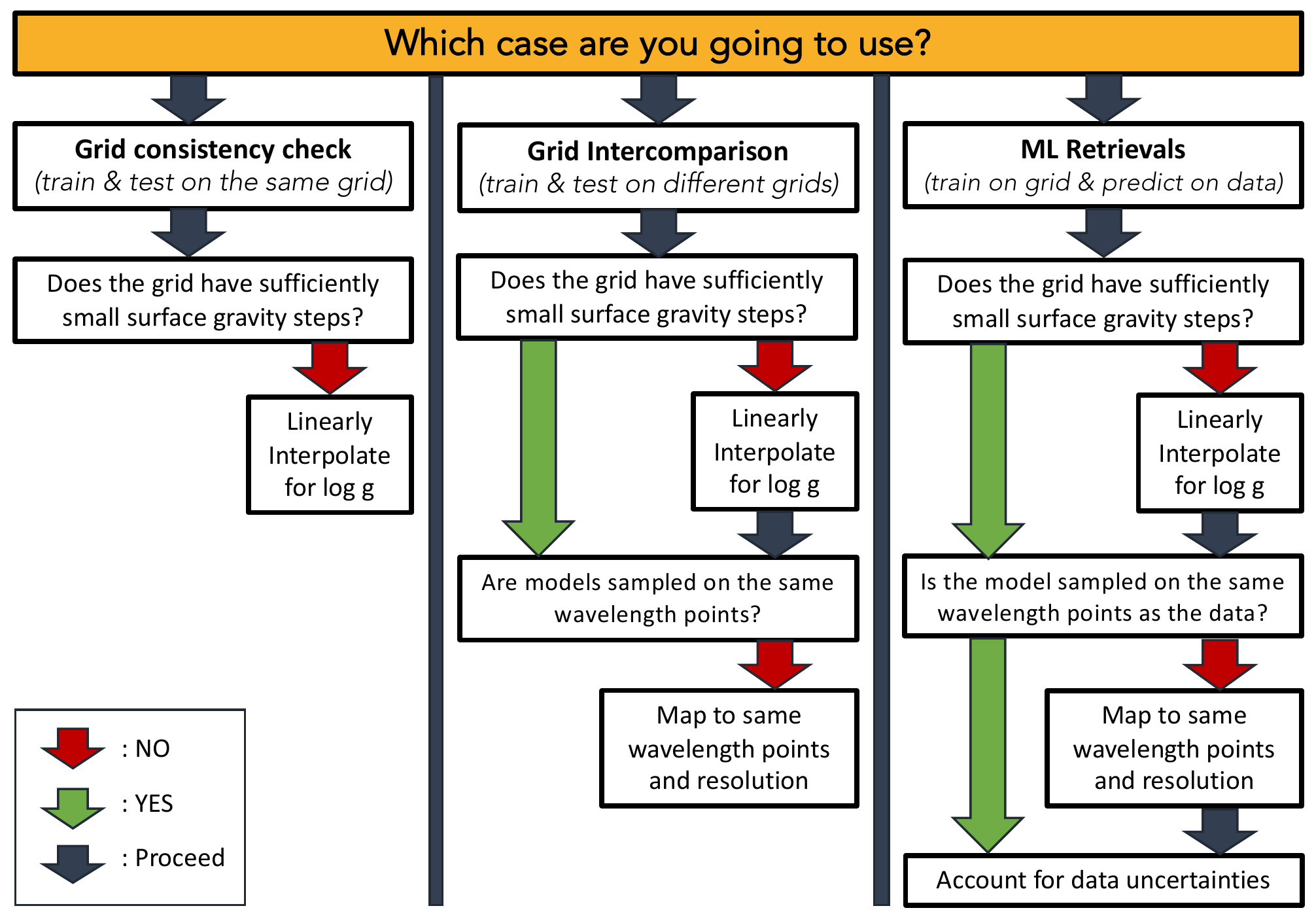}
\caption{Schematic flowchart describing the three applications of our supervised machine learning method (random forest) in the current study: examining the information content of each model grid in isolation, comparing the information content of a pair of model grids, and finally performing atmospheric retrieval on measured spectra in the framework of Approximate Bayesian Computation.}
\label{fig:Schematic}         
\end{figure}

In this section we first briefly introduce the machine learning code \texttt{HELA} that we use to perform the retrieval calculations. We also present the results of testing the different grids and discuss the necessity of increasing their sizes using interpolation. Figure~\ref{fig:Schematic} represents a schematic flowchart, which describes the different application areas of \texttt{HELA} and the corresponding treatment of the model grids.

\subsection{Random forest machine learning algorithm \texttt{HELA}}
\label{sect:HELA}

All atmospheric retrieval calculations for this study were done using \texttt{HELA}\footnote{\url{https://github.com/exoclime/HELA}} \citep{Marquez2018NatAs...2..719M}. This retrieval code applies the supervised machine learning random forest algorithm \citep{Ho1998random, Breiman2001random} to a pre-computed grid of atmosphere model spectra to generate a statistical model that infers parameters such as temperature, surface gravity, or composition. The random forest method is a form of Approximate Bayesian Computation \citep{Sisson2018handbook}.  A main advantage of a random forest retrieval is its speed of computation, as well as the determination of \textit{feature importance}, a quantity that ranks the importance of each spectral data point (the ``features" in the jargon of machine learning) in constraining a model parameter. \texttt{HELA} has been used to analyze the information content of exoplanet transmission spectra  \citet{Marquez2018NatAs...2..719M} and brown dwarf spectra \citet{Oreshenko2020AJ}, as well as other applications \citep{Fisher2020AJ....159..192F, Guzman2020AJ....160...15G, Fisher2022ApJ...934...31F}. 

Machine learning retrievals necessarily split the model grid into training and testing sets.  The outcome of this testing can be evaluated using \textit{real versus predicted} (RvP) assessments, quantified by the coefficient of determination ($R^2$). $R^2$ can take on negative and positive values depending on the degree of correlation and offset in inferred values: 
\begin{equation}
    R^{2}= 1 - S_\mathrm{res} / S_\mathrm{tot}\hspace{1cm}\,    
\end{equation}
where $S_\mathrm{res}$ is the sum of squares of the residual errors, and $S_\mathrm{tot}$ is the total sum of the errors. Comparisons between real and predicted parameters that align perfectly have $R^2$~=~1 (perfect correlation). If the relationship between two parameters is random, then one obtains $R^2=0$. $R^2<0$ indicate worse-than-random relationships between two parameters; $R^2$ is unbounded from below.  In practice, the presence of noise leads to $R^2$ values that are less than unity even if the noise-free RvP is associated with $R^2=1$. For this study, we used a ratio of approximately 80:20 between training and testing sets, following previous studies \citep{Marquez2018NatAs...2..719M, Fisher2020AJ....159..192F}. Our random forests were set up to have 1000 regression trees with no restriction on the maximum tree depth. Instead, each tree grows by splitting the space of the models until the decrease in variance of further splits is smaller than 0.01. These hyper-parameters associated with the random forest calculations are taken from \cite{Marquez2018NatAs...2..719M}. Additional calculations with 3000 instead of 1000 trees produced essentially identical posterior outcomes.

\subsection{Grid consistency and the effect of grid size}
\label{sect:Gridsize}

One disadvantage of forward-modeling analyses using model grids is the typically coarse sampling of model parameters and corresponding large parameter step sizes, as well as limitations in the overall parameter space. To investigate the impact of grid size on parameter inference, we first performed an internal consistency analysis of individual grids by training and testing on the same grid using \texttt{HELA}. 

\begin{figure*}[ht!]
\centering
\includegraphics[width=0.45\textwidth]{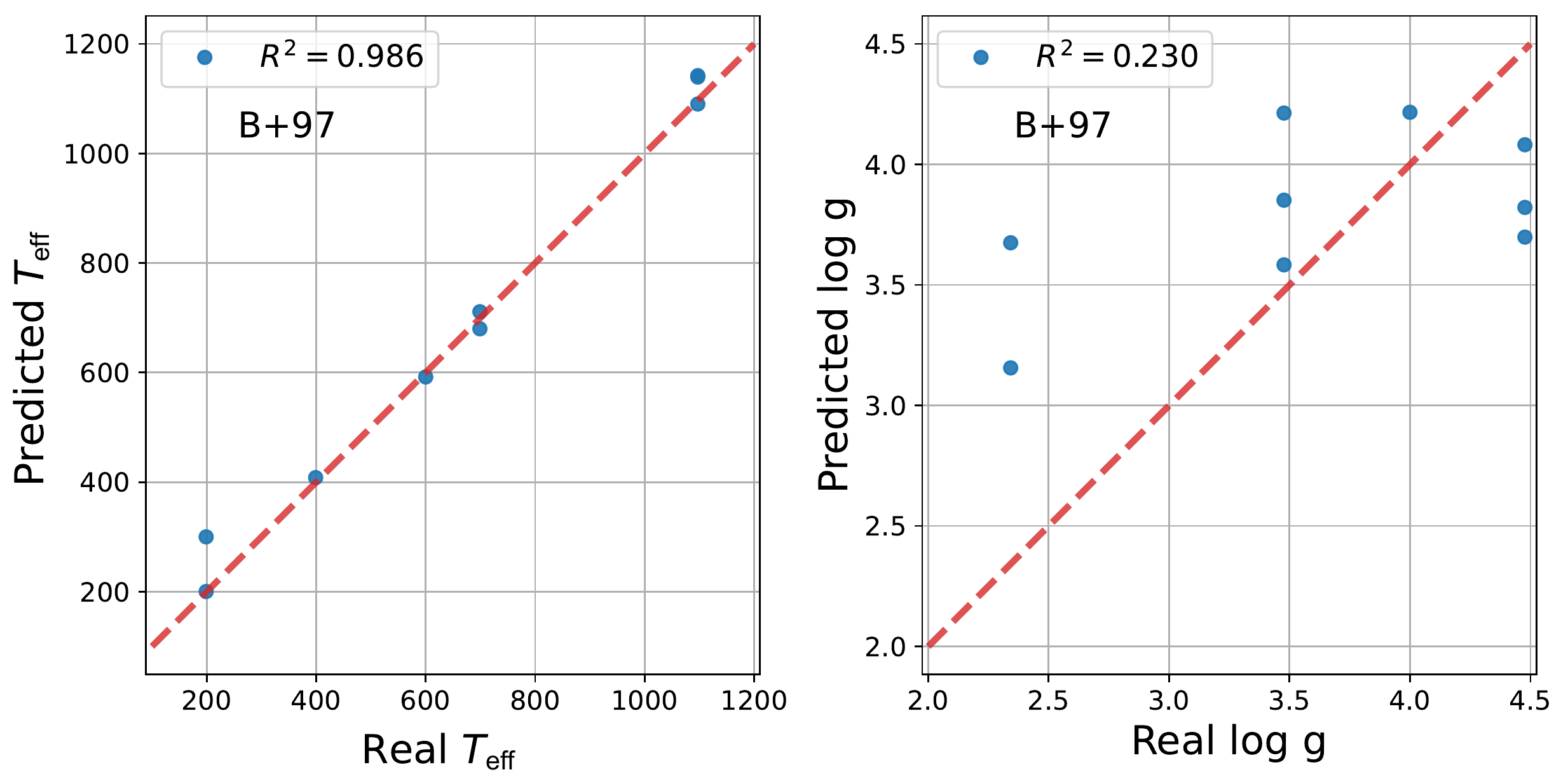}
\includegraphics[width=0.45\textwidth]{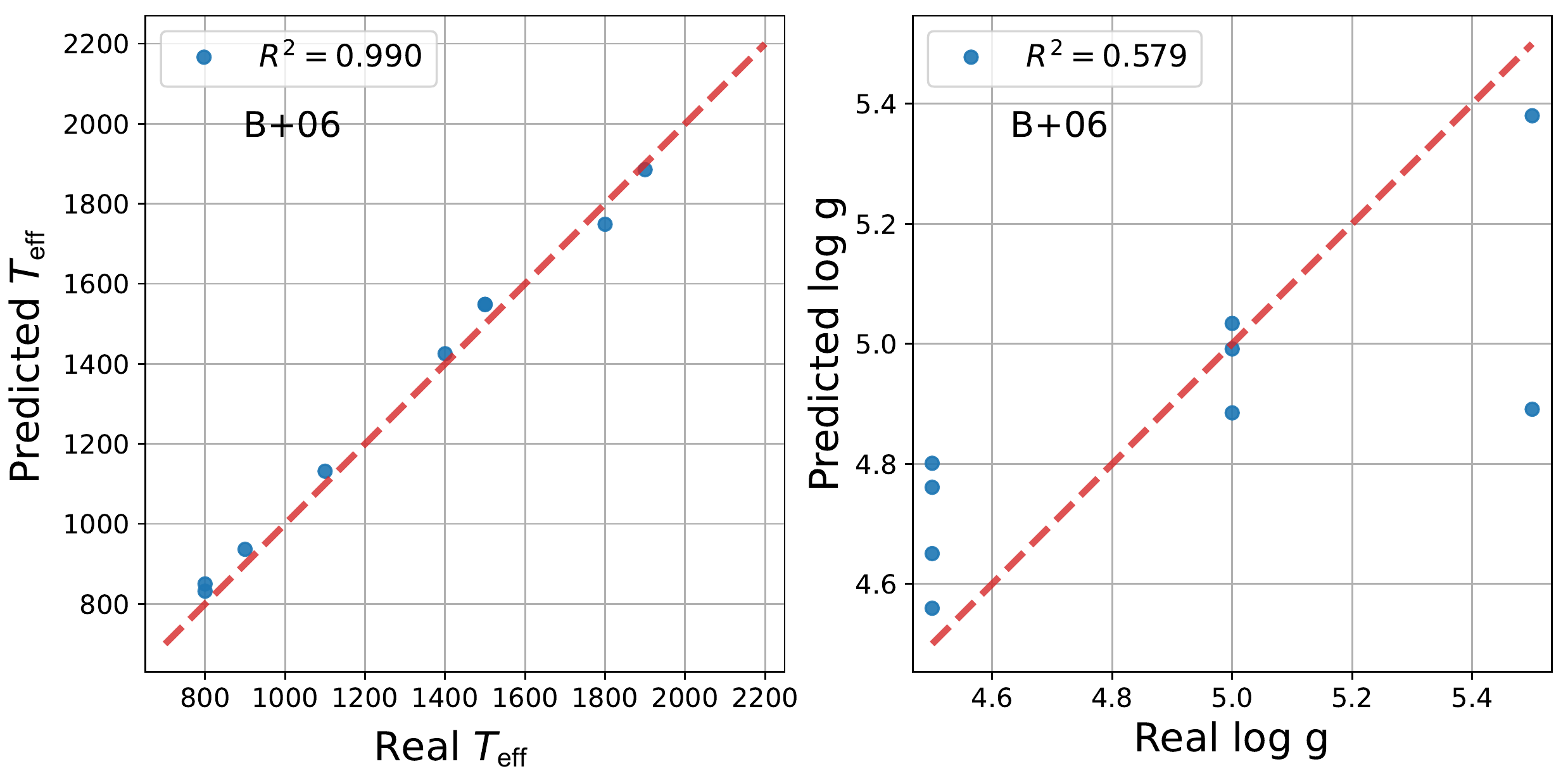}
\includegraphics[width=0.45\textwidth]{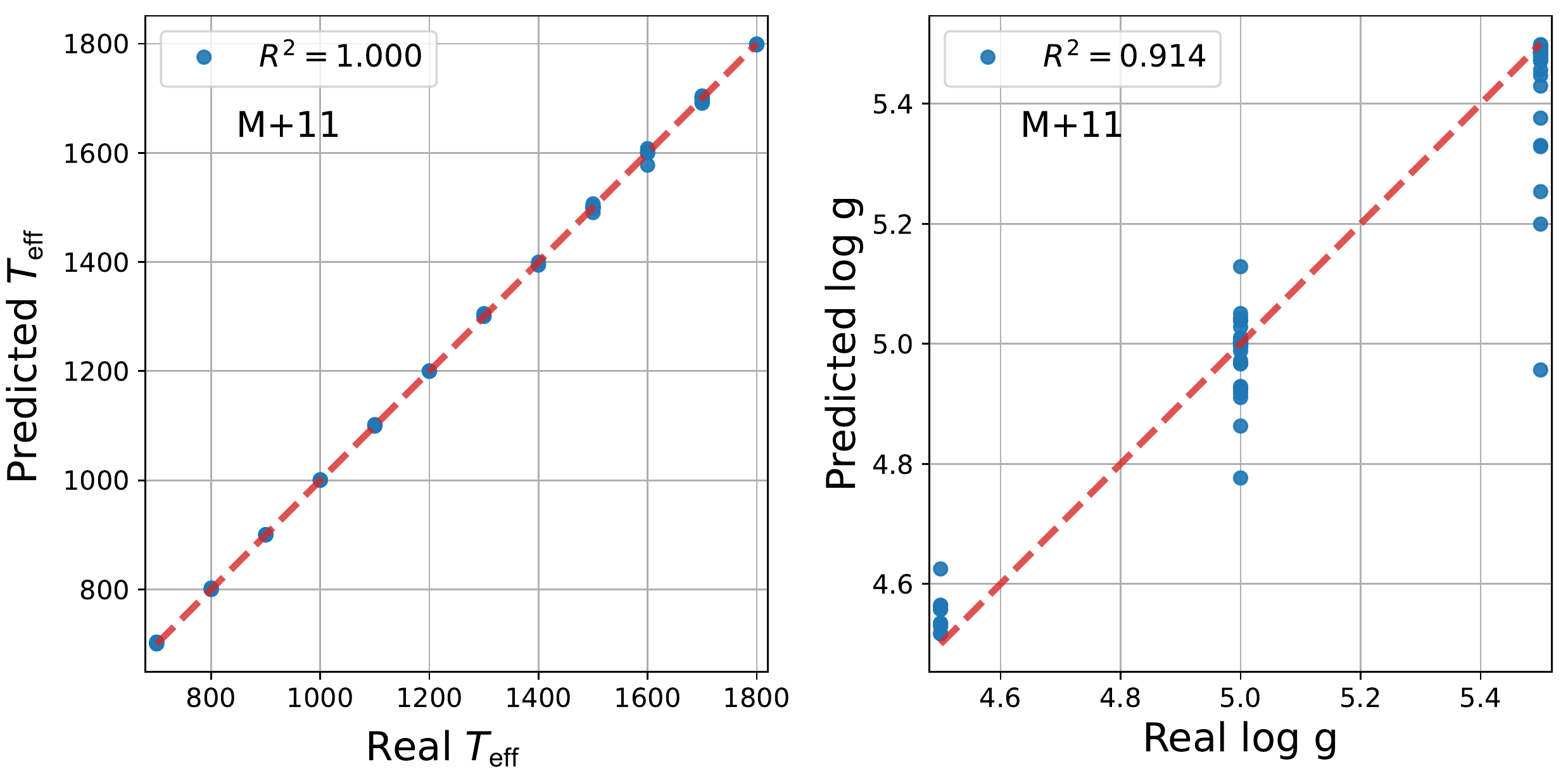}
\includegraphics[width=0.45\textwidth]{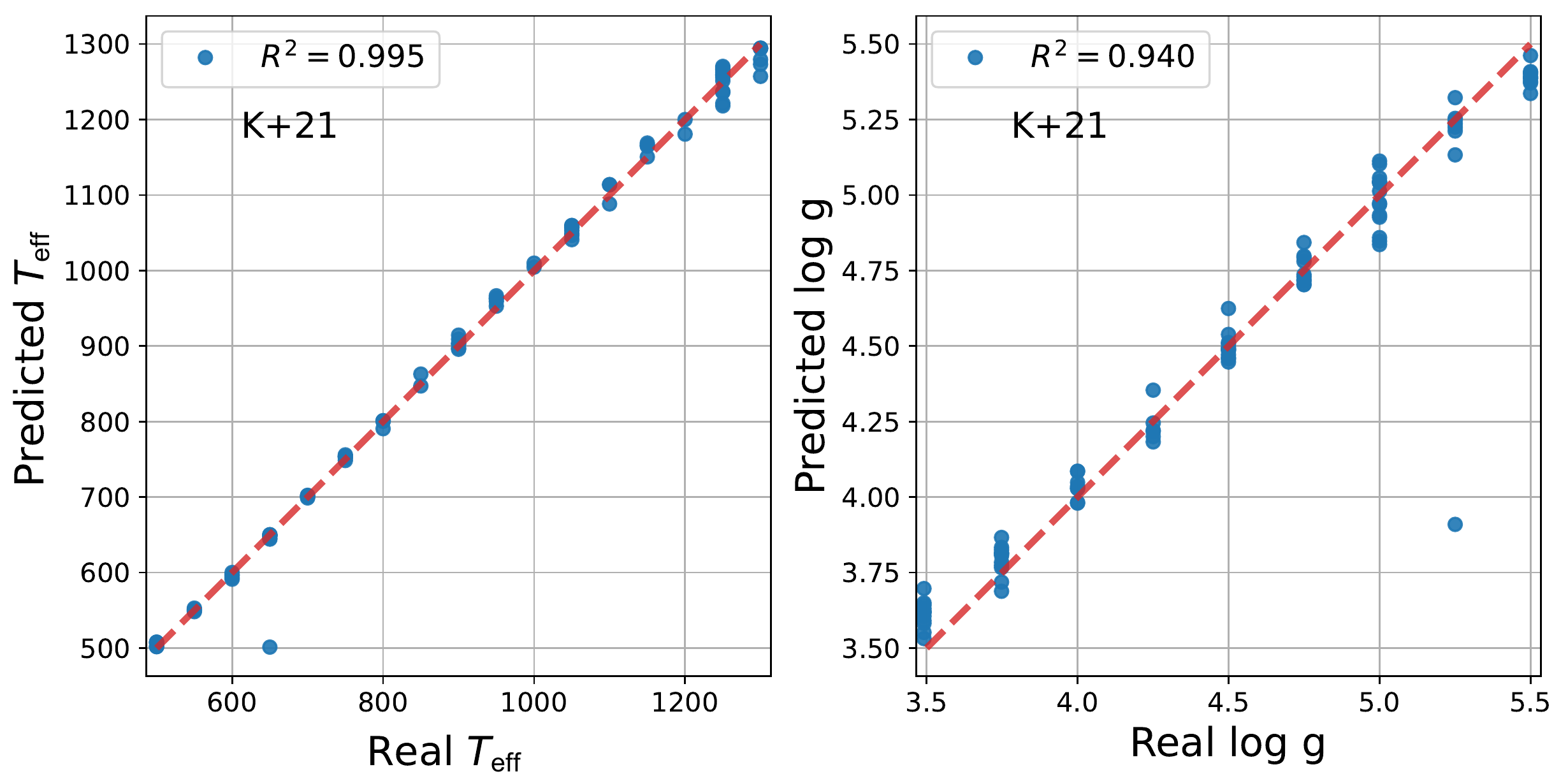}
\caption{Real versus predicted (RvP) comparison for effective temperature $T_\mathrm{eff}$ and surface gravity $\log{g}$
for random forest training and testing on a single grid. No noise is assumed. Here, we show analysis for the grids B+97, B+06, M+11, and K+21. In each panel, the red dashed line indicates perfect agreement. Differences in the grid sizes, particularly for surface gravity, and corresponding effect on the accuracy of predictions can be seen.}
\label{fig:Effect_of_grid_size}         
\end{figure*}

Figure~\ref{fig:Effect_of_grid_size} shows the RvP plots for the two model parameters $T_\mathrm{eff}$ and $\log{g}$ for four of the model grids: B+97, B+06, M+11, and K+21. These results show that the effective temperature has high self-consistency, with $R^2 > 0.98$ for all of the grids, even for sparse grids such as B+97 that contains only 44 models. Surface gravity, on the other hand, is problematic in forms of predictability with $R^2$ as low as 0.230 for the B+97 grid. The problem arises largely from the coarse sampling and small number of models available in some grids such that there is not enough information to properly train the random forest. Grids that are larger and less sparse, such as the M+11 and K+21 grids, show better internal consistency with $R^2$ values exceeding 0.9.

To address this issue, we artificially increased grid sizes and sampling through interpolation of the spectral models. This approach had previously been established by \citet{Oreshenko2020AJ}. They found that training the random forest without grid enlargement would lead to non-convergence of the random forest algorithm. However, adding additional model spectra can cause additional issues. For example, \citet{Zhang2020ApJ...891..171Z} showed that linear interpolation of models produced artificially small uncertainties for inferred physical parameters, while \citet{Cottaar2014ApJ...794..125C} and \citet{Czekala2015ApJ...812..128C} found that posterior distributions can be biased towards the original model grid points. \citet{Arnould2022arXiv220203688A} and \citet{Fisher2022ApJ...934...31F} similarly found that interpolated grids could result in inferred parameters that deviated significantly from the original grid in cases where parameter effects were non-linear.

\begin{figure*}[ht!]
\centering
\includegraphics[width=0.45\textwidth]{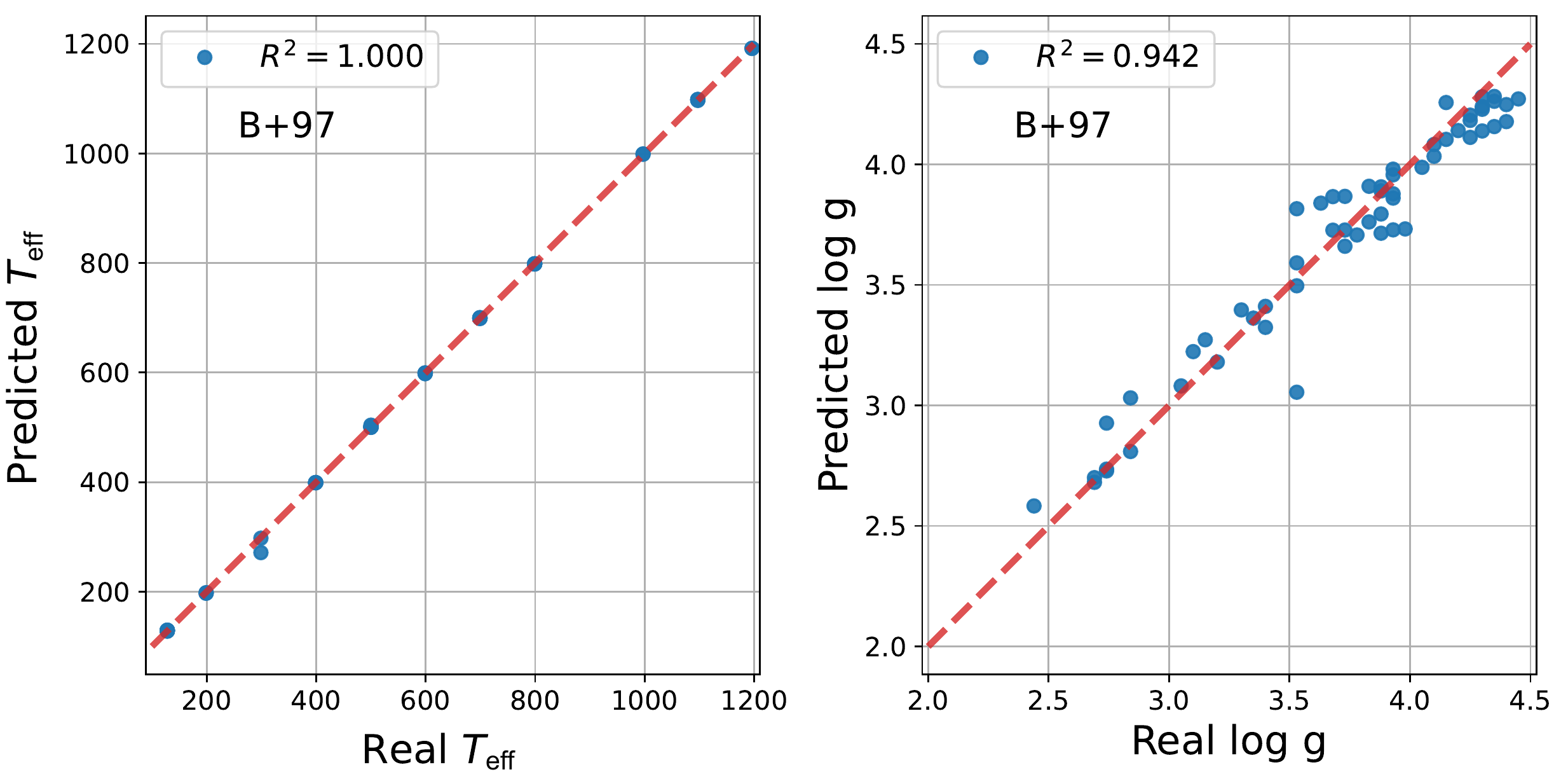}
\includegraphics[width=0.45\textwidth]{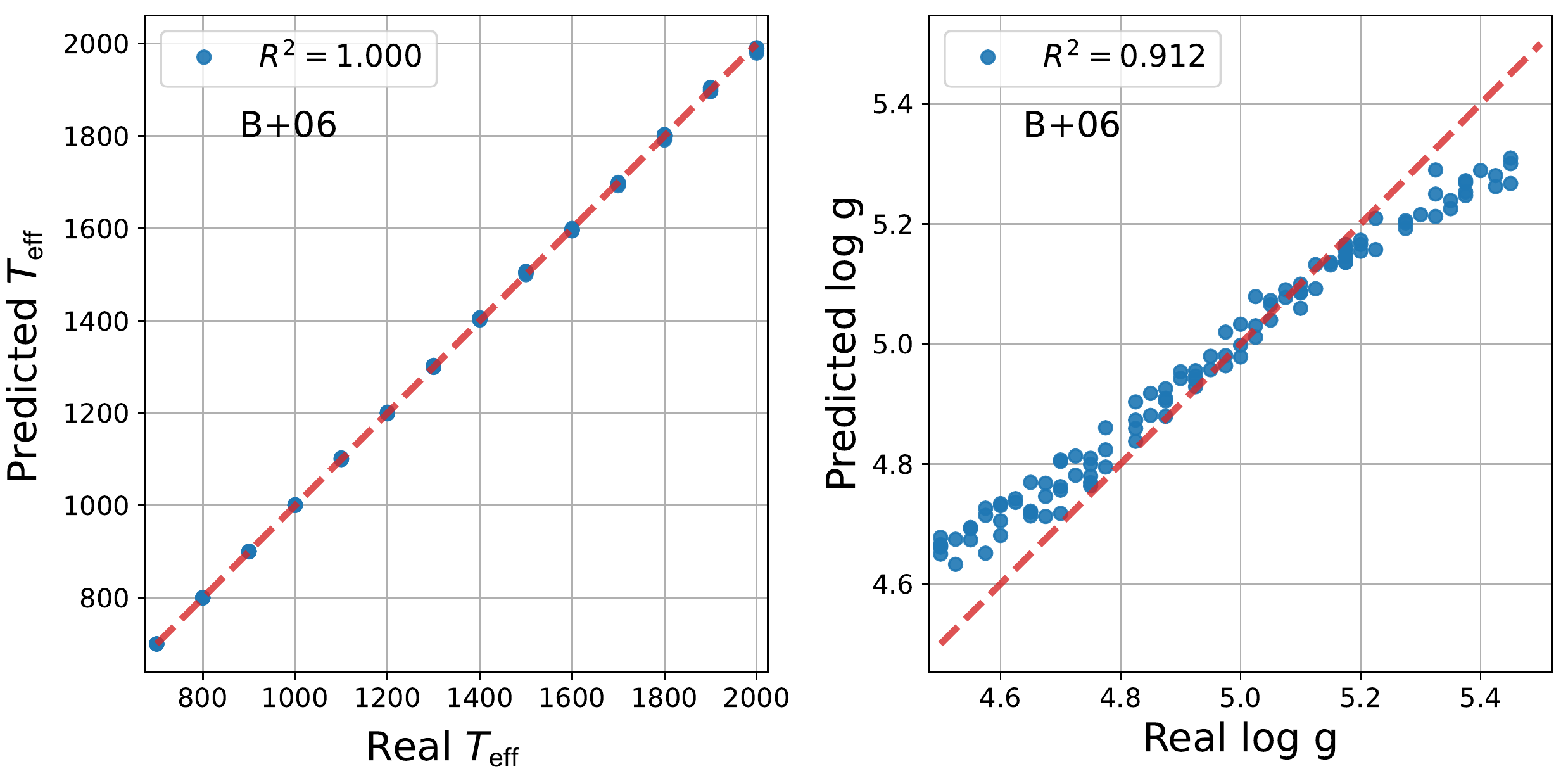}
\includegraphics[width=0.45\textwidth]{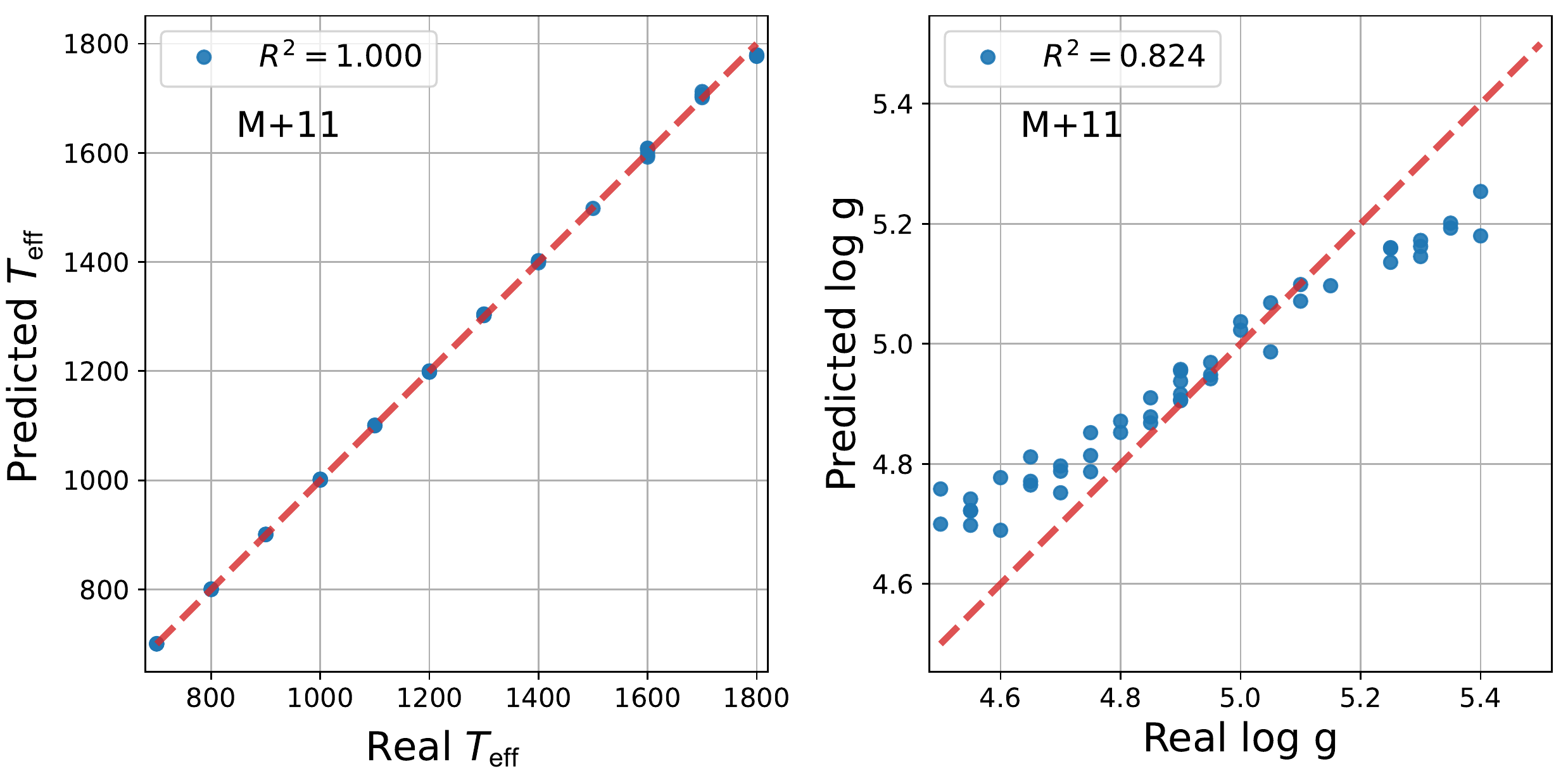}
\includegraphics[width=0.45\textwidth]{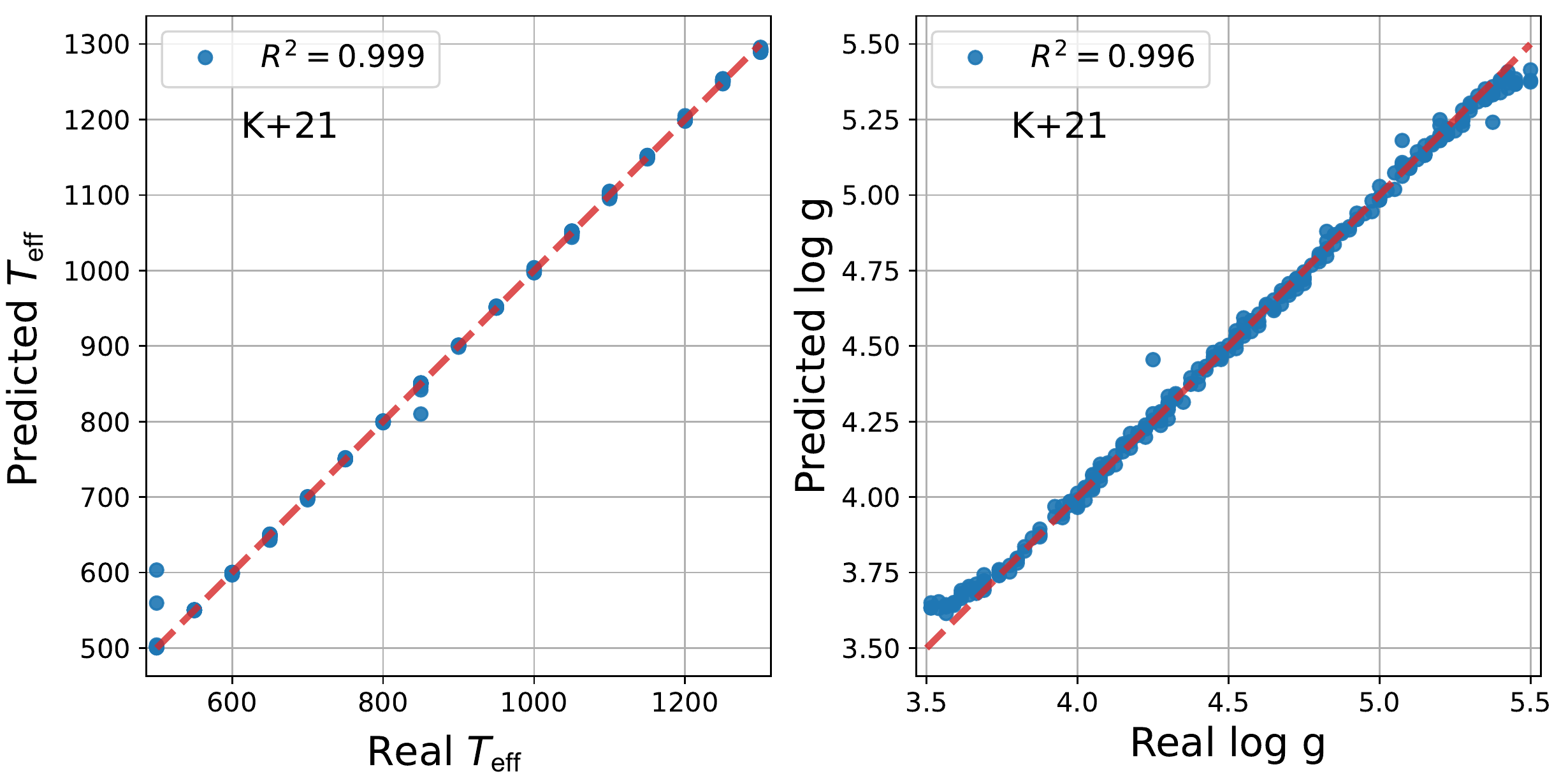}
\caption{Same as Figure~\ref{fig:Effect_of_grid_size} but with grids enlarged through interpolation of surface gravity. No noise is assumed. Note the significant improvement in $R^2$ scores for both effective temperature ($T_\mathrm{eff}$) and surface gravity ($\log{g}$).}
\label{fig:Effect_interpolated_grid}         
\end{figure*}

We performed a one-dimensional linear interpolation of the measured linear flux in small logarithmic surface gravity steps (0.025 to 0.05 dex) to increase the resolution and total number of all model grids. Note that interpolation is a convenient approach for estimating inter-grid models, and ignores key aspects of the underlying physics for intermediate surface gravity parameters. It is a reasonable approach when parameters effects are smoothly varying, but can produce issues in non-linear regimes (e.g., a sharp chemical abundance transition at a given atmospheric temperature and pressure). Interpolating the log fluxes instead of the linear fluxes did not improve the generated $R^2$ scores. A superior option to this approach would be the calculation of a finer model grid, which is generally restricted by the high computational cost of atmosphere modeling and/or the proprietary (non-open-source) status of the computer codes. Discussion of optimally-sampled model grids for exoplanet and brown dwarf spectra can be found in \cite{Fisher2022ApJ...934...31F}. 

Figure~\ref{fig:Effect_interpolated_grid} shows the same RvP plots as in Figure \ref{fig:Effect_of_grid_size} but with the enlarged grids. There is a pronounced improvement in $R^2$ scores, especially for the surface gravity. In the B+97 case, for example, the $R^2$ value for $\log{g}$ increases from 0.230 to 0.942. Figure ~\ref{fig:Effect_interpolated_grid} also shows that surface gravity interpolation also improves the internal consistency for effective temperature. As no new information is added in these interpolations, the most likely explanation for the improvement is that the larger grid provides the machine learning algorithm a larger training set for which to model the set of decision trees, making the overall random forest model more robust.

\section{Intercomparison of Model Grids}
\label{sect:Grid_comparison}

In the previous section, we explored self-consistency in individual model grids, and the importance of expanding grid sampling and sizes to improve random forest model training. Here, we examine intercomparisons between pairs of grids to highlight differences that arise from different modeling approaches and assumptions.

\subsection{Grid preparation}
\label{sect:Comparison_grid_prep}

The information content of spectra, which serve as our training set for the random forest, varies with spectral resolution. The spectral resolution of the model grids is heterogeneous and ranges from $\sim 10$ to $\sim 10^5$ (Table~\ref{tab:properties}).  When comparing a pair of model grids, we require them to share a common binning across wavelength.  In order to make a fair comparison between model grids and later transition to using these grids as training sets for interpreting measured spectra, we adopt the resolution of spectral data acquired with the prism mode of the SpeX spectrograph, mounted on the 3m Infrared Telescope Facility \citep{Rayner2003PASP}. Many brown dwarf observations have been conducted with this instrument, which covers a wavelength range of 0.85 $\,\mu$m to 2.45$\,\mu$m at a (slit-dependent) spectral resolution ranging from approximately 80 to about 300. In cases where the spectral resolution of the model grid exceeds that of the measured spectra, this process of binning down corresponds to a loss of information. This near-infrared region covers strong absorption bands of important molecules, such as \ch{H2O}, \ch{CO}, and \ch{CH4}, as well as spectral continuum variations shaped by collision-induced H$_2$ absorption and cloud opacity. We linearly interpolated all model grid spectra to match the wavelength range and resolution of SpeX prism spectra obtained with the 0$\farcs$8 slit, and retrained our grid random forest models using \texttt{HELA}. 

At the shortest wavelengths, SpeX spectra sample the blue line wings of the K~I resonant doublet centered at 0.77~$\mu$m. In brown dwarf spectra, these lines are heavily pressure-broadened, with broad, non-Lorentzian line wings that are challenging to model theoretically \citep{Tsuji1999ApJ...520L.119T, Burrows2000ApJ...531..438B, Burrows2003ApJ...583..985B, Allard2012EAS....57....3A, Allard2016A&A}. Indeed, the model grids in this study use different approaches to modeling these lines, and one expects that this part of the spectrum will show greater variance when comparing across model grids. We therefore included a second modification to our model grids, following \citet{Oreshenko2020AJ}, in which we removed data at wavelengths $\leq$1.2$\,\mu$m, excising the strongest impact of the alkali line wings. A second set of random forests models were computed with these ``cut'' model grids using \texttt{HELA}. In the Appendix, Figure 13 shows examples of mean tree depths for random forest calculations with full versus restricted wavelength ranges.  For reasons that are not easy to diagnose, the runs associated with spectra with restricted wavelength ranges (cut off blueward of 1.2$\,\mu$m) have slightly larger tree depths.

With the grid models consistently mapped onto the same wavelength range and resolution, we evaluated parameter agreement between them by training on one grid model and testing on a second. For 14 grids, this amounts to 182 distinct comparisons, as swapping between training and testing grids constitutes independent tests. We therefore present here an analysis of a subset of 24 representative cases.

\subsection{Accounting for parameter space coverage}
\label{sect:Parameter_coverage}

\begin{figure} [ht!]
\gridline{\fig{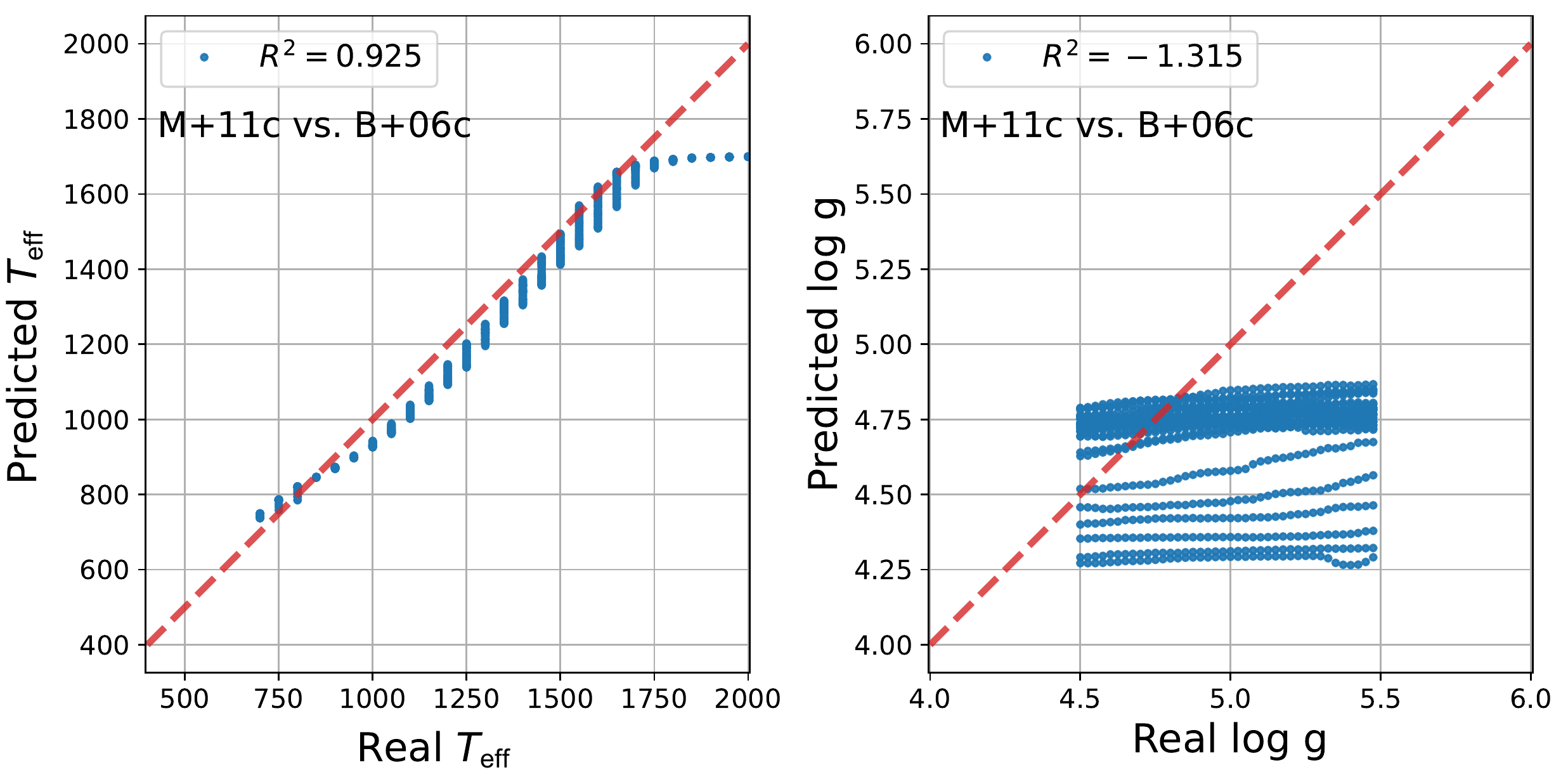}{0.99\columnwidth}{No restrictions}}
\gridline{\fig{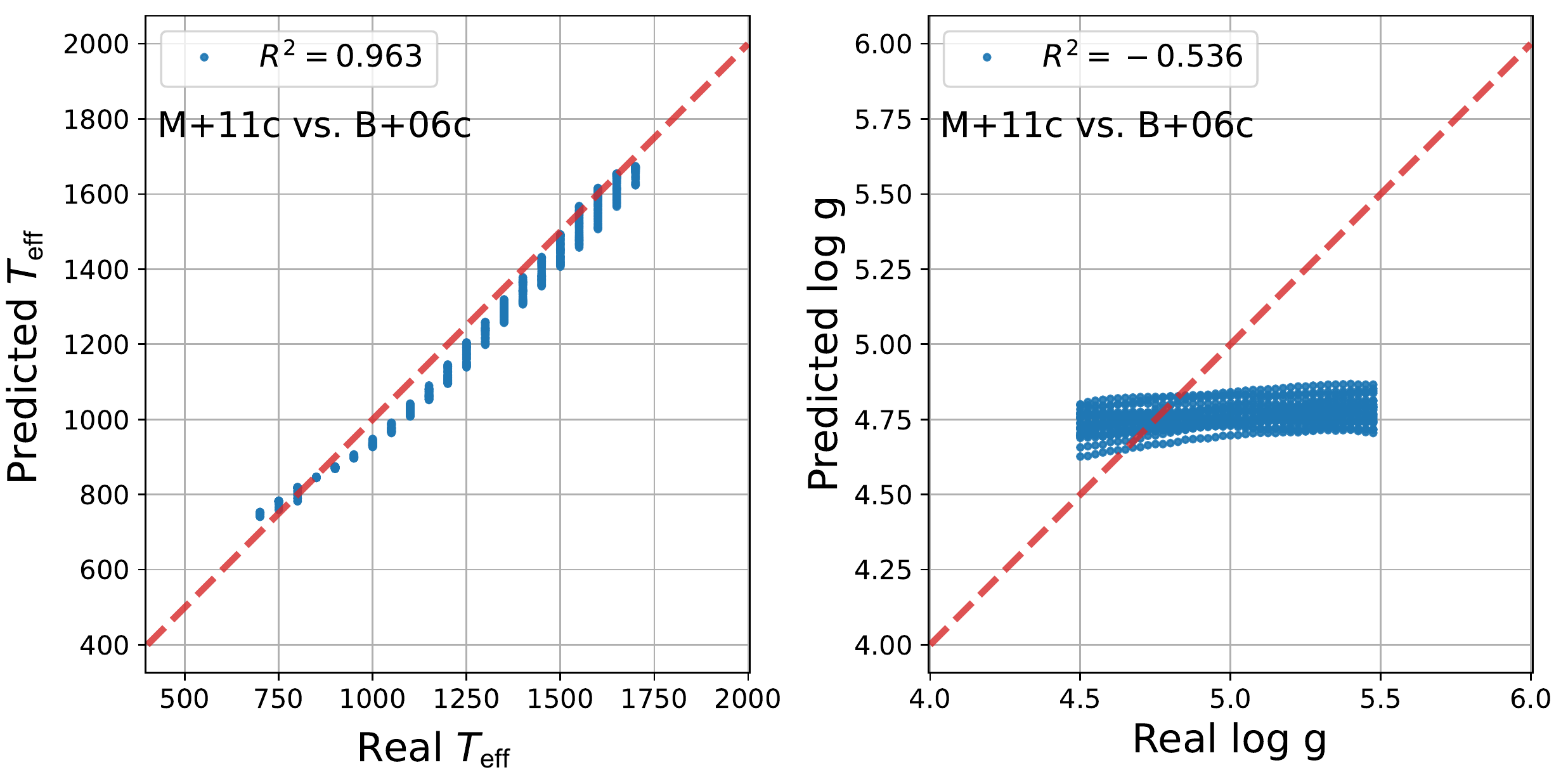}{0.99\columnwidth}{$700 \leq T \leq 1700$K}}
\caption{RvP comparison for M+11c versus B+06c grids, focusing on the influence of effective temperature range. The top panels show the RvPs when the full range of models are trained and tested; the bottom panels show the RvPs when models are restricted to overlapping effective temperatures. No noise is assumed.}
\label{fig:Effect_of_Teff}          
\end{figure}

Before proceeding to a full analysis of grid comparison, one aspect of the fits that needs to be accounted for is the differing ranges of parameters encompassed by the models, as listed in Table~\ref{tab:properties}. The influence of this can be seen in the RvP plots shown in Figure~\ref{fig:Effect_of_Teff} comparing the M+11c and the B+06c grids. The effective temperature RvP plot in particular displays a deviation from perfect agreement at $T_\mathrm{eff} \approx$~1700~K, saturating at this value in the predictions and creating an ``S-shaped'' curve. This deviation is a consequence of the limited temperature range of the M+11c grid, which only extends to $T_\mathrm{eff}$~=~1700~K, whereas the B+06c grid extends to $T_\mathrm{eff}$~=~2000~K. Since the random forest model has been trained on the M+11c models, it cannot correctly predict $T_\mathrm{eff}$ values higher than the 1700~K limit.

Note that the RvP plots of brown dwarf surface gravity often exhibit lines corresponding to different effective temperatures. A positive correlation between effective temperature and predicted surface gravity is typically observed, with higher effective temperatures showing higher predicted surface gravity values. The evolution of surface gravity and effective temperature in brown dwarfs results from the cooling and contracting of the interior. However, the decline in surface gravity is not consistently uniform, leading to a wide range in predicted values.
If we instead restrict the testing of the M+11c random forest model with B+06c models that are in the same effective temperature range (700~K to 1700~K), the most deviant comparisons in both temperature and surface gravity are eliminated, improving the overall $R^2$ scores for both parameters. Therefore, for the subsequent analysis we compare models only in the effective temperature and surface gravity ranges that encompass both model sets. Note that in many cases this limits our surface gravity comparison to the range 4.5 $\leq \log{g} \leq$ 5.5 based on the B+06c, H+07c, and M+11 models.

\subsection{Comparison results}
\label{sect:intercomparison}

With the constraints described above, Figure~\ref{fig:Effect_of_wavelength_coverage} shows RvP comparisons for $T_\mathrm{eff}$ and $\log{g}$ for our 24 selected cases. We include comparison of both the full SpeX wavelength range, and for spectra cut below 1.2$\,\mu$m. Figure \ref{fig:R2_Grid_Comparisons} summarizes the $R^2$ scores for the cut spectra. These figures illustrate very different behaviors in the random forest retrieval in terms of predicting effective temperatures and surface gravities. For most cases, the former are predicted accurately across model grids, with $R^2$ scores of 0.9 and above. 

Much larger and more consistent discrepancies are found in the predictions of surface gravities between the grids. The RvP curves for these are often shifted well off of the perfect agreement line, producing highly negative $R^2$ scores (cf.\ M+11c versus B+06c or M+11c versus H+07c cases). These deviations suggest that the different modeling approaches produce distinct spectral variations with respect to changes in surface gravity. Alternately, this behavior may indicate that the spectra in some grids have differing sensitivities to surface gravity variations, resulting in different degrees of gravity segregation.

Part of the surface gravity disagreements could be caused by differing treatments of the alkali lines, which are pressure-sensitive and thus influenced by atmospheric surface gravity. Comparing between random forest models trained on full versus restricted wavelength ranges, we find improvement in surface gravity inference in several cases. For example in the M+21 vs H+07 comparison, the $R^2$ score for $\log{g}$ increases from -3.670 to -0.067 between full and restricted spectra. While the latter is far from indicating good agreement, it does indicate improvement. However, not all cases yield better surface gravity predictions for the restricted wavelength range. For example, in the K+21 versus H+07 comparison, the $R^2$ score decreases from 0.206 to -1.348 when the spectrum is cut at 1.2 $\mu$m as surface gravities are systematically more underestimated. The $R^2$ scores for effective temperature, on the other hand, barely change when the cut spectrum is used, suggesting that differences in alkali line treatment are not essential for determining this parameter.

The grid comparisons also demonstrate how modifications to assumptions in the atmosphere model calculations manifest in temperature and surface gravity inference. Not surprisingly, grids that are produced by the same computational framework (e.g., \texttt{COOLTLUSTY}; see M+11 versus H+07 comparison) agree better than those computed with differing frameworks. Nevertheless, significant modifications to the modeling assumptions can result in large differences in inferred parameters even within the same computational framework. An illustrative case is the M+21 and K+21 grid comparison. Both models use the \texttt{Sonora} atmosphere model code, but the former assumes chemical equilibrium while the latter assumes disequilibrium chemistry.

While this change has minimal effect on the inferred effective temperatures, there are significant differences in the inferred surface gravities, with $R^2$ = -0.373 for the full wavelength range and $R^2$ = 0.474 for the restricted wavelength range. This deviation suggests that chemistry considerations may be degenerate with surface gravity variations, and that both may need to be considered in a full retrieval.

\begin{figure*}[ht!]
\centering
\gridline{\fig{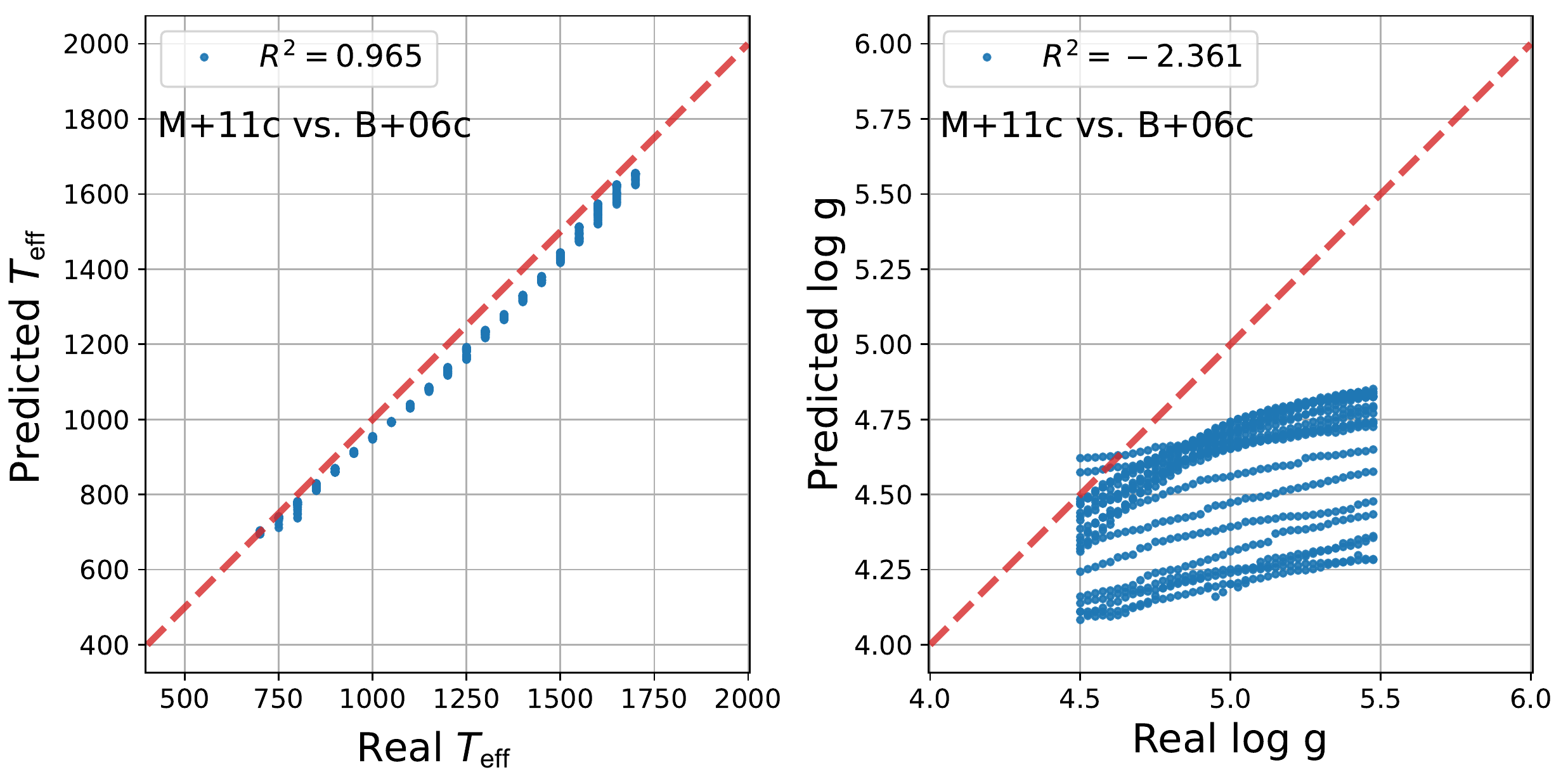}{0.4\textwidth}{M+11c vs B+06c: No cut}
\fig{RvP_comparisons/RvP_M11cvsB06c_short_teffcut.pdf}{0.4\textwidth}{M+11c vs B+06c: Cut at 1.2$\,\mu$m}}
\gridline{\fig{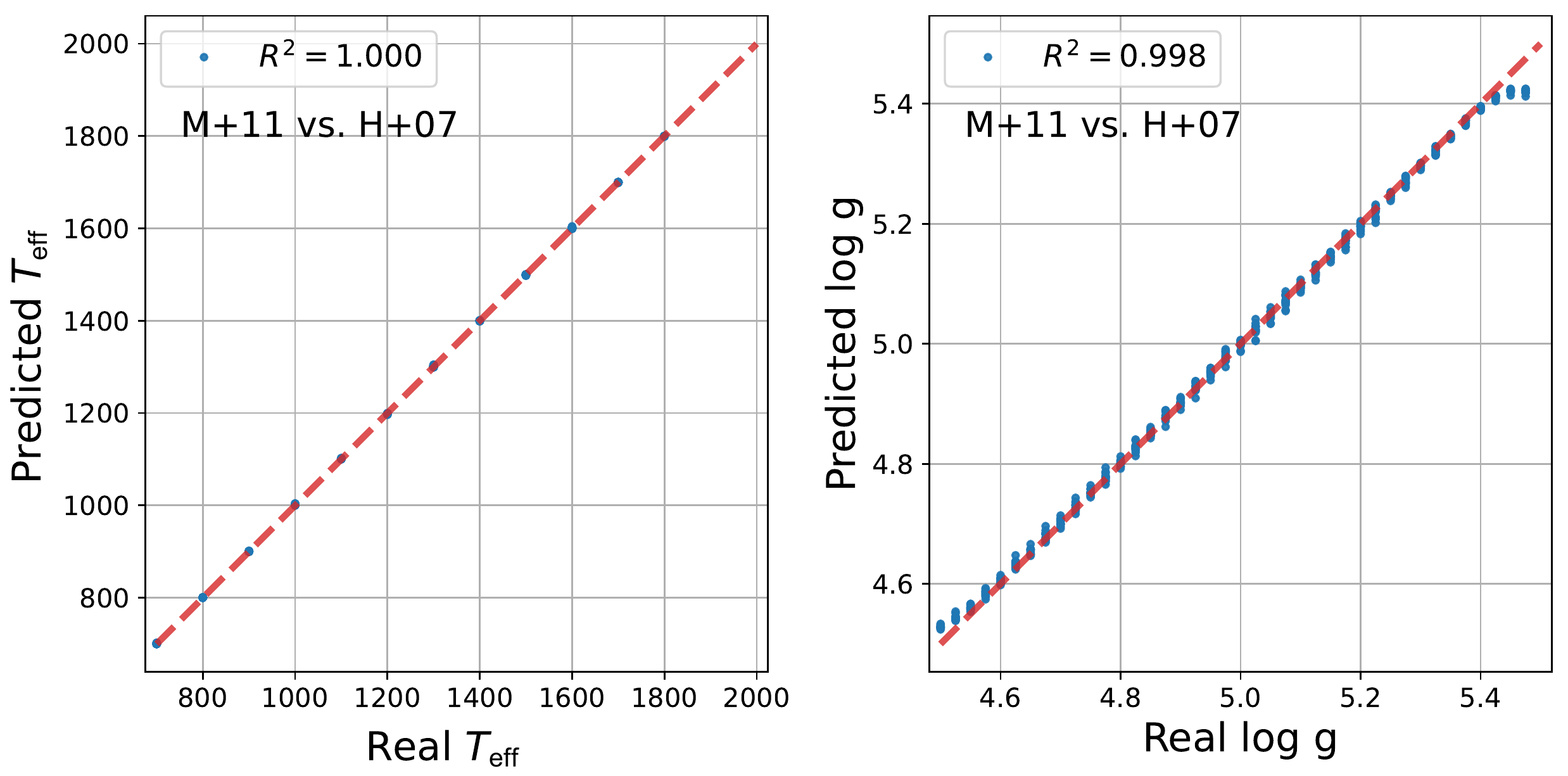}{0.4\textwidth}{M+11 vs H+07: No cut}
\fig{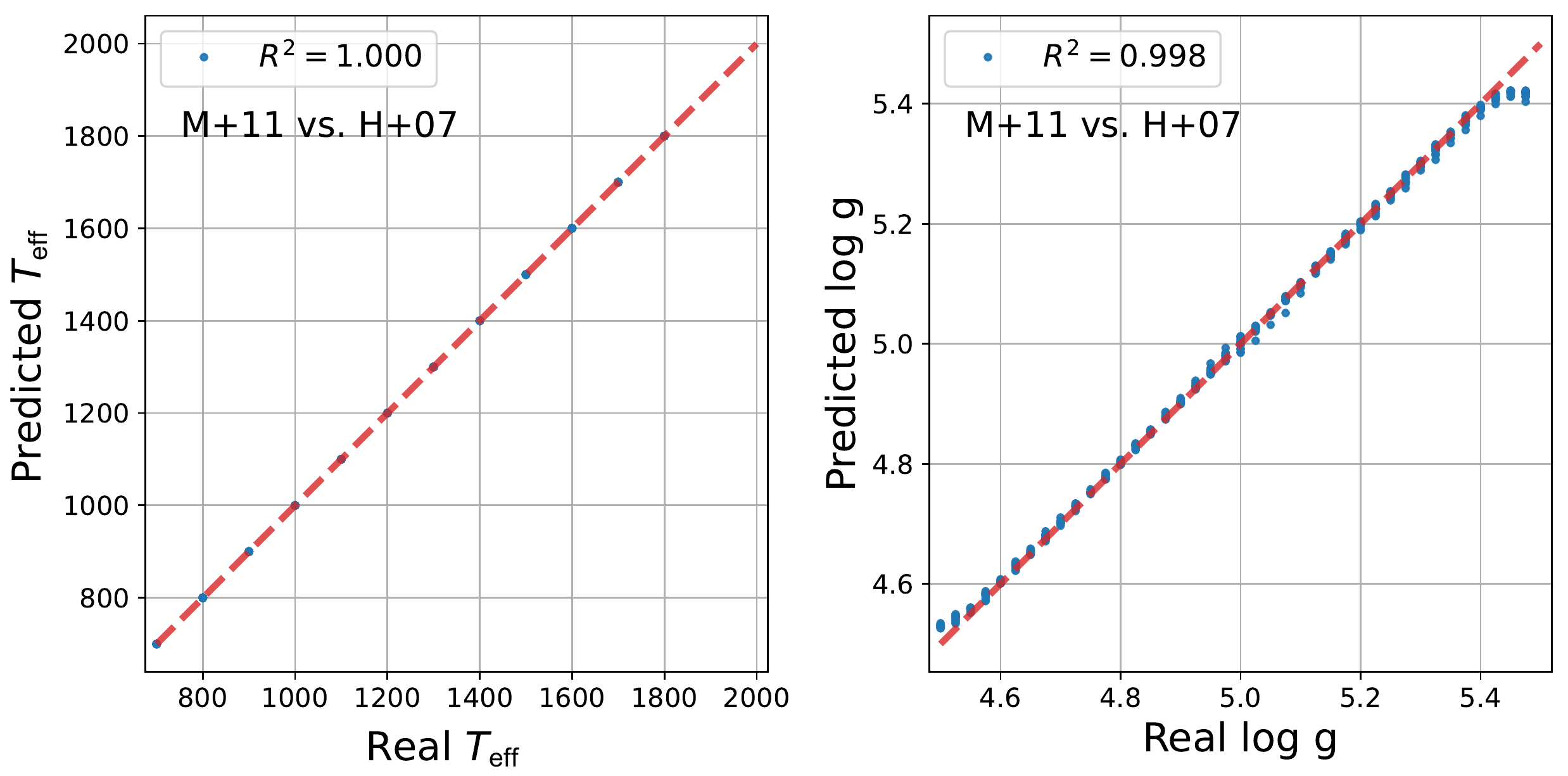}{0.4\textwidth}{M+11 vs H+07: Cut at 1.2$\,\mu$m}}
\gridline{\fig{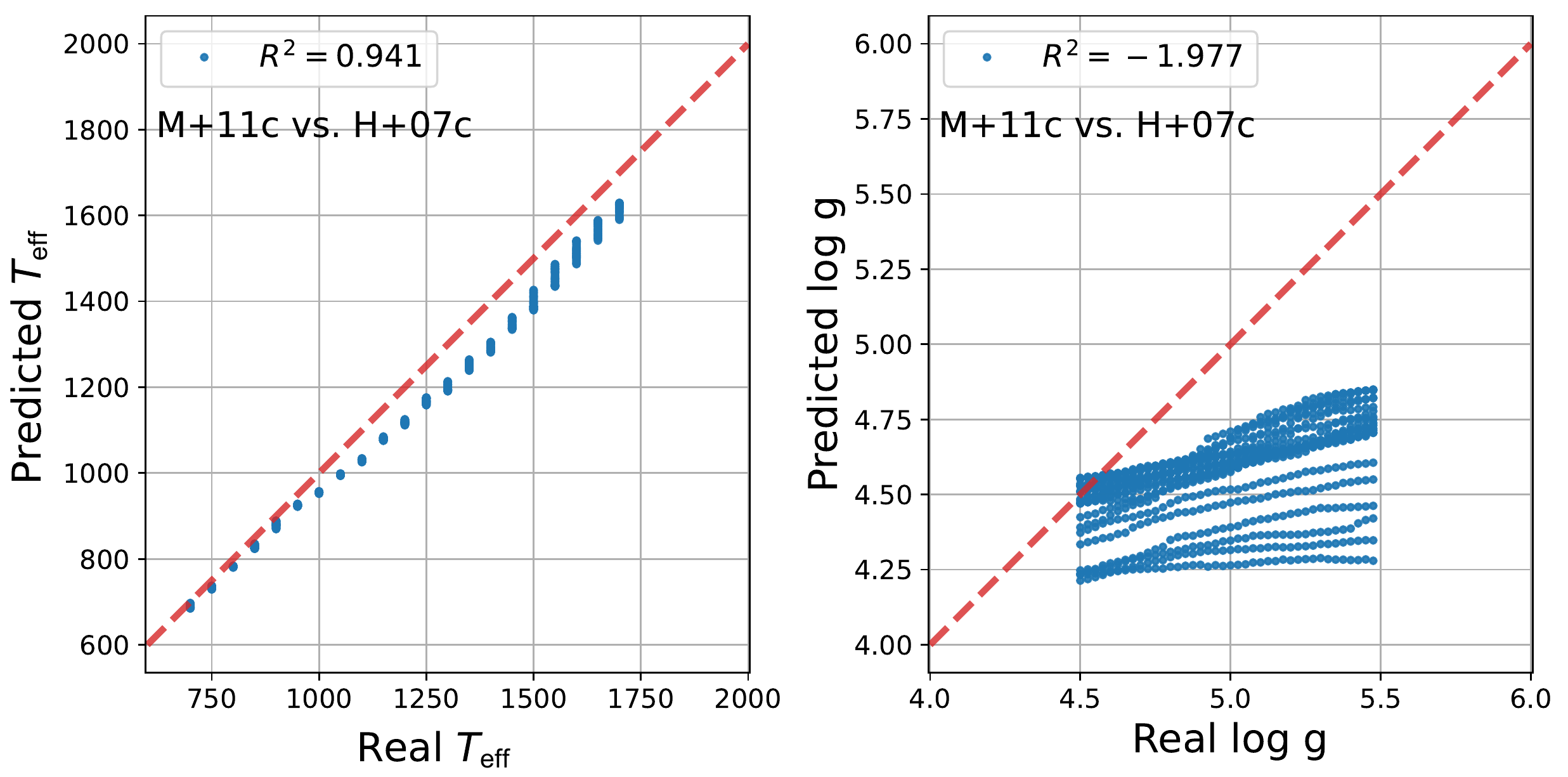}{0.4\textwidth}{M+11c vs H+07c: No cut}
\fig{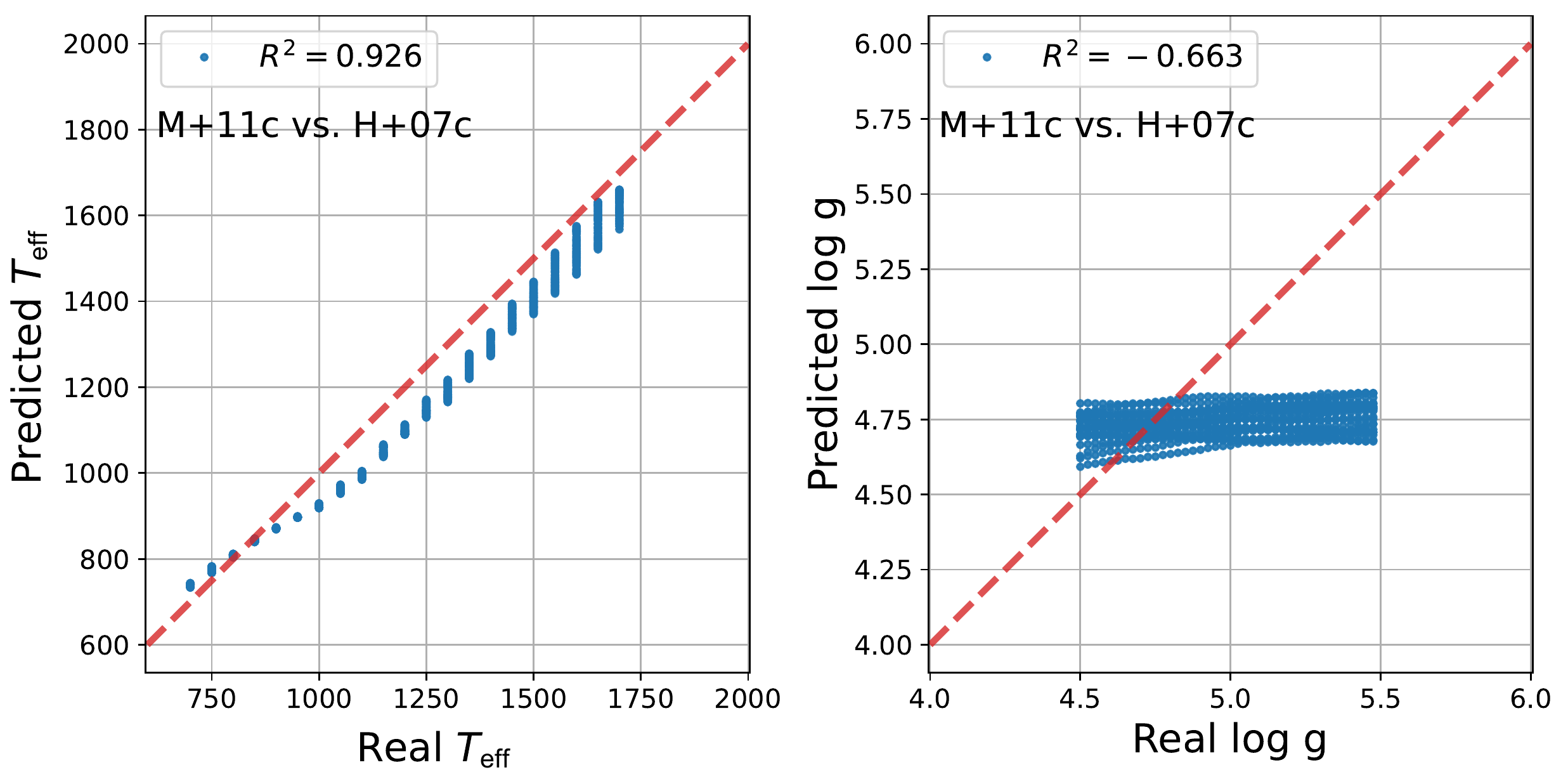}{0.4\textwidth}{M+11c vs H+07c: Cut at 1.2$\,\mu$m}}
\gridline{\fig{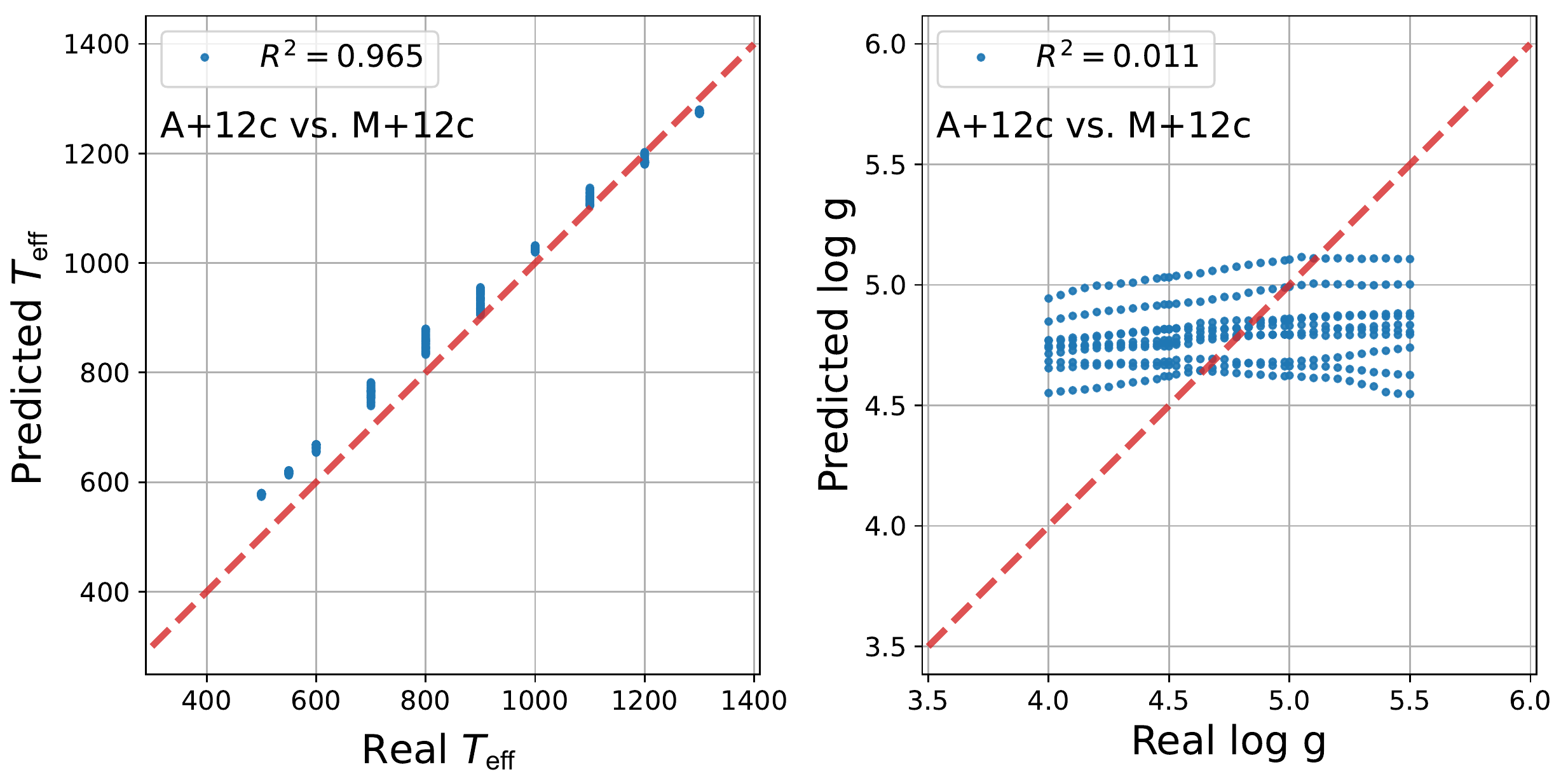}{0.4\textwidth}{A+12c vs M+12c: No cut}
\fig{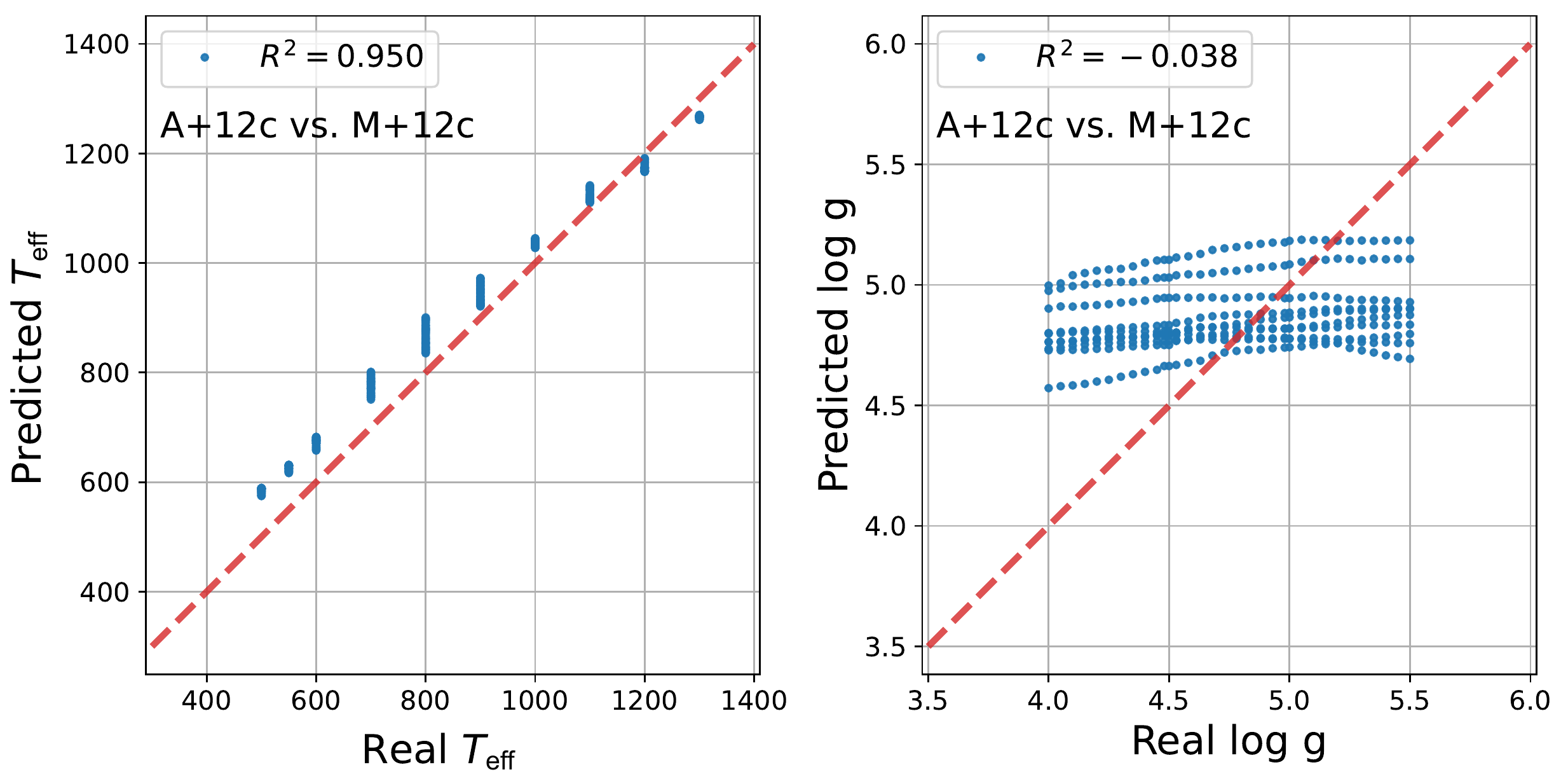}{0.4\textwidth}{A+12c vs M+12c: Cut at 1.2$\,\mu$m}}
\caption{Real versus predicted (RvP) comparison for temperature ($T_\mathrm{eff}$) and surface gravity ($\log{g}$) in the case of training on one model grid (first model listed) and testing on a second model grid (second model listed). The leftmost two columns of the figures display analyses for the full spectral range, the rightmost two columns of the figures display analyses for a spectral range $\geq$1.2~$\mu$m. In each figure, the red dashed line indicates perfect agreement. Here we show analysis of the model grid pairs 
M+11c versus B+06c,
M+11 versus H+07,
M+11c versus H+07c, and
A+12c versus M+12c.
No noise is assumed.
The complete figure set (48 images) for the $T_\mathrm{eff}$ and $\log{g}$ RvP comparisons of both the full SpeX wavelength range, and for spectra cut below 1.2$\,\mu$m for our 24 selected cases is available in the online journal.
}
\label{fig:Effect_of_wavelength_coverage}          
\end{figure*}

\begin{figure*}[ht!]
\gridline{\fig{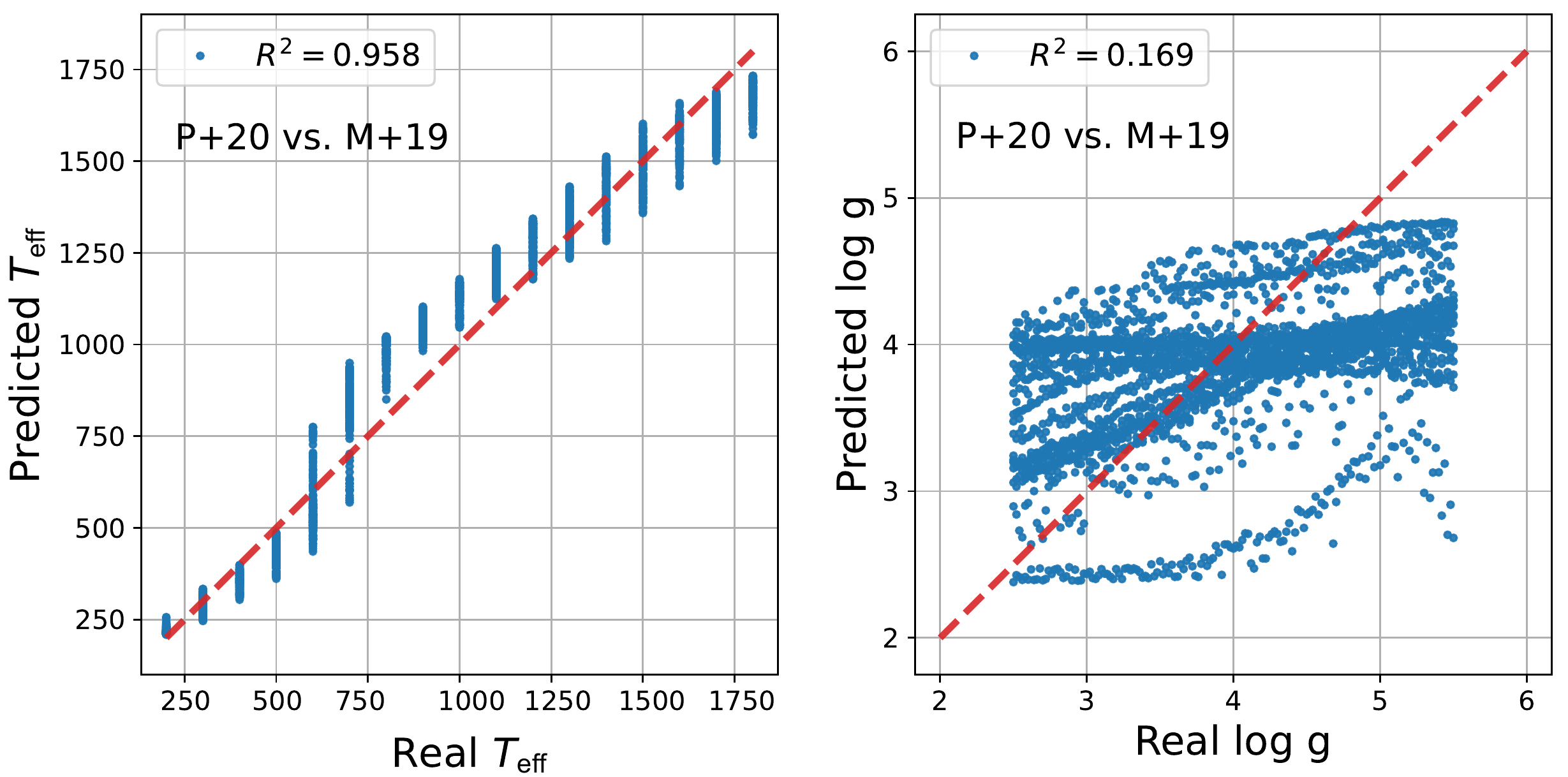}{0.4\textwidth}{P+20 vs M+19: No cut}
\fig{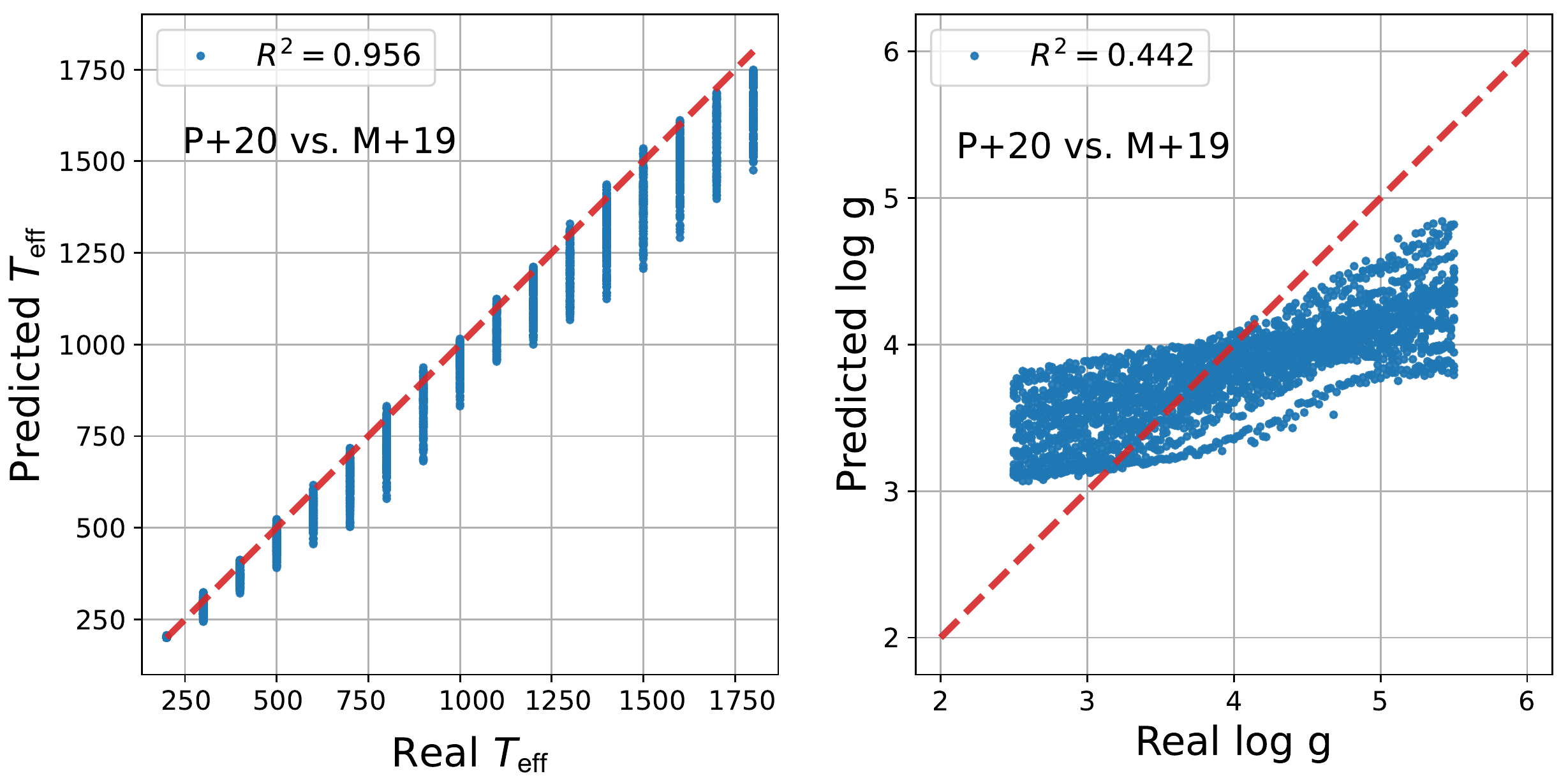}{0.4\textwidth}{P+20 vs M+19: Cut at 1.2$\,\mu$m}}
\gridline{\fig{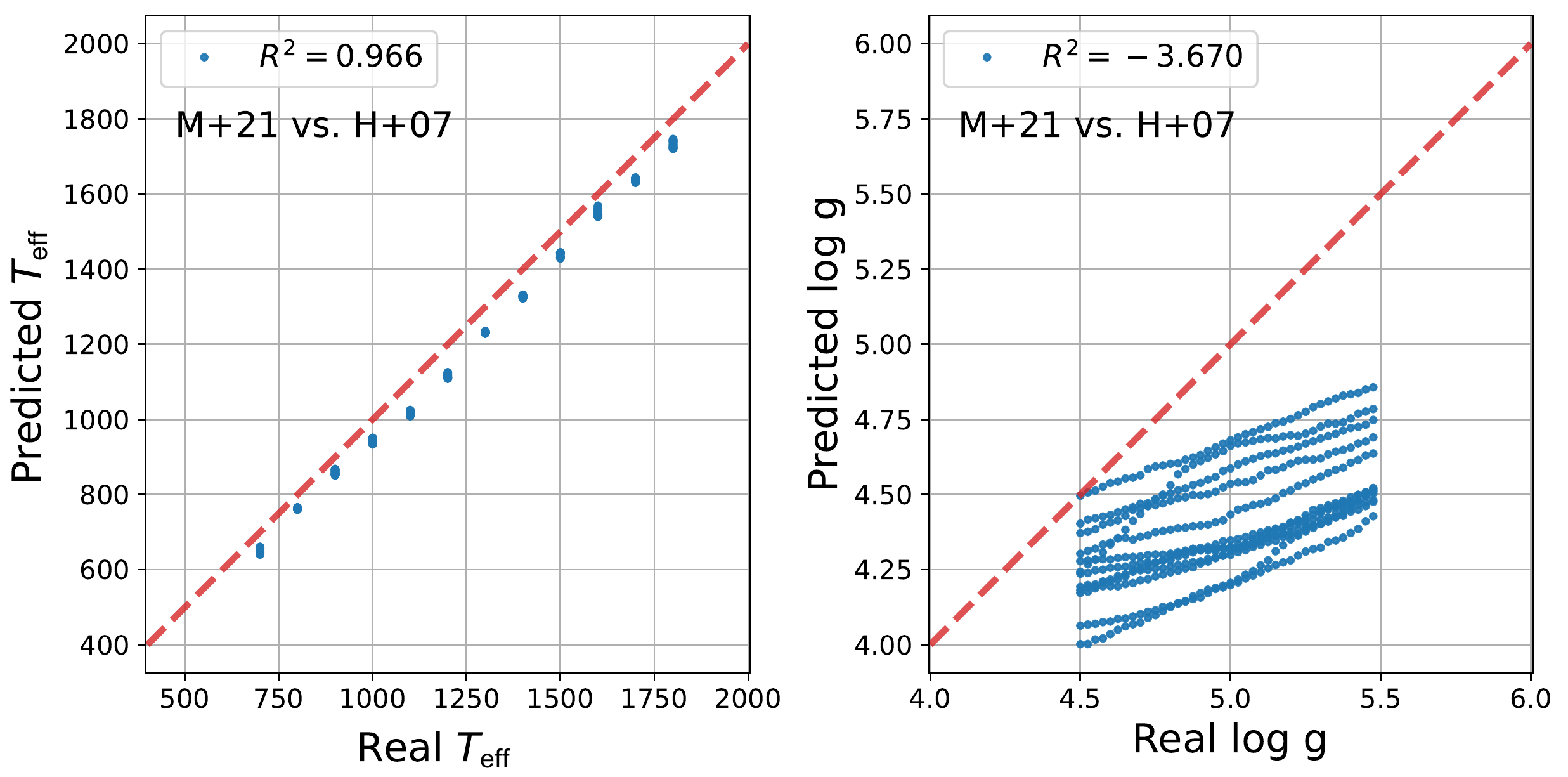}{0.4\textwidth}{M+21 vs H+07: No cut}
\fig{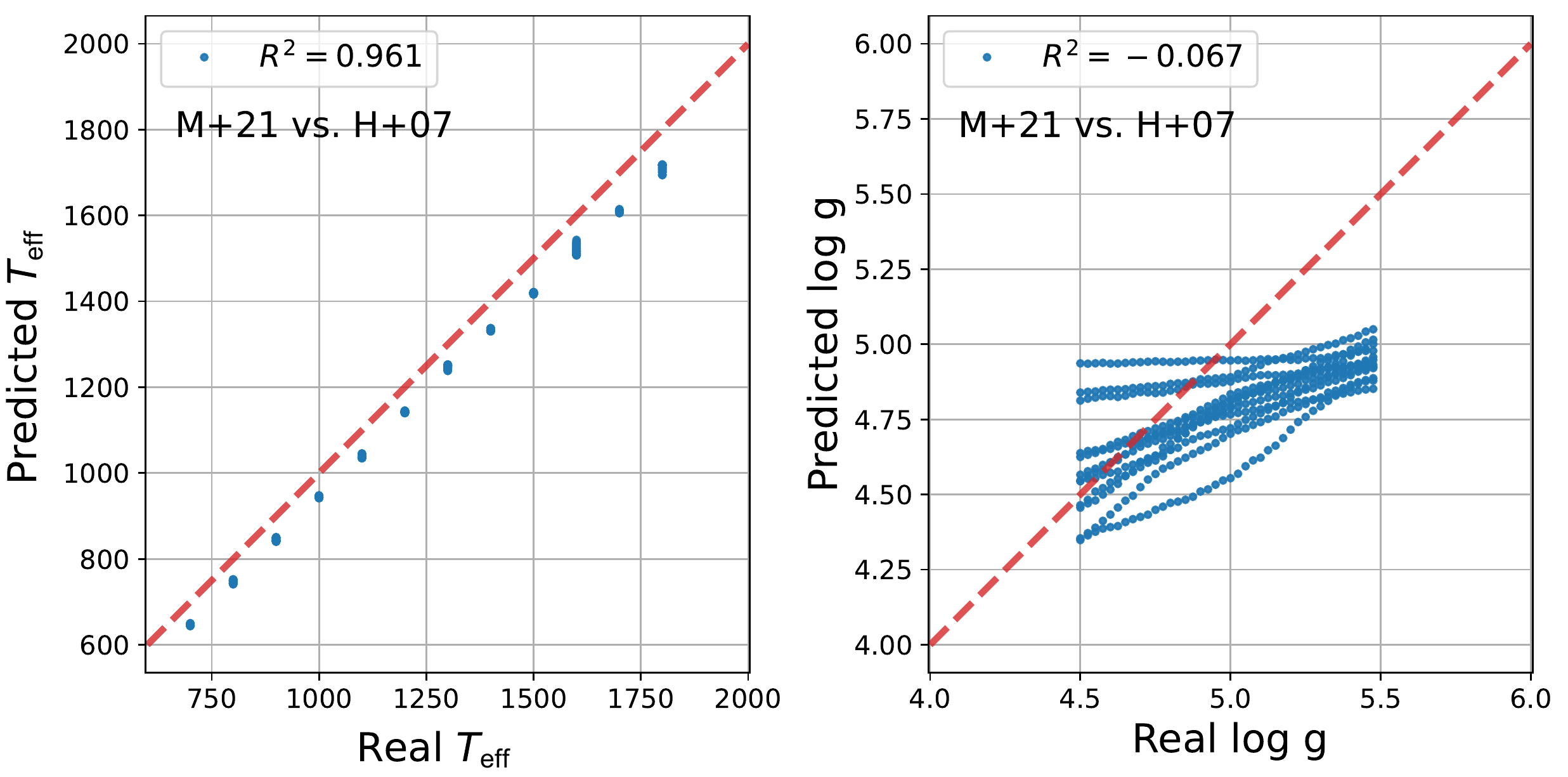}{0.4\textwidth}{M+21 vs H+07: Cut at 1.2$\,\mu$m}}
\gridline{\fig{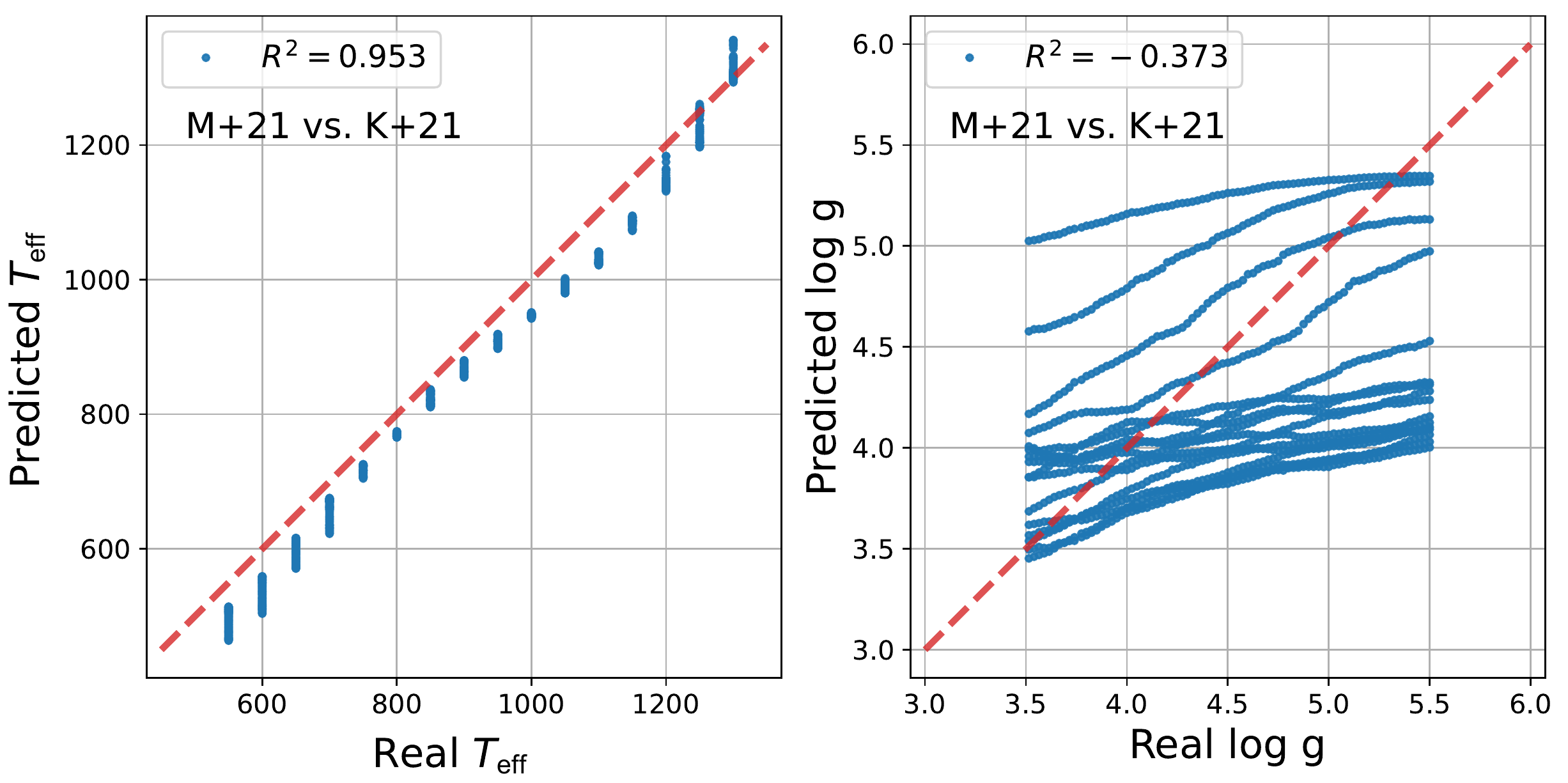}{0.4\textwidth}{M+21 vs K+21: No cut}
\fig{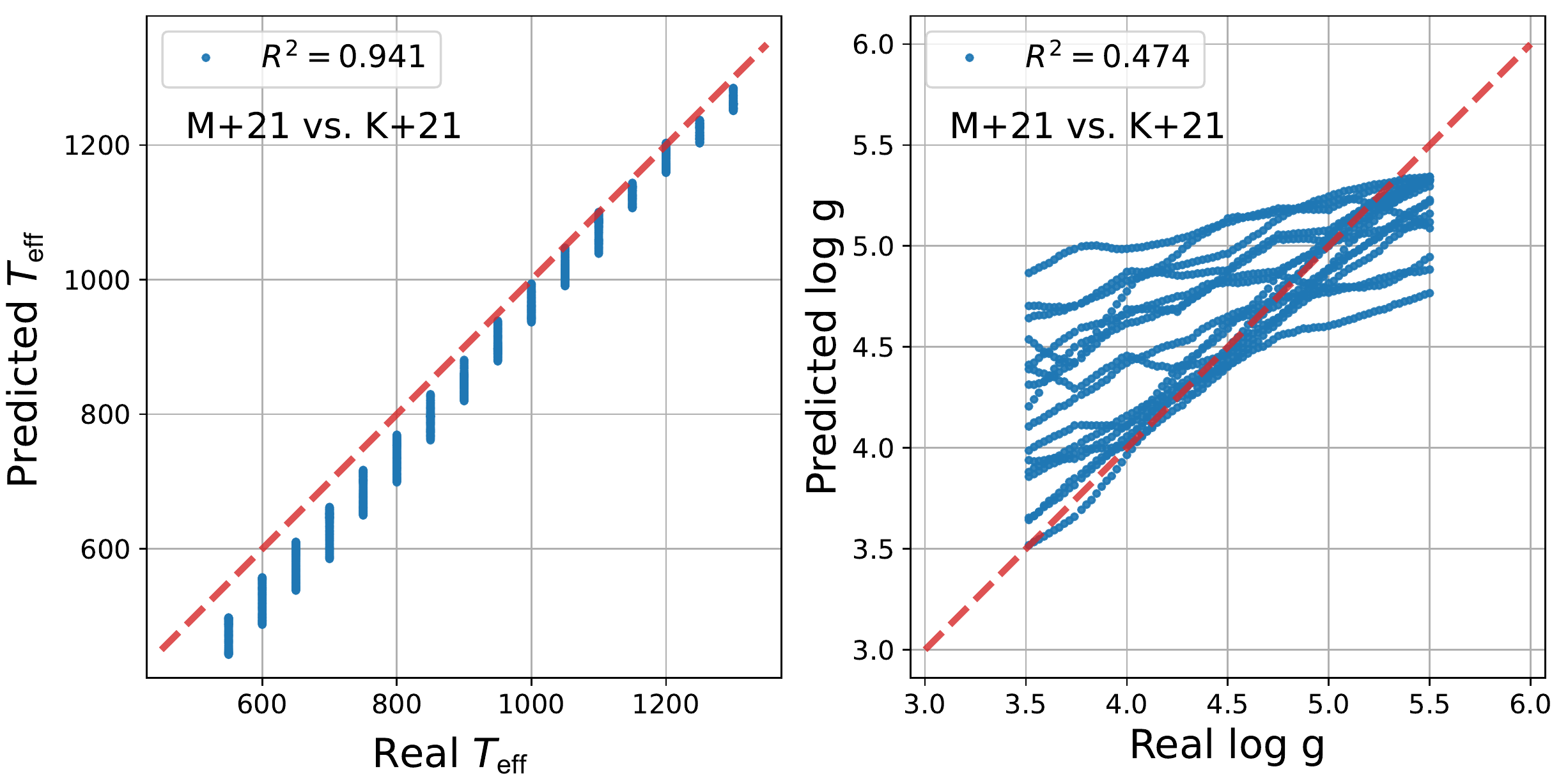}{0.4\textwidth}{M+21 vs K+21: Cut at 1.2$\,\mu$m}}
\gridline{\fig{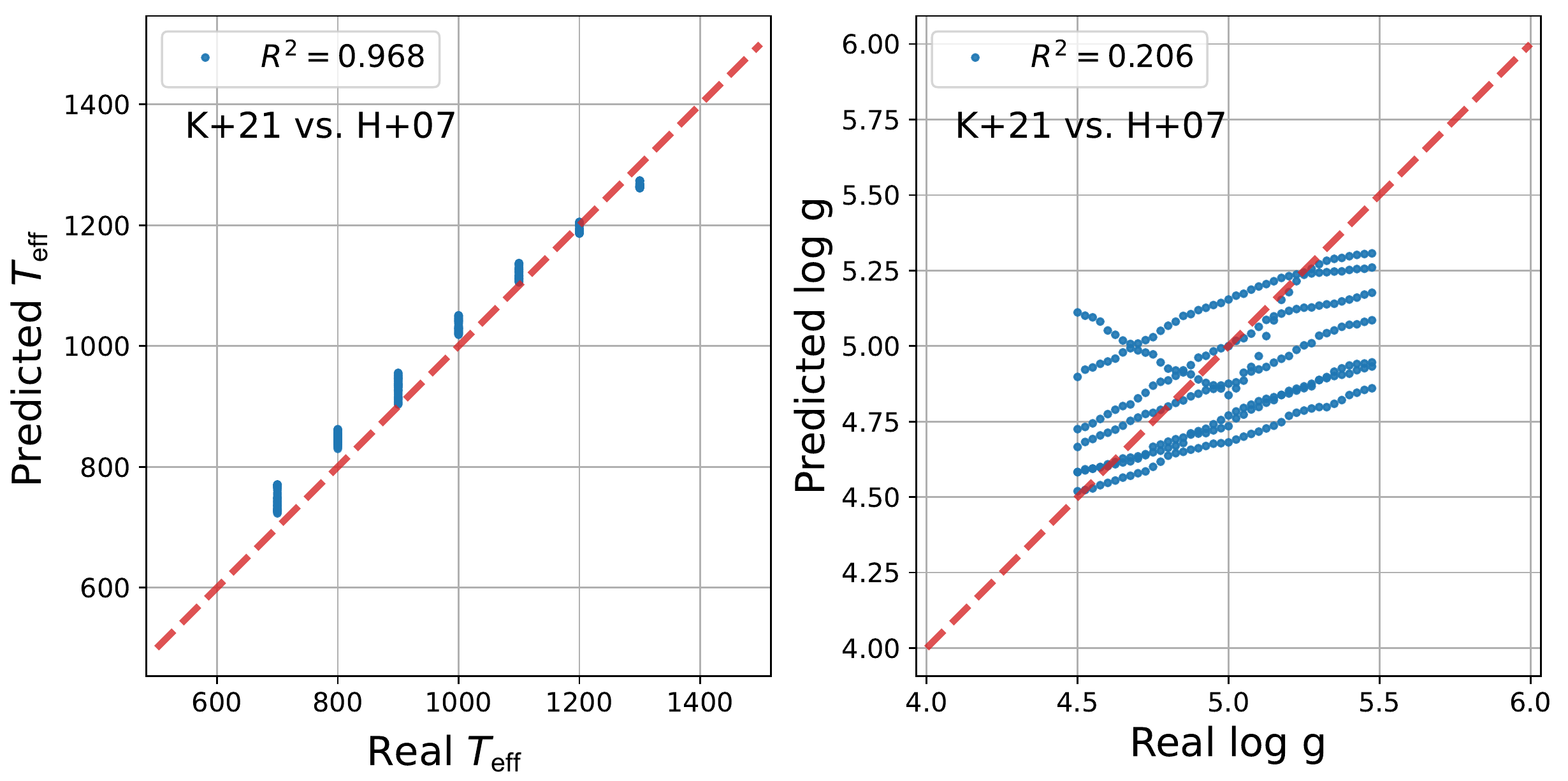}{0.4\textwidth}{K+21 vs H+07: No cut}
\fig{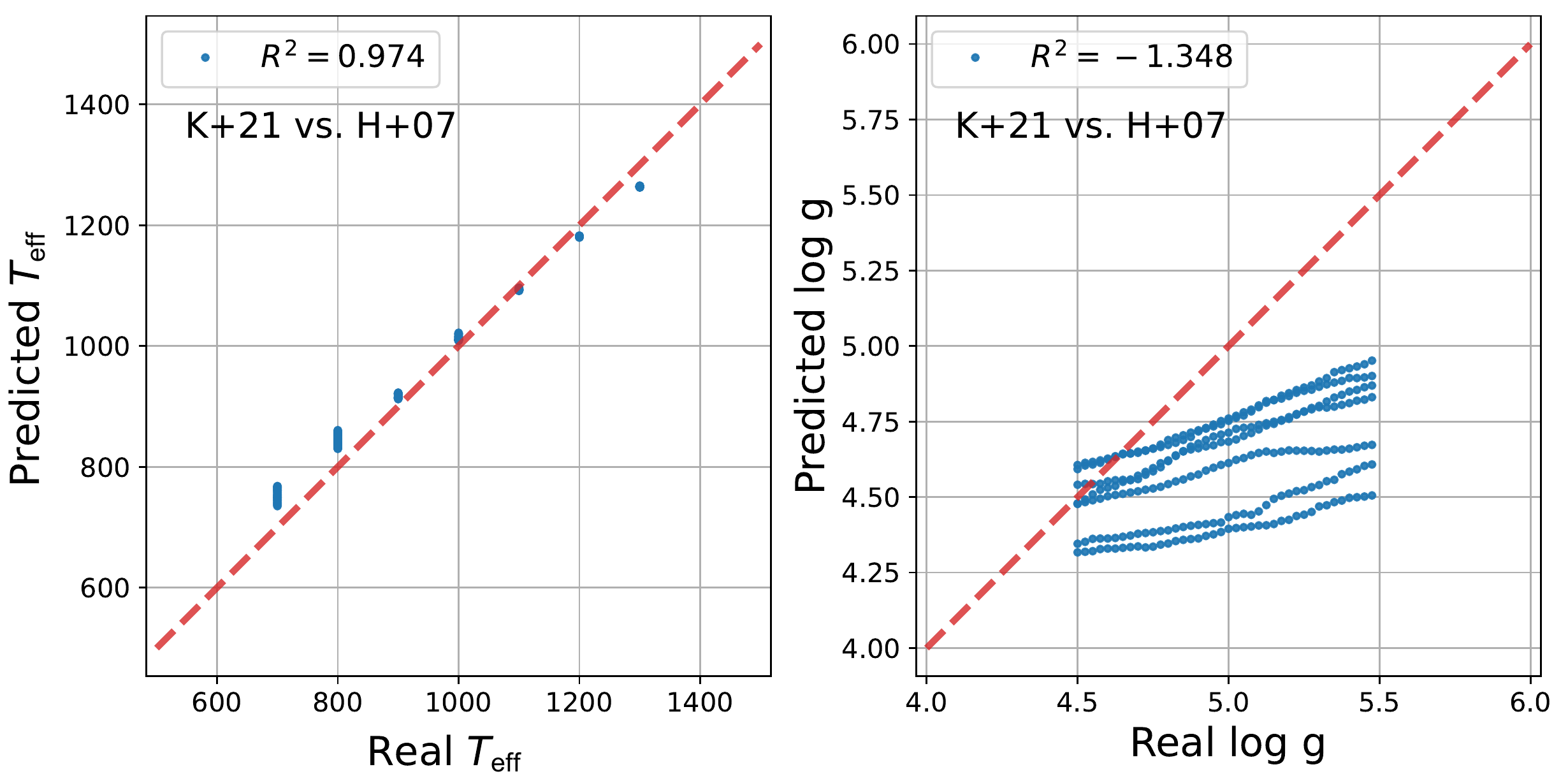}{0.4\textwidth}{K+21 vs H+07: Cut at 1.2$\,\mu$m}}
\figurenum{\ref{fig:Effect_of_wavelength_coverage}}
\caption{continued, for 
model grid pairs 
P+20 versus M+19,
M+21 versus H+07,
M+21 versus K+21, and
K+21 versus H+07.
No noise is assumed.}  
\end{figure*}

\begin{figure}[ht!]
    \centering
    \includegraphics[width=\columnwidth]{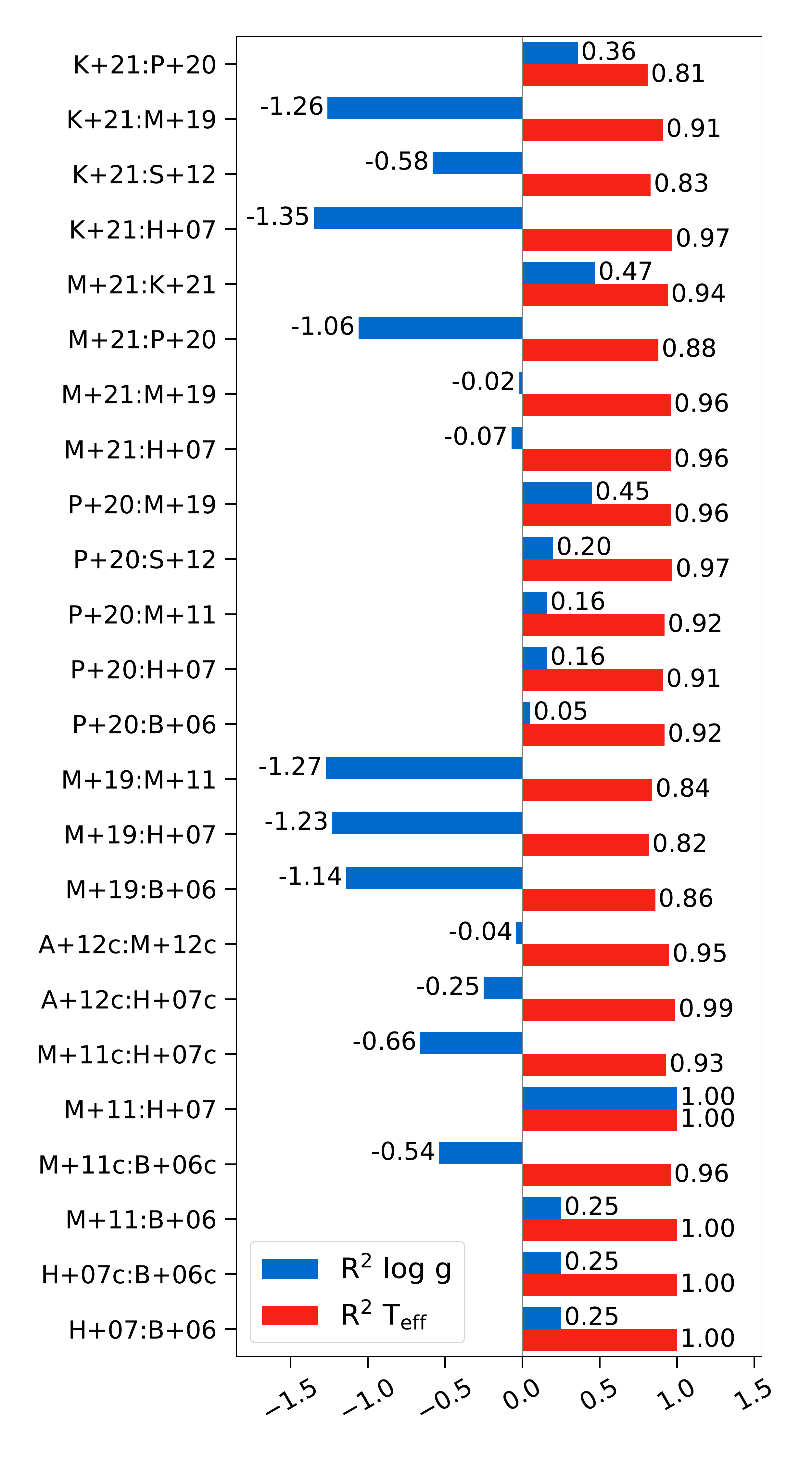}
    \caption{$R^2$ scores for $\log{g}$ (blue bands) and $T_\mathrm{eff}$ (red bands) for random forests trained on the first model listed and tested on the second model listed. Values are shown for the restricted wavelength range of 1.2~$\mu$m to 2.45~$\mu$m. No noise is assumed.}
    \label{fig:R2_Grid_Comparisons}
\end{figure}

\subsection{Feature importance}
\label{sect:Feature_importance}

Decomposing how different wavelength regions may influence parameter determination, such as alkali effects on surface gravity inference, can be accomplished by evaluation of feature importance plots. Figure \ref{fig:Effect_of_wavelength_range} shows these plots for all grids, including those not shown in Figure \ref{fig:Effect_of_wavelength_coverage}, based on the full SpeX wavelength range.

\begin{figure*}[ht!]
\centering
\includegraphics[width=0.8\textwidth]{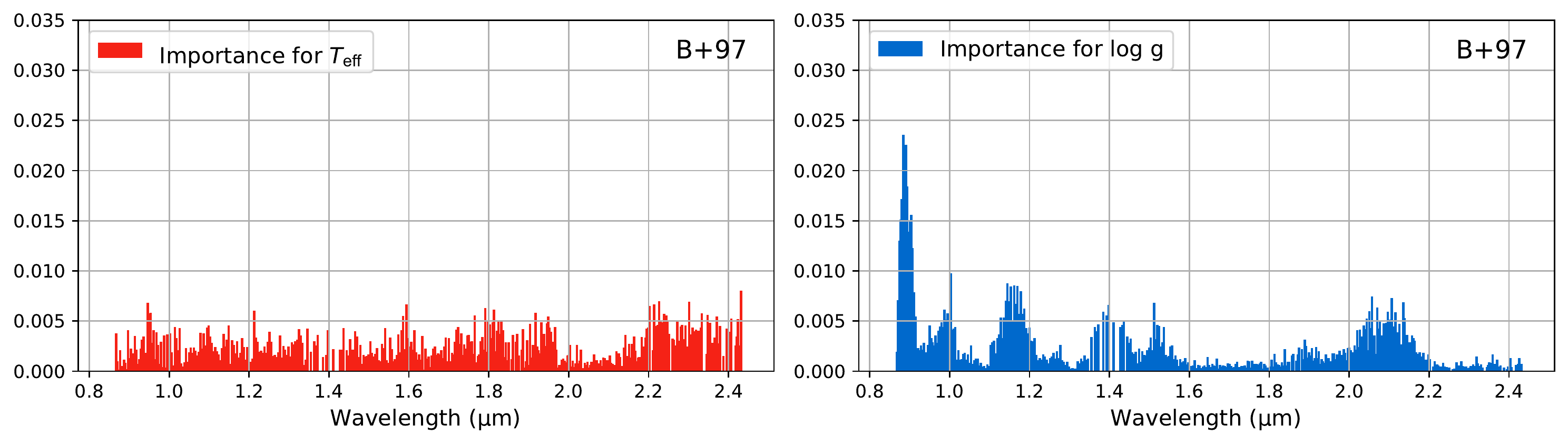}
\includegraphics[width=0.8\textwidth]{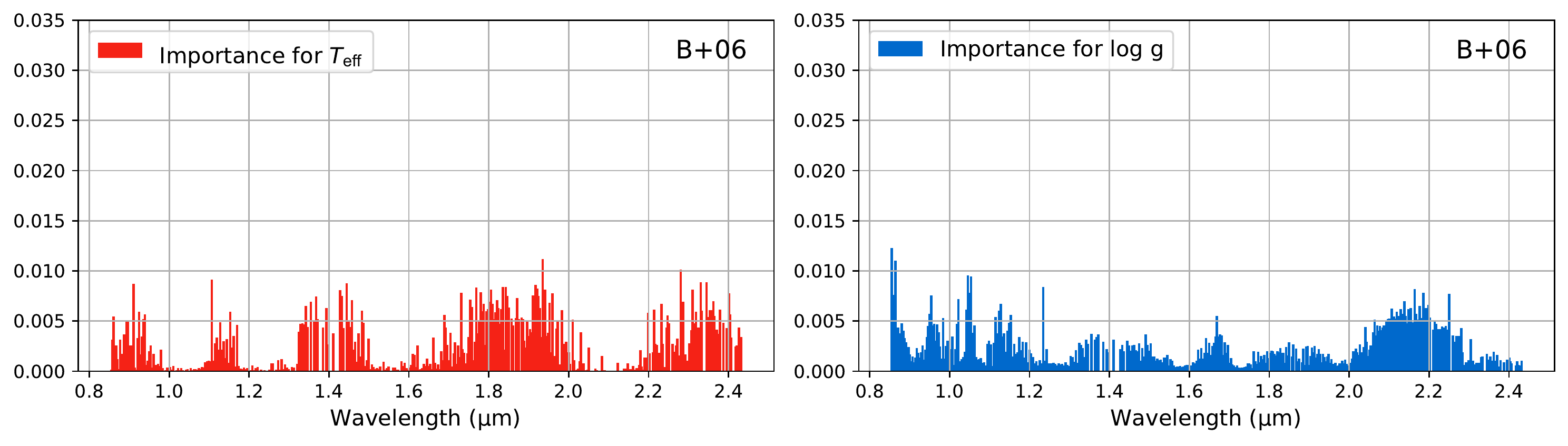}
\includegraphics[width=0.8\textwidth]{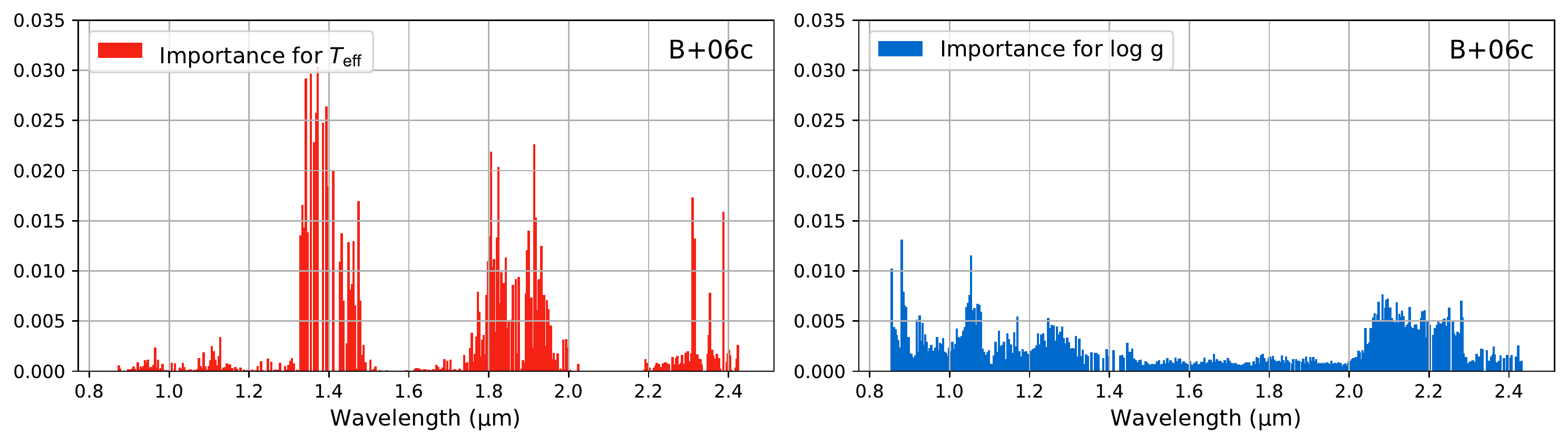}
\includegraphics[width=0.8\textwidth]{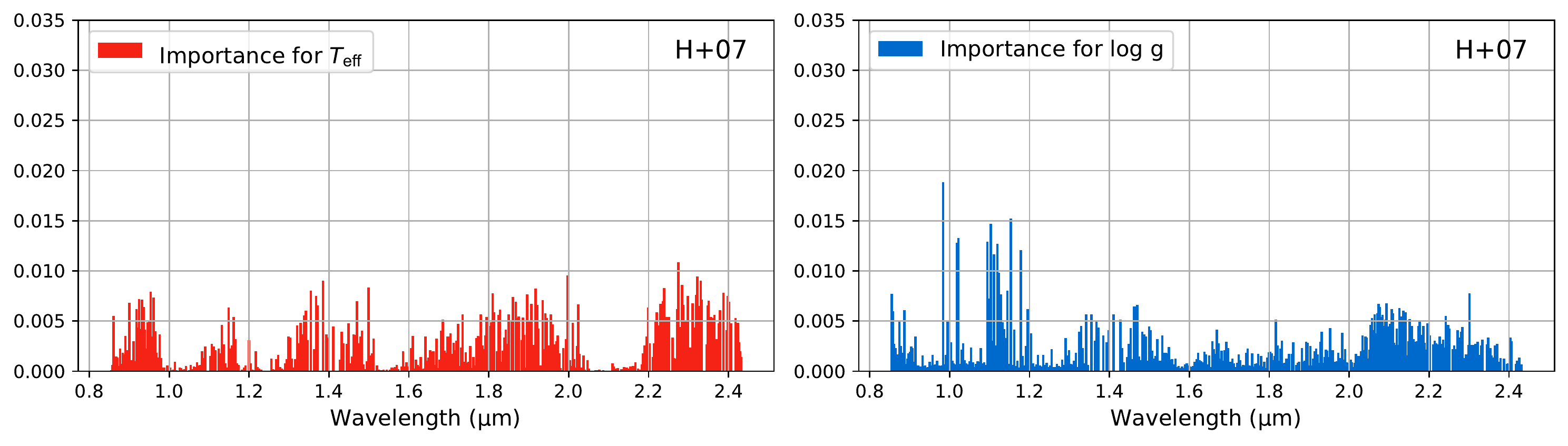}
\includegraphics[width=0.8\textwidth]{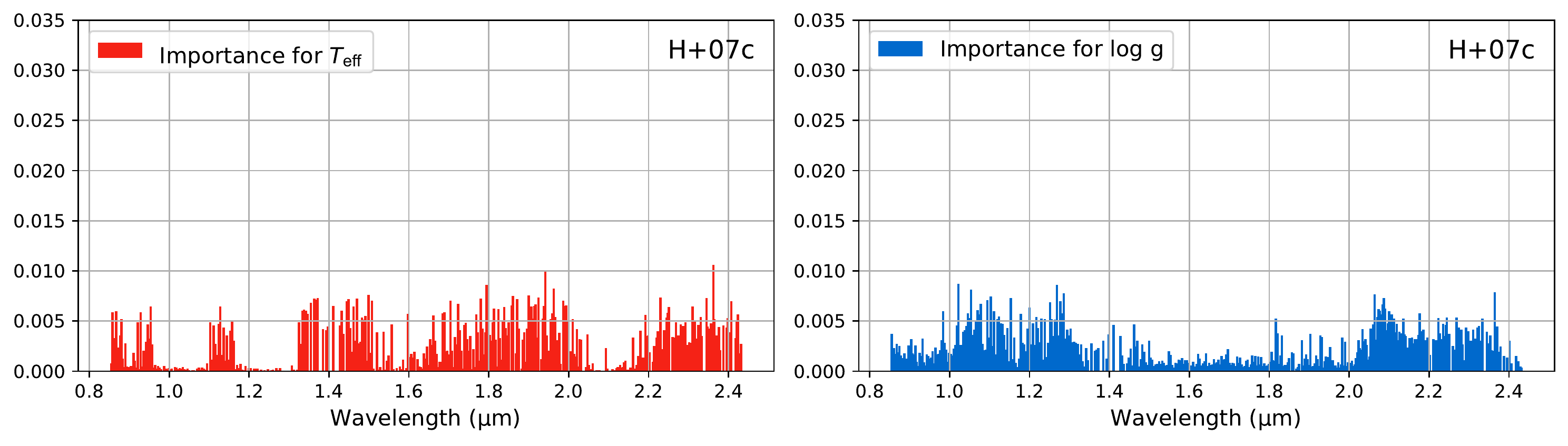}
\caption{Feature importance plots are shown for the temperature $T_{\rm eff}$ (red) and the surface gravity $\log{g}$ (blue). All 14 different grids are shown. No noise is assumed.}
\label{fig:Effect_of_wavelength_range}          
\end{figure*}

\begin{figure*}[ht!]
\centering
\includegraphics[width=0.8\textwidth]{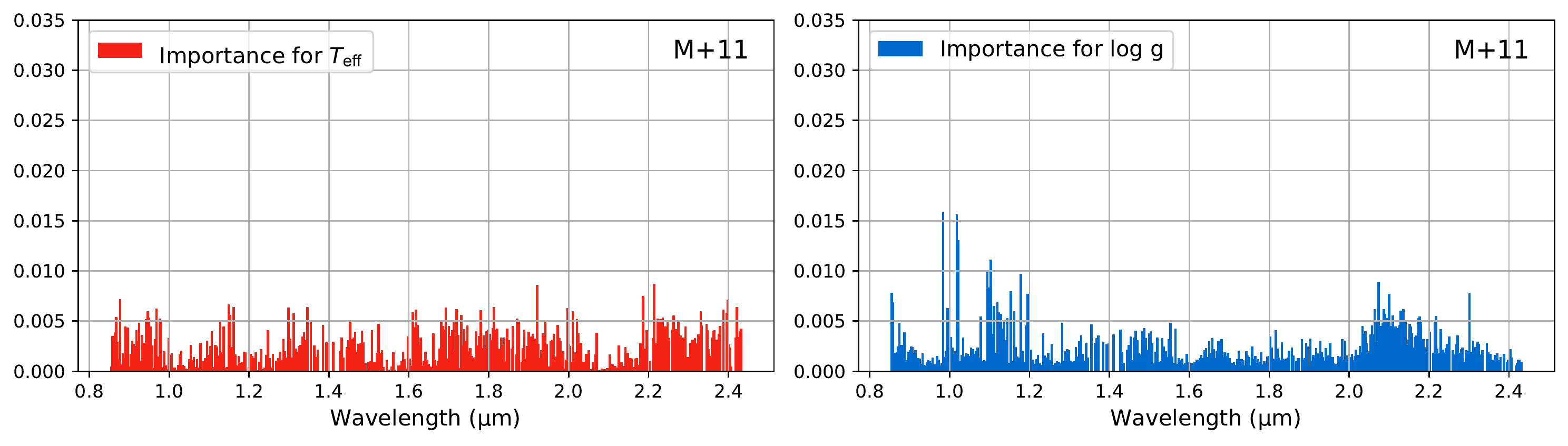}
\includegraphics[width=0.8\textwidth]{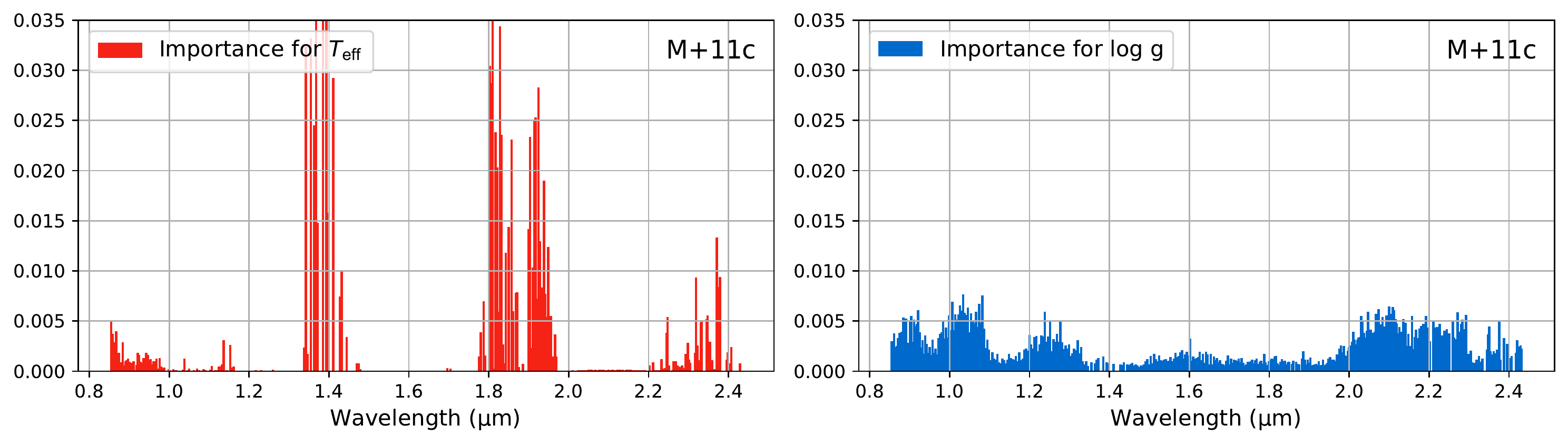}
\includegraphics[width=0.8\textwidth]{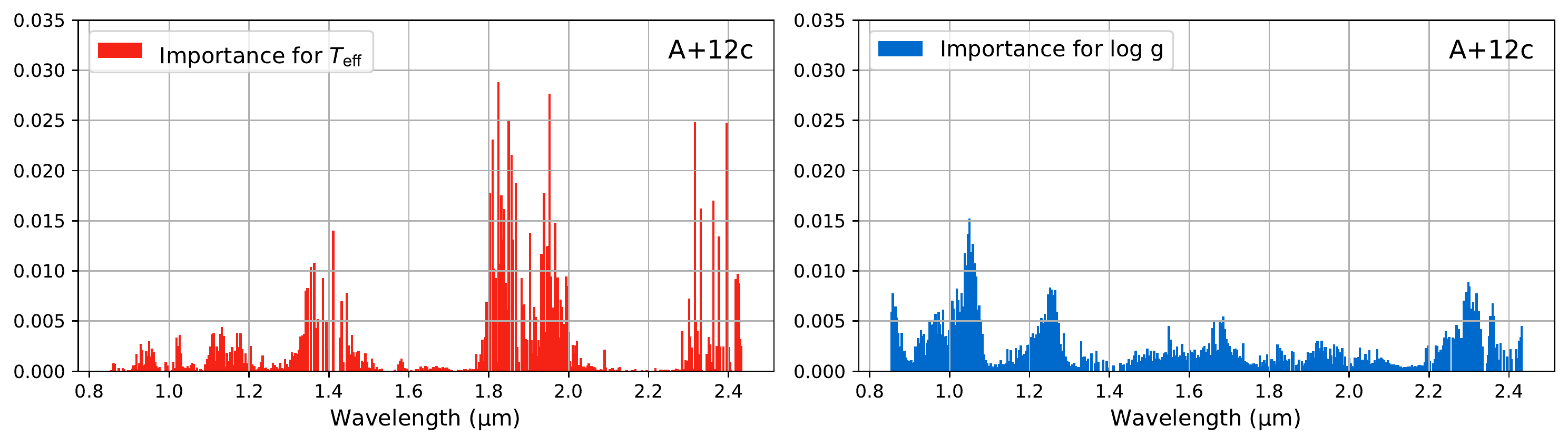}
\includegraphics[width=0.8\textwidth]{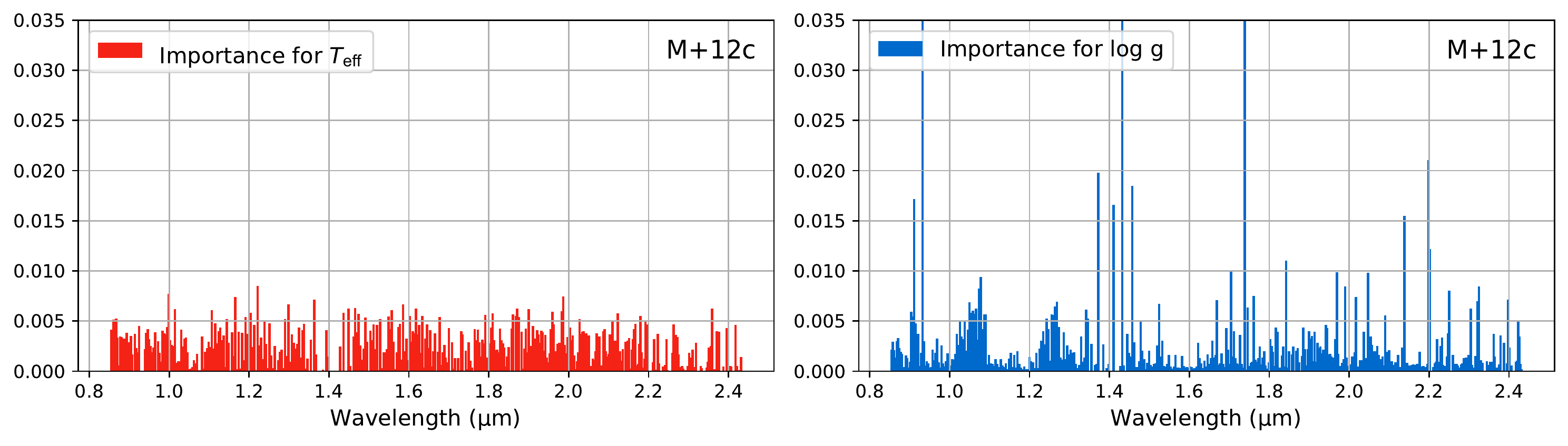}
\includegraphics[width=0.8\textwidth]{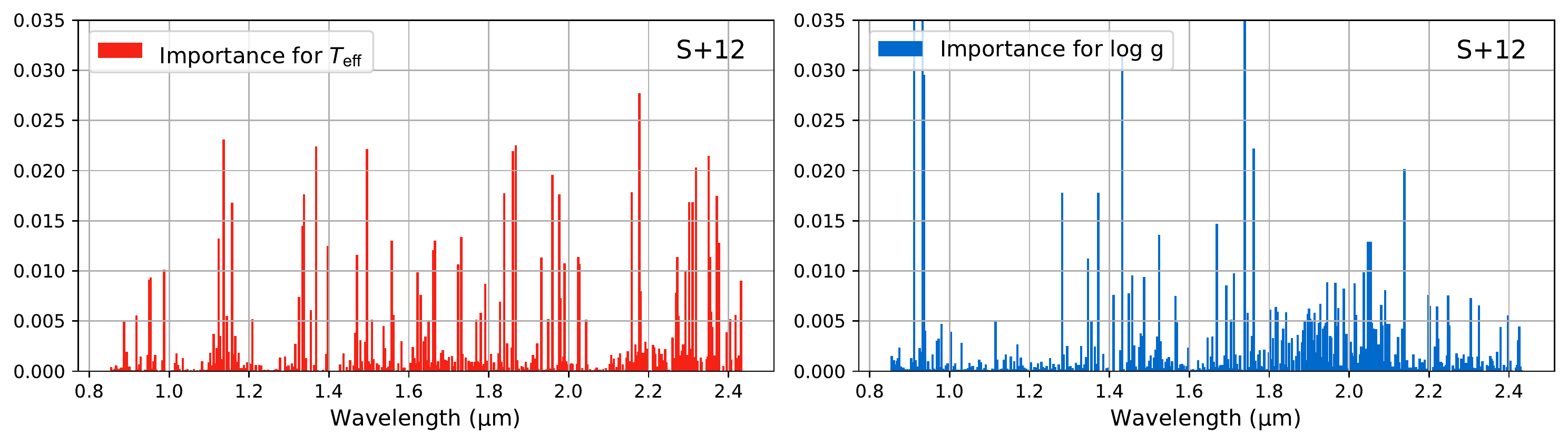}
\figurenum{\ref{fig:Effect_of_wavelength_range}}
\caption{continued. No noise is assumed.}     
\end{figure*}

\begin{figure*}[ht!]
\centering
\includegraphics[width=0.8\textwidth]{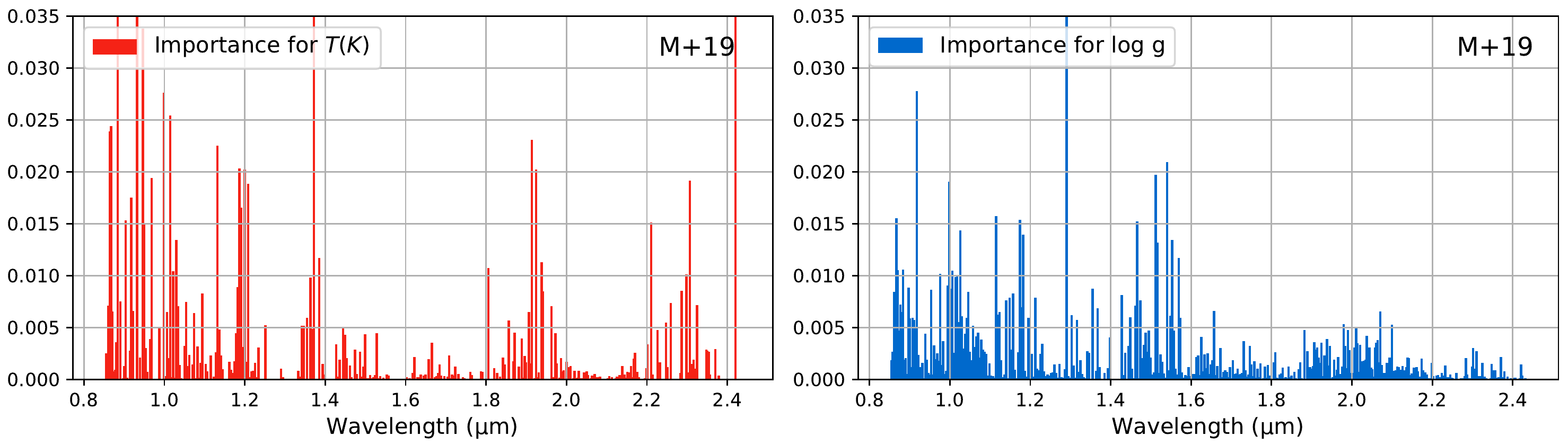}
\includegraphics[width=0.8\textwidth]{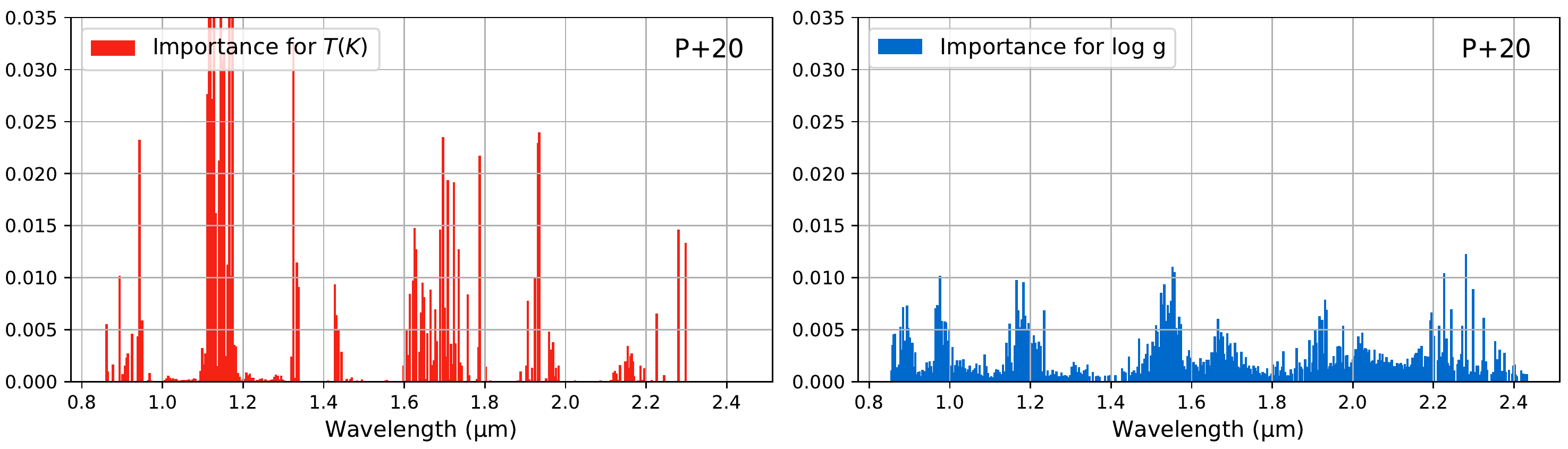}
\includegraphics[width=0.8\textwidth]{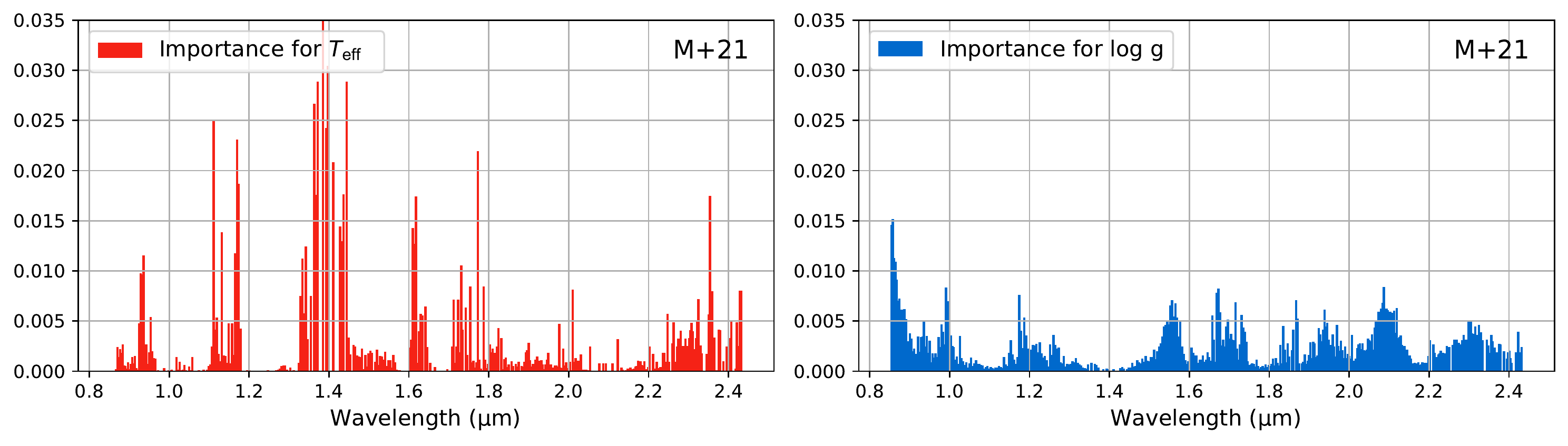}
\includegraphics[width=0.8\textwidth]{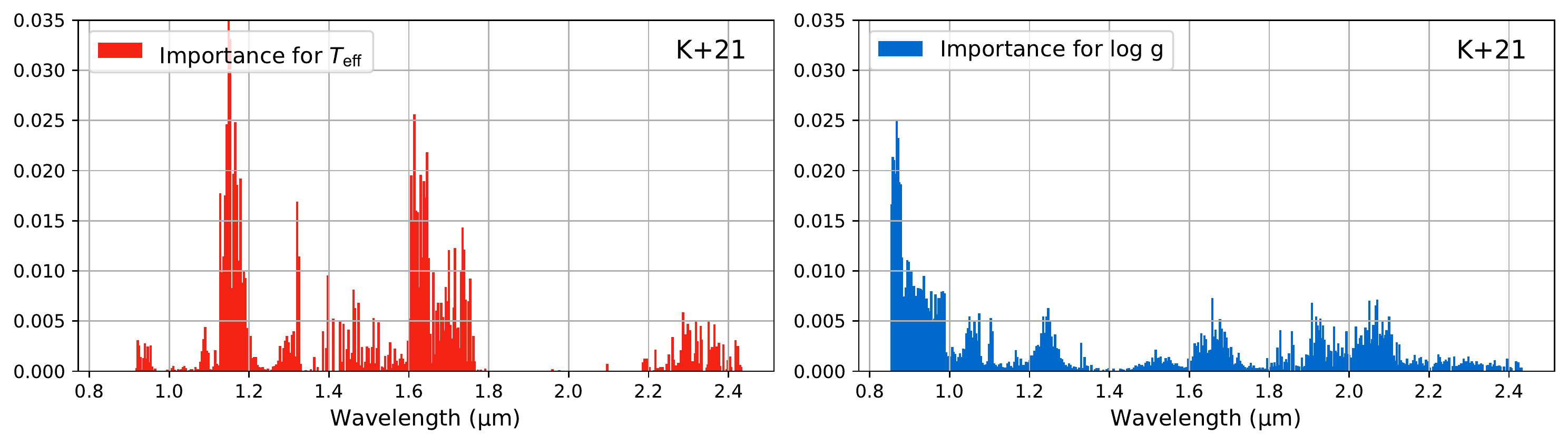}
\figurenum{\ref{fig:Effect_of_wavelength_range}}
\caption{continued. No noise is assumed.}     
\end{figure*}

As anticipated, most of these grids indicate that the wavelength range below 1.2~$\mu$m is important for the prediction of $\log{g}$ (cf.\ B+97, H+07, M+21, and K+21). Since the grids use different treatments for the alkali lines, the importance of this region explains in part why excising this region leads to better agreement in the surface gravity $R^2$ scores between models.  Another region that seems to be particularly important for determining $\log{g}$ is the 2--2.3~$\mu$m region. The majority of grids (with the exception of the M+12c, S+12, and K+21) show a peak in feature importance here. This behavior can likely be attributed to the role of collision-induced H$_2$ absorption \textbf{\citep{1969ApJ...156..989L,2002ApJ...564..421B}}, which is highly sensitive to atmospheric pressure and hence surface gravity. The differing sensitivity of this feature between grids with similar atmosphere model frameworks, for example M+21 and K+21, suggests that other parameters such as chemical equilibrium versus disequilibrium could mask the surface gravity sensitivity, and explains why the surface gravity $R^2$ score between these models is notably low.

For effective temperature, most model grids show similar overall structure in feature importance, with bands aligned with strong molecular absorption bands of H$_2$O, CO, and CH$_4$ (cf.\ Figure~\ref{fig:Spectra}). There are, however, notable exceptions to this trend. The B+97 and M+12c grids show a relatively flat feature importance distribution with respect to surface temperature, potentially connected to the underlying opacity sources in the case of B+97, or the impact of salt and sulfide clouds that mute gas absorption features in the case of M+12c. The agreement in feature importance plot structure, for example between the H+07 and M+11 grids, explains why their effective temperature $R^2$ scores are close to unity. As explained above, since these two grids essentially use the same atmosphere model, this result is not too surprising. 

It is worth noting that the wavelength ranges of 1.3--1.5 $\mu$m and 1.8--2 $\mu$m are affected by strong telluric absorption due to water.  Therefore, feature importance in these wavelength ranges should be interpreted with caution.

It is potentially misleading to compare feature importance plots between two model grids, as feature importance quantifies the relative importance of a parameter, across wavelength, for a specific model grid.  Nevertheless, it is instructive to see that the relative importance across wavelength is being altered when a model is cloudy versus cloud-free.  Physically, this occurs because clouds are selectively muting spectral features and hence altering their relative importance across a specific model spectrum.

\newpage
\section{Random forest retrievals}
\label{sect:random_forest_retrievals}

Having evaluated the self-consistency and inter-comparisons of the various model grids, we now turn to applying random forest retrievals to brown dwarf observations. We make comparison to two sets of spectra: benchmark objects with independently-derived parameter values, and spectral standards that span the L and T dwarf sequence.

\subsection{Data sources}
\label{sect:data_source}

For continuity with previous studies, we use the same three brown dwarf benchmarks examined in \citet{Oreshenko2020AJ} and \citet{Kitzmann2020ApJ}: the late T dwarf companion Gl 570 D and the binary brown dwarf companion system $\epsilon$ Indi Ba and Bb. Both are companions to more massive, nearby stars with detailed analyses of their ages and chemical compositions. Gl 570 D \citep{Burgasser2000ApJ...531L..57B} is a T8 dwarf and a wide companion ($\sim$1500 AU) to a K4 plus close M1+M3 hierarchical triple with an estimated age of 1--5~Gyr and a near-solar metallicity. The brown dwarf Gl 570 D has been studied extensively as a benchmark brown dwarf \citep{Burgasser2006ApJ...639.1095B, Saumon2006ApJ...647..552S, DelBurgo2009AA...501.1059D, Testi2009AA...503..639T, Line2015ApJ, Filippazzo2015ApJ, Oreshenko2020AJ, Kitzmann2020ApJ, Zhang2021ApJ, Mukherjee2022ApJ...938..107M}. $\epsilon$ Indi Ba+b \citep{2003AA...398L..29S,2004A&A...413.1029M} is a closely-separated T1 + T6 brown dwarf binary that is widely-separated (also $\sim$1500 AU) from the slightly metal-poor, $\sim$4~Gyr-old K4/5 dwarf $\epsilon$ Indi A, which itself hosts a 2--5~M$_\mathrm{Jup}$-mass giant planet ($\epsilon$ Indi Ab; \citealt{2019MNRAS.490.5002F}). The $\sim$11~yr orbit period of the binary, plus its gravitational perturbation on the host star, has permitted component masses of the brown dwarf binary to be determined astrometrically as 68--75~M$_\mathrm{Jup}$ for the primary and 53--70~M$_\mathrm{Jup}$ for the secondary, with recent results favoring the lower ends of these ranges \citep{2012PhDT.......463C, Dieterich2018ApJ...865...28D, 2022AJ....163..288C}. Given the direct inference of masses, both components of this system have been well-studied as benchmark brown dwarfs  \citep{Smith2003ApJ...599L.107S, Roellig2004ApJS..154..418R, Kasper2009ApJ...695..788K, King2010, Oreshenko2020AJ, Kitzmann2020ApJ}. 

For Gl 570 D, we use the SpeX spectrum from \cite{Burgasser2004AJ} covering the wavelength range 0.8--2.5~$\mu$m with average spectral resolution of $\sim$100. For $\epsilon$ Indi Ba \& Bb, we use spectra from \citet{King2010} obtained with the Very Large Telescope ISAAC spectrograph \citep{1998Msngr..94....7M} which spans 0.6--5~$\mu$m at a resolution of $\lambda/\Delta\lambda \approx$~5,000, considerably higher than that provided by SpeX prism data. The calibration of these spectra to apparent flux densities was done using broadband photometry as described in \citet{Kitzmann2020ApJ}.

We also conducted random forest retrievals on a sequence of 19 spectral standards spanning spectral classes from L0 to T8, as previously used in \citet{Lueber2022ApJ...930..136L} for traditional Bayesian nested-sampling retrievals. All 19 spectra were extracted from the SpeX Prism Library \citep{Burgasser2014}; a comprehensive description of the individual objects and their spectral properties can be found in \citet{Lueber2022ApJ...930..136L}.\footnote{Spectral data corresponding to each object -- L0 : \citet{Burgasser2006AJ}, L1: \citet{Kirkpatrick2010ApJS}, L2: \citet{Marocco2013AJ}, L3: \citet{Burgasser2007ApJ...659..655B}, L4: \citet{Kirkpatrick2010ApJS}, L5: \citet{Chiu2006AJ}, L6: \citet{Reid2006ApJ}, L7: \citet{Cruz2004ApJ}, L8: \citet{Burgasser2007ApJ...659..655B}, L9: \citet{Burgasser2006ApJ...637.1067B}, T0: \citet{Looper2007AJ},  T1: \citet{Burgasser2004AJ}, T2: \citet{Burgasser2004AJ}, T3: \citet{Chiu2006AJ}, T4: \citet{Burgasser2004AJ}, T5: \citet{Burgasser2004AJ}, T6: \citet{Burgasser2006ApJ...639.1095B}, T7: \citet{Burgasser2006ApJ...639.1095B}, T8: \citet{Burgasser2004AJ}.}

\subsection{Adding noise to training sets}
\label{sect:RF_grid_preparation}

As a first step, each spectrum in a given model grid is binned to match the binning in the measured spectrum.  Adding noise to the training set is now object-specific and depends on the measured spectrum of each object in our sample.

Defining the wavelength-dependent flux in each bin to be $F_\lambda$, and the measured uncertainty in each bin to be $\delta F_\lambda$, following \cite{Oreshenko2020AJ}, we adopt $\delta F_\lambda/F_\lambda$ as the standard deviation of a Gaussian and randomly draw from it.  This flux excess or deficit is then added to the binned model flux.  It implies that, even for the same model spectrum, it is possible to produce different realizations of it with noise added.

\subsection{Scaling factor}

In addition to effective temperature and surface gravity, we add a flux scaling factor $f$ as a third parameter to our retrievals in order to scale the outgoing, photospheric flux $F_{\nu}^+$ of the brown dwarf model to that measured by the observer, $F_\nu$. These are simply related by the radius–distance relation:
\begin{equation}\label{radius-distance relation}
    F_{\nu}=F_{\nu}^+ f\left(\frac{R_\mathrm{p}}{d}\right)^2 \ .
\end{equation}
Here, $d$ is the distance between the observer and the brown dwarf and $R_\mathrm{p}$ the assumed radius of the latter. For distances, we use values of $d = 5.84 \pm 0.03$ pc for Gl 570 D based on Hipparcos parallax measurements \citep{vanLeeuwen2007A&A...474..653V}, and $d = 3.638 \pm 0.001$ pc for $\epsilon$ Indi B \citep{Lindegren2021A&A...649A...2L} based on Gaia Early Data Release 3 parallax measurements, respectively.
For the radius, we assume an {\it a priori} value of $R_\mathrm{p} = 1\,R_\mathrm{Jup}$, based on the evolutionary models of \citet{Burrows1993RvMP}, a value that is appropriate for an sufficiently evolved brown dwarf and is only weakly dependent on mass (see their Figure 1). In principle, the retrieved value for $f$ can be converted into the actual brown dwarf radius, $R_\mathrm{d}$, as
\begin{equation}
    R_\mathrm{d} = \sqrt{f} R_\mathrm{Jup}\ .
    \label{eq:derived_radius}
\end{equation}
In practice $f$ contains not only the flux scaling between radius and distance, but also accounts for inaccuracies in the observed spectral flux calibration and shortcomings of the atmospheric models \citep{Kitzmann2020ApJ}. 

To incorporate the scaling factor into our training sets, we assume a uniform distribution of $0.5 \le f \le 2.0$.  In tandem with randomly drawing for the noise in each bin, we randomly draw an $f$ for each model spectrum and multiply the entire set of fluxes by it.

\subsection{Results for benchmark brown dwarfs}
\label{sect:Real data}

\begin{figure*}[ht!]
\includegraphics[width=0.5\textwidth]{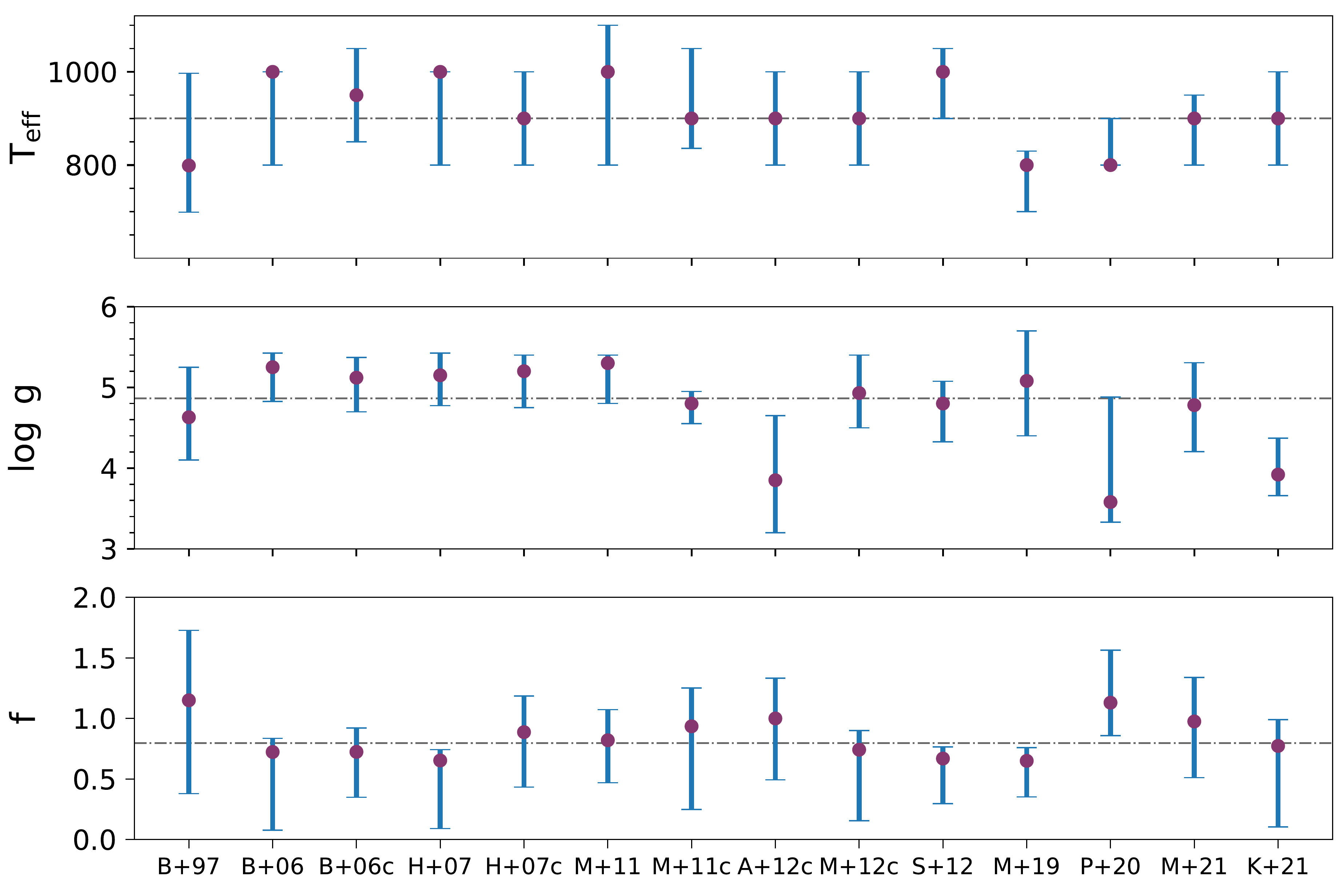}
\includegraphics[width=0.5\textwidth]{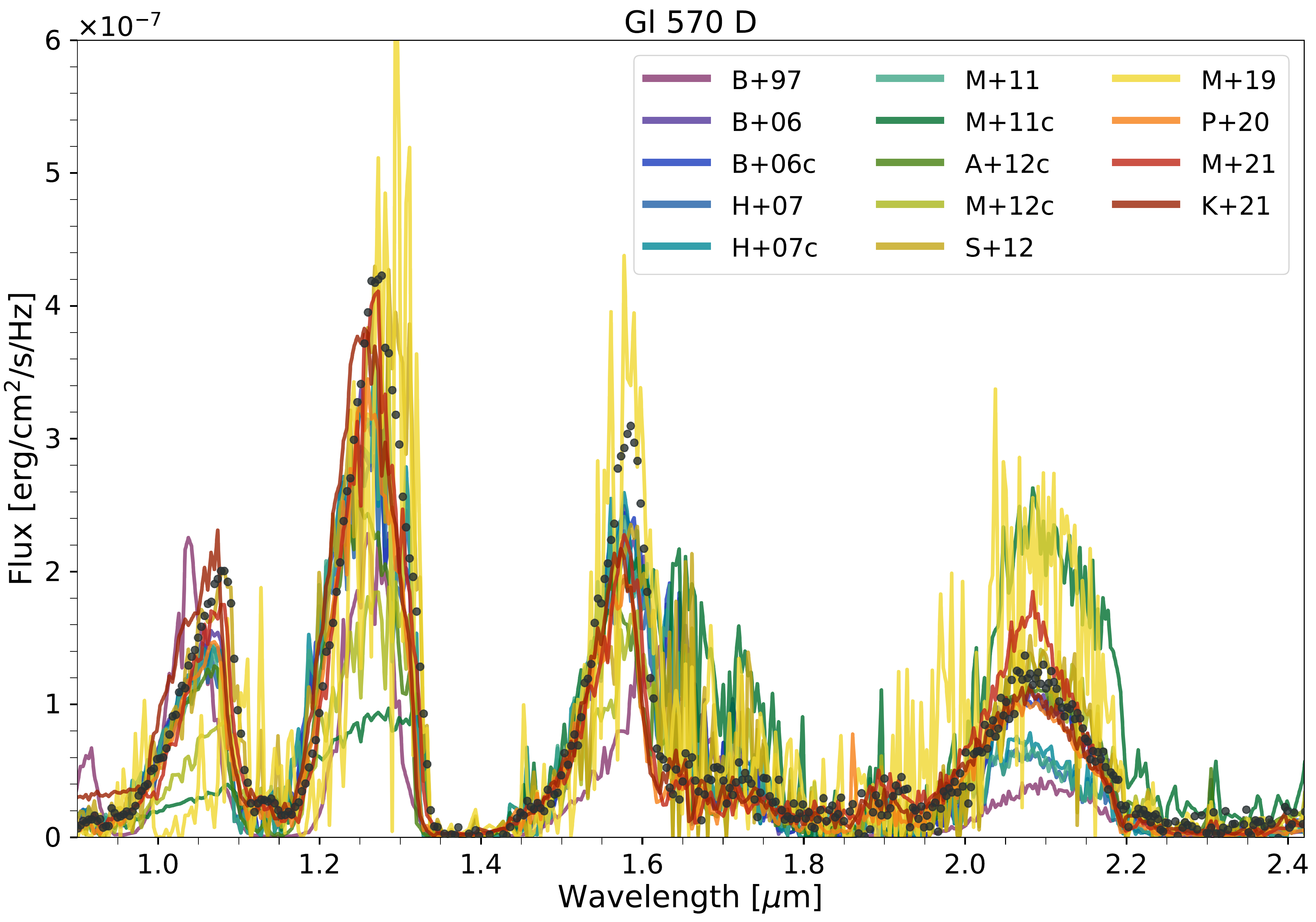} \hfill
\includegraphics[width=0.5\textwidth]{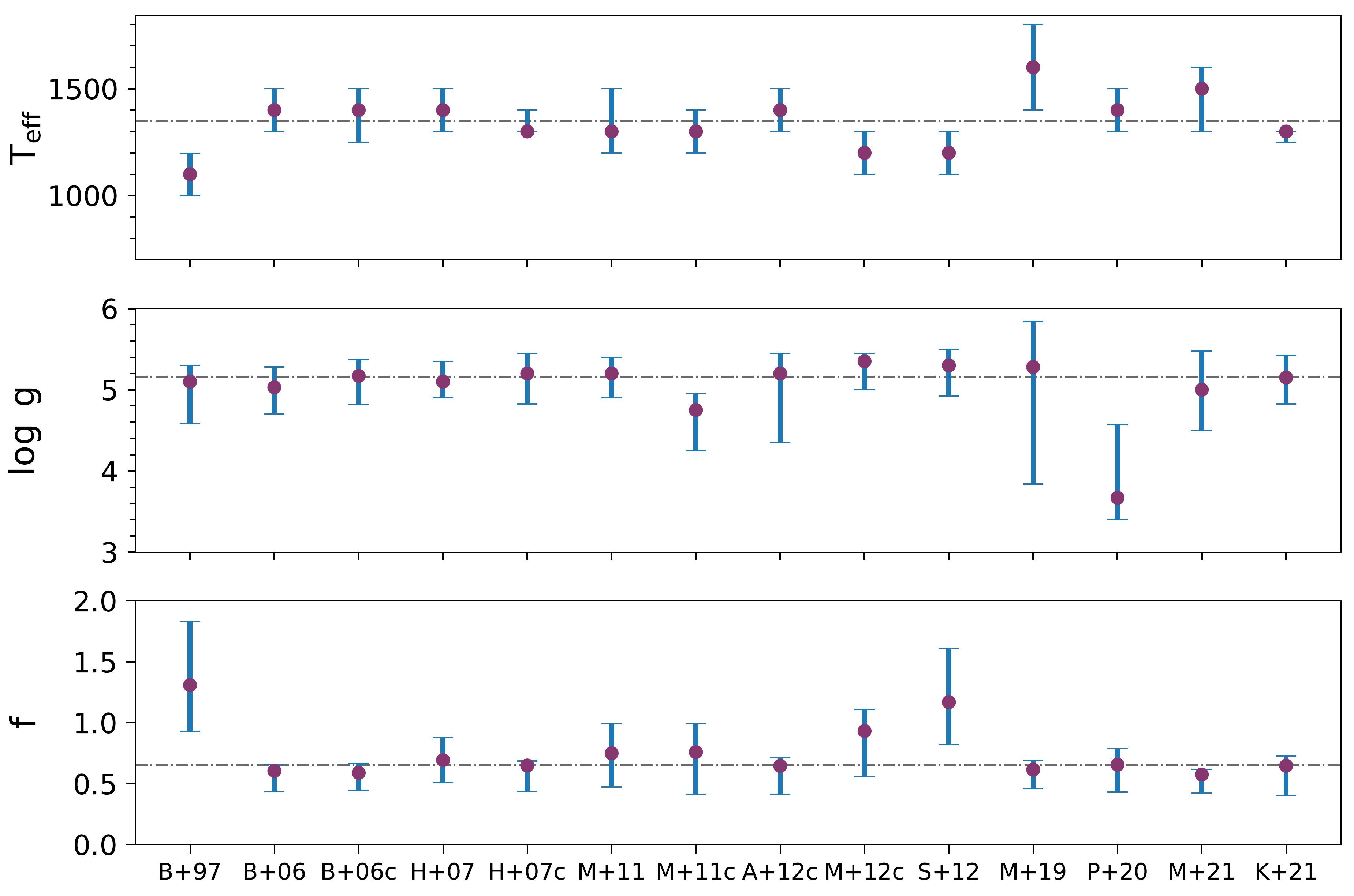}
\includegraphics[width=0.5\textwidth]{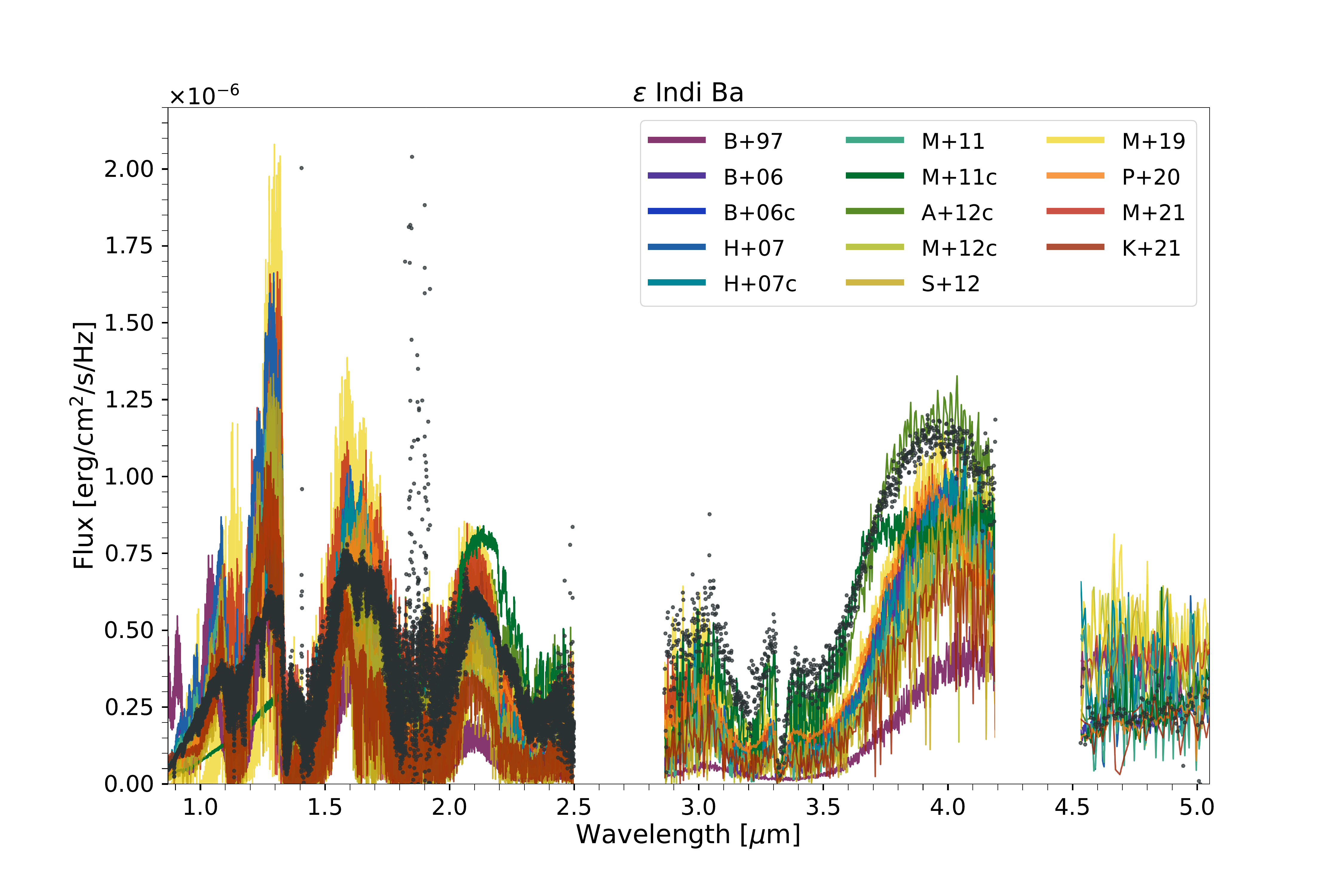} \hfill
\includegraphics[width=0.5\textwidth]{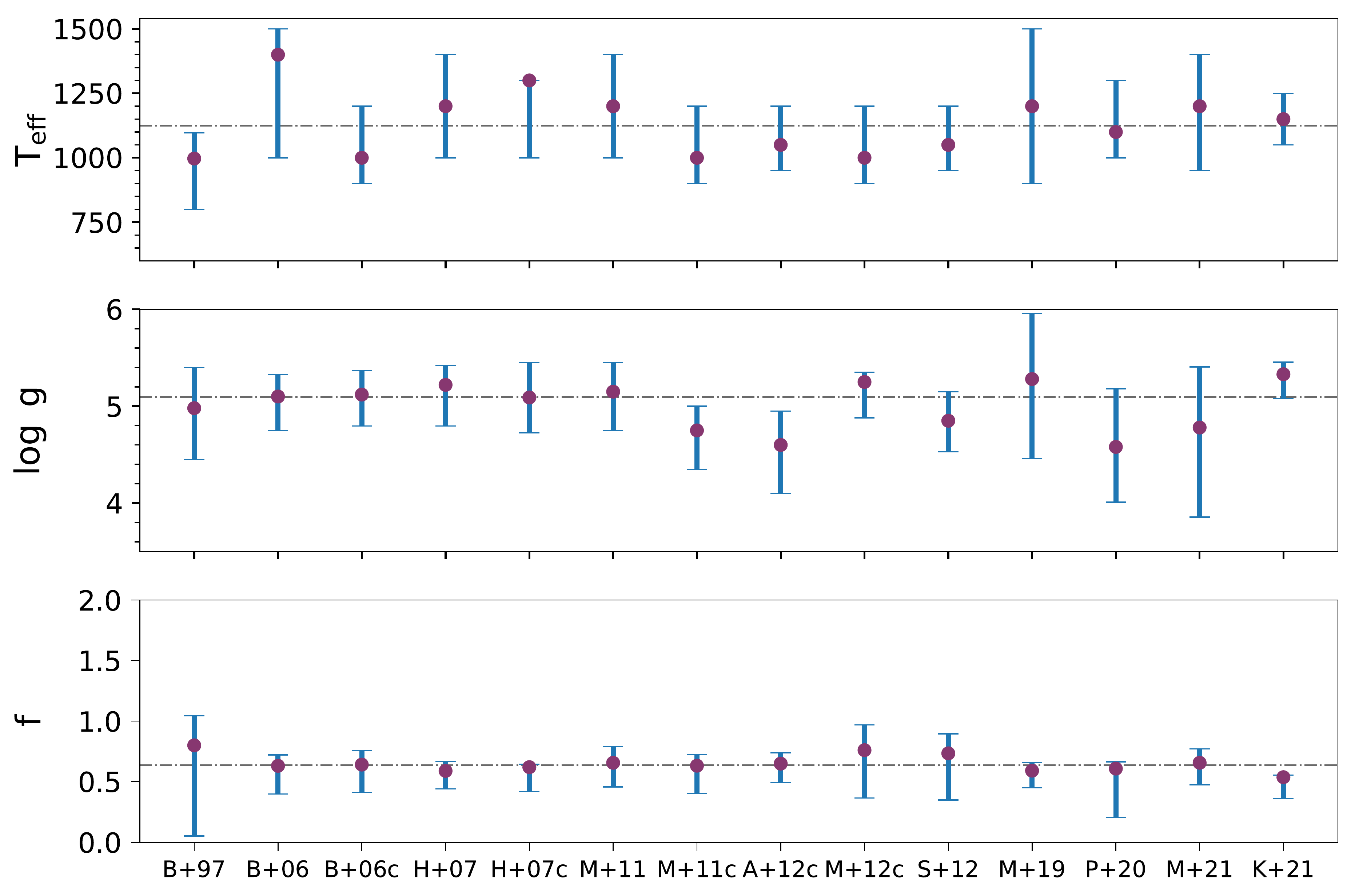}
\includegraphics[width=0.5\textwidth]{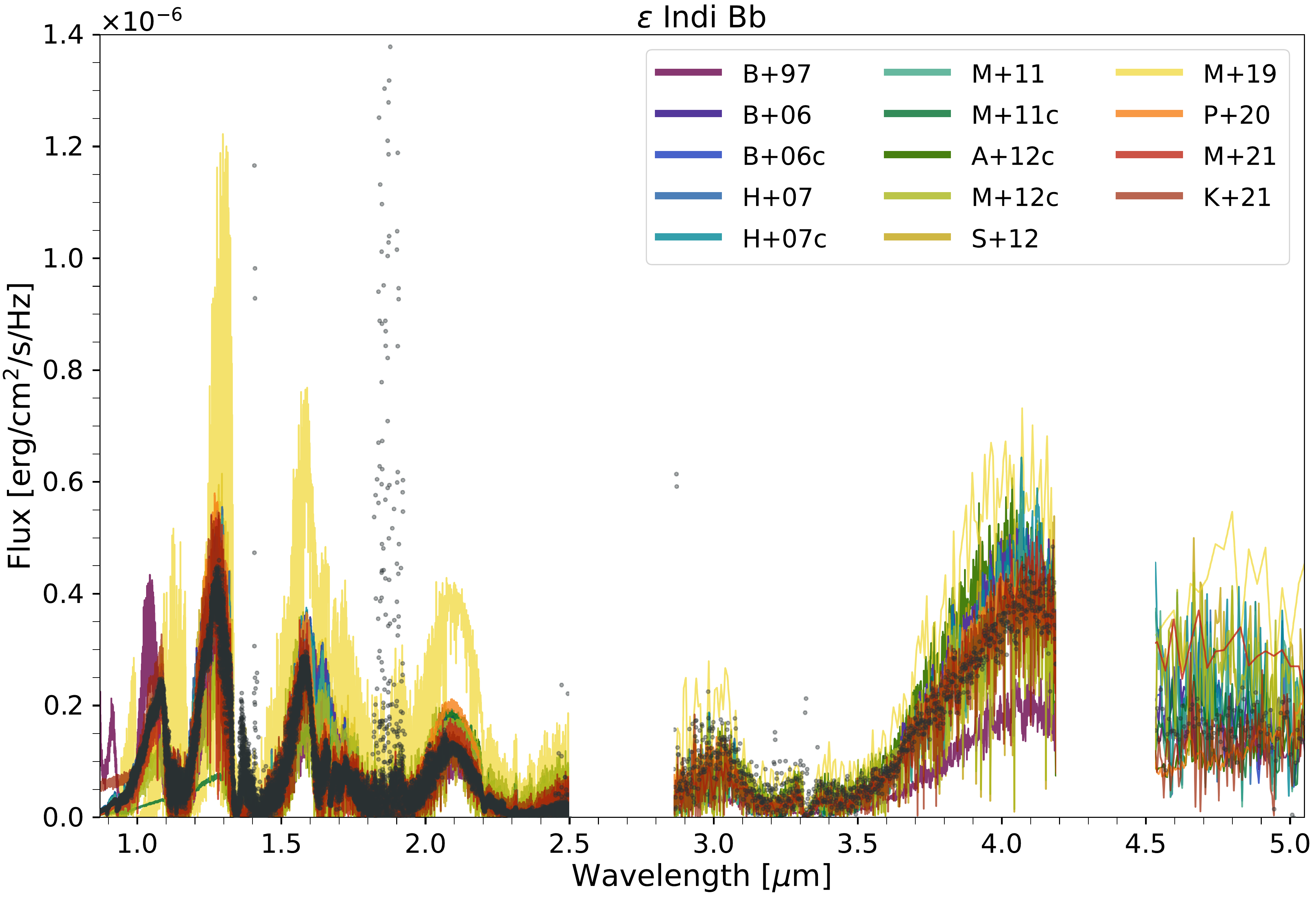}
\caption{Retrieval outcomes for Gl 570 D (top), $\epsilon$ Indi Ba (middle), and $\epsilon$ Indi Bb (bottom). Left panels show the retrieved posterior median values (points) and 1 $\sigma$ uncertainties (error bars) for each of the model grids. The flat line in each panel is a fit across the retrieved parameter values.  Right panels show the median spectra for each individual grid compared to the original data (black dots). Note that for Gl 570 D, we use the SpeX spectrum from \cite{Burgasser2004AJ} covering the wavelength range 0.8--2.5~$\mu$m with average spectral resolution of $\sim$100; For $\epsilon$ Indi Ba \& Bb, we use spectra from \citet{King2010} obtained with the Very Large Telescope ISAAC spectrograph \citep{1998Msngr..94....7M} which spans 0.6--5~$\mu$m at a resolution of $\lambda/\Delta\lambda \approx$~5,000, considerably higher than that provided by SpeX prism data. It should be noted that object-specific uncertainties have been considered according to Section~\ref{sect:RF_grid_preparation}, and are therefore data-driven.}
\label{fig:DataApplication}          
\end{figure*}

After the spectral resolution of the model grids has been degraded using a Gaussian kernel to match each spectrum, we trained random forest models for each of the 14 grids, and then performed retrievals on each of the benchmark spectra. Corresponding real versus predicted (RvP) plots can be found in the Appendix (Figure~\ref{fig:Effect of interpolation}). Figure~\ref{fig:DataApplication} shows the parameter posterior values and the median spectrum for all of the model grids for each source. The median spectrum is obtained by calculating the wavelength-dependent flux median from the collected training spectra of the predicted leaf of each tree. Posterior values are summarized in Table~\ref{tab:benchmark posteriors} and compared to corresponding values from previous studies.

\begin{deluxetable*}{ccccccccccccc}
\tabletypesize{\ssmall}
\tablecolumns{13}
\tablewidth{0.8\columnwidth}
\tablecaption{Results of the random forest retrieval algorithm for the benchmark brown dwarfs Gl 570 D, $\epsilon$ Indi Ba \& Bb. Previously published values are shown for comparison.}
\label{tab:benchmark posteriors}
\tablehead{
 \colhead{} & \colhead{} & \colhead{\textbf{\scriptsize Gl 570 D}} & \colhead{} & \colhead{} & \colhead{} & \colhead{\textbf{\scriptsize $\epsilon$ Indi Ba}} & \colhead{} & \colhead{} & \colhead{} & \colhead{\textbf{\scriptsize $\epsilon$ Indi Bb}} & \colhead{} & \colhead{}\\
 \hline
 \colhead{\scriptsize Model} & \colhead{\scriptsize $T_\mathrm{eff}$ (K)} & \colhead{\scriptsize $\log{g}$ (\text{cm s}$^{−2}$)} & \colhead{\scriptsize $f$} & \colhead{\scriptsize $\chi_r^2$} & \colhead{\scriptsize $T_\mathrm{eff}$ (K)} & \colhead{\scriptsize $\log{g}$ (\text{cm s}$^{−2}$)} & \colhead{\scriptsize $f$} & \colhead{\scriptsize $\chi_r^2$} & \colhead{\scriptsize $T_\mathrm{eff}$ (K)} & \colhead{\scriptsize $\log{g}$ (\text{cm s}$^{−2}$)} & \colhead{\scriptsize $f$} & \colhead{\scriptsize $\chi_r^2$} 
 }
\startdata 
\hline
\textbf{\scriptsize This work} &   &  &  & &  &  &  &  & & & \\
\hline
\text{\scriptsize B+97} &  $799^{+198}_{-100}$ &  $4.63^{+0.62}_{-0.53}$ & $1.150^{+0.770}_{-0.578}$ & 88.05 &
        $1100^{+99}_{-100}$ &  $5.10^{+0.20}_{-0.52}$ & $1.310^{+0.380}_{-0.526}$ & 551.23 &
        $997^{+100}_{-198}$ &  $4.98^{+0.42}_{-0.53}$ & $0.800^{+0.749}_{-0.246}$ &633.50\\
\text{\scriptsize B+06} &  $1000^{+0}_{-200}$ & $5.25^{+0.18}_{-0.43}$  & $0.723^{+0.646}_{-0.113}$ & 20.01 &
        $1400^{+100}_{-100}$ & $5.03^{+0.25}_{-0.33}$  & $0.606^{+0.172}_{-0.053}$ & 14.56 &
        $1400^{+100}_{-400}$ & $5.10^{+0.23}_{-0.35}$  & $0.631^{+0.233}_{-0.091}$ & 83.72\\
\text{\scriptsize B+06c} & $950^{+100}_{-100}$ & $5.12^{+0.25}_{-0.42}$  & $0.724^{+0.374}_{-0.197}$ & 19.90 &
        $1400^{+100}_{-150}$ & $5.17^{+0.20}_{-0.35}$  & $0.590^{+0.143}_{-0.076}$ & 10.92 &
        $1000^{+200}_{-100}$ & $5.12^{+0.25}_{-0.33}$  & $0.641^{+0.230}_{-0.118}$ & 27.96\\
\text{\scriptsize H+07} &  $1000^{+0}_{-200}$ & $5.15^{+0.28}_{-0.38}$  & $0.653^{+0.562}_{-0.090}$ & 21.53&
        $1400^{+100}_{-100}$ & $5.10^{+0.25}_{-0.20}$  & $0.694^{+0.185}_{-0.184}$ & 14.82&
        $1200^{+200}_{-200}$ & $5.22^{+0.20}_{-0.43}$  & $0.589^{+0.149}_{-0.078}$ &28.25\\
\text{\scriptsize H+07c} & $900^{+100}_{-100}$ & $5.20^{+0.20}_{-0.45}$  & $0.887^{+0.454}_{-0.298}$ & 21.43 &
        $1300^{+100}_{-0}$ & $5.20^{+0.25}_{-0.38}$  & $0.650^{+0.213}_{-0.038}$ & 13.70 &
        $1300^{+0}_{-300}$ & $5.09^{+0.36}_{-0.36}$  & $0.620^{+0.200}_{-0.024}$ & 53.16\\
\text{\scriptsize M+11} &  $1000^{+100}_{-200}$ & $5.30^{+0.10}_{-0.50}$  & $0.820^{+0.350}_{-0.252}$ & 19.78 &
        $1300^{+200}_{-100}$ & $5.20^{+0.20}_{-0.30}$  & $0.750^{+0.276}_{-0.241}$ &16.79 &
        $1200^{+200}_{-200}$ & $5.15^{+0.30}_{-0.40}$  & $0.656^{+0.199}_{-0.132}$ & 27.40\\
\text{\scriptsize M+11c} & $900^{+150}_{-64}$ & $4.80^{+0.15}_{-0.25}$  & $0.935^{+0.687}_{-0.317}$ &28.91 &
        $1300^{+100}_{-100}$ & $4.75^{+0.20}_{-0.50}$  & $0.760^{+0.344}_{-0.232}$ & 11.76 &
        $1100^{+200}_{-100}$ & $4.75^{+0.25}_{-0.40}$  & $0.632^{+0.227}_{-0.094}$ & 36.10\\
\text{\scriptsize A+12c} &  $900^{+100}_{-100}$ & $3.85^{+0.80}_{-0.65}$  & $1.000^{+0.506}_{-0.332}$ & 12.59 &
        $1400^{+100}_{-100}$ & $5.20^{+0.25}_{-0.85}$  & $0.647^{+0.213}_{-0.066}$ & 6.50 &
        $1050^{+150}_{-100}$ & $4.60^{+0.35}_{-0.50}$  & $0.649^{+0.157}_{-0.091}$ & 29.97\\
\text{\scriptsize M+12c} &  $900^{+100}_{-100}$ & $4.93^{+0.47}_{-0.43}$  & $0.742^{+0.587}_{-0.158}$ & 36.07 &
        $1200^{+100}_{-100}$ & $5.35^{+0.10}_{-0.35}$  & $0.933^{+0.373}_{-0.178}$ & 18.68 &
        $1000^{+200}_{-100}$ & $5.25^{+0.10}_{-0.37}$  & $0.760^{+0.394}_{-0.208}$ & 30.57\\
\text{\scriptsize S+12} &  $1000^{+50}_{-100}$ & $4.80^{+0.28}_{-0.48}$  & $0.669^{+0.372}_{-0.096}$ & 41.84 &
        $1200^{+100}_{-100}$ & $5.30^{+0.20}_{-0.38}$  & $1.170^{+0.349}_{-0.443}$ &25.20&
        $1050^{+150}_{-100}$ & $4.85^{+0.30}_{-0.32}$  & $0.734^{+0.385}_{-0.161}$ & 40.32\\
\text{\scriptsize M+19} &  $800^{+30}_{-100}$ & $5.08^{+0.62}_{-0.68}$  & $0.650^{+0.297}_{-0.109}$ & 85.79 &
        $1600^{+200}_{-200}$ & $5.28^{+0.56}_{-1.44}$  & $0.616^{+0.155}_{-0.079}$ & 38.64 &
        $1200^{+300}_{-300}$ & $5.28^{+0.68}_{-0.82}$  & $0.591^{+0.140}_{-0.065}$ & 106.05\\
\text{\scriptsize P+20} &  $800^{+100}_{-0}$ & $3.58^{+1.30}_{-0.25}$  & $1.130^{+0.272}_{-0.434}$ & 13.96 &
        $1400^{+100}_{-100}$ & $3.67^{+0.90}_{-0.27}$  & $0.657^{+0.225}_{-0.130}$ & 3.55 &
        $1100^{+200}_{-100}$ & $4.58^{+0.60}_{-0.57}$  & $0.608^{+0.403}_{-0.056}$ & 17.44\\   
\text{\scriptsize M+21} &  $900^{+50}_{-100}$ & $4.78^{+0.53}_{-0.58}$  & $0.975^{+0.464}_{-0.363}$ & 8.00 &
        $1500^{+100}_{-200}$ & $5.00^{+0.48}_{-0.50}$  & $0.576^{+0.152}_{-0.044}$ & 9.65&
        $1200^{+200}_{-250}$ & $4.78^{+0.63}_{-0.93}$  & $0.656^{+0.181}_{-0.115}$ &28.88\\
\text{\scriptsize K+21} &  $900^{+100}_{-100}$ & $3.92^{+0.45}_{-0.26}$  & $0.773^{+0.668}_{-0.217}$ & 7.46 &
        $1300^{+0}_{-50}$ & $5.15^{+0.28}_{-0.33}$  & $0.648^{+0.245}_{-0.081}$ & 51.93 &
        $1150^{+100}_{-100}$ & $5.33^{+0.13}_{-0.25}$  & $0.537^{+0.177}_{-0.018}$ & 68.88\\
\hline
\text{\scriptsize Weighted Mean} & $900^{+88}_{-0}$ & $4.87^{+0.28}_{-0.20}$ & $0.797^{+0.167}_{-0.073}$ & & $1350^{+50}_{-50}$ & $5.16^{+0.04}_{-0.11}$ & $0.654^{+0.104}_{-0.030}$ & & $1125^{+75}_{-113}$ & $5.10^{+0.11}_{-0.30}$ & $0.637^{+0.020}_{-0.025}$ &\\
\hline
\textbf{\scriptsize Previous work} &  &  &  & &  &  &  &  & & & \\
\hline
\text{\citet{Burgasser2006ApJ...639.1095B}} & $780-820$  & $5.1$ & $0.717-0.750$ \tablenotemark{\ssmall a} & & -& - & - & &  - & - & - &\\
\text{\citet{Saumon2006ApJ...647..552S}} & $800-820$  & $5.09 - 5.23$ & $0.769^{+0.115}_{-0.080}$ \tablenotemark{\ssmall a} & & -& - & - & & - & - & - & \\
\text{\citet{DelBurgo2009AA...501.1059D}} & $948^{+53}_{-53}$  & $4.5^{+0.5}_{-0.5}$ & - & & -& - & - & & - & - & - &\\
\text{\citet{Testi2009AA...503..639T}} & $900$  & $5.0$ & - & & - & - & - & & - & - & - &\\
\text{\citet{Line2015ApJ}} & $714^{+20}_{-23}$  & $4.76^{+0.27}_{-0.28}$ & $1.300^{+0.010}_{-0.008}$ \tablenotemark{\ssmall a}& & - & - & - & & - & - & - &\\
\text{\citet{Filippazzo2015ApJ}} & $759^{+63}_{-63}$  & $4.90^{+0.5}_{-0.5}$ & $0.884 \pm 0.026$ \tablenotemark{\ssmall a} & & -& - & - & & - & - & -& \\
\text{\citet{Zhang2021ApJ}} & $786^{+20}_{-20}$  & $5.04^{+0.13}_{-0.13}$ & $0.792^{+0.003}_{-0.002}$ \tablenotemark{\ssmall a} & & -& - & -& & - & - & -& \\
\text{\citet{Mukherjee2022ApJ...938..107M}} & $725$  & $5.0$ & $1.082$ \tablenotemark{\ssmall a} & & -& - & - & & - & - & -& \\
\text{\citet{King2010}, atm.} & - & - & - & & $1300-1340$ & $5.25$ & - & & $880-940$ & $5.5$ & -& \\
\text{\citet{King2010}, evo.} & - & - & - & & $1352–1385$ & $5.25$ & $0.606-0.621$ \tablenotemark{\ssmall a} & & $880-940$ & $5.5$ & $0.637-0.652$ \tablenotemark{\ssmall a}& \\
\text{\citet{Smith2003ApJ...599L.107S}} & - & - & - & & $1400-1600$ & $5.2 \pm 0.3$ & $0.364$ \tablenotemark{\ssmall a} & & - & - & -& \\
\text{\citet{Roellig2004ApJS..154..418R}} & - & - & -&  & $1250$ & $5.13$ & $0.837$ \tablenotemark{\ssmall a} & & $840$ & $4.89$ & $0.947$ \tablenotemark{\ssmall a} & \\
\text{\citet{Kasper2009ApJ...695..788K}} & - & - & - & & $1250-1300$ & $5.2-5.3$ & $0.689$ \tablenotemark{\ssmall a} & &  $875–925$ & $4.9-5.1$ & $0.845$ \tablenotemark{\ssmall a} &\\
\text{\citet{Dieterich2018ApJ...865...28D}} & - & - & - & & - & $5.27 \pm 0.09$ \tablenotemark{\ssmall c} & - & & - & $5.24 \pm 0.09$ \tablenotemark{\ssmall c} & - & \\
\text{\citet{Oreshenko2020AJ} \texttt{Sonora}} & $808^{+43}_{-27}$ & $4.93^{+0.38}_{-0.55}$ & $0.618^{+0.156}_{-0.053}$ & & $1530^{+173}_{-127}$ & $5.17^{+0.24}_{-0.52}$ & $0.582^{+0.052}_{-0.030}$ & & $1130^{+352}_{-157}$ & $5.19^{+0.199}_{-1.02}$ & $0.593^{+0.0907}_{-0.0419}$ & \\
\text{\citet{Oreshenko2020AJ} \texttt{Sonora}} \tablenotemark{\ssmall b} & - & - & - & & $1300^{+100}_{-100}$ & $4.39^{+0.748}_{-0.603}$ & -&  & $900^{+76.7}_{-26.6}$ & $5.45^{+0.0176}_{-0.0995}$ & - & \\
\text{\citet{Oreshenko2020AJ} \texttt{AMES-cond}} & $878^{+23}_{-78}$ & $5.27^{+0.43}_{-0.67}$ & $0.633^{+0.147}_{-0.070}$ & & $1600^{+122}_{-100}$ & $5.68^{+0.20}_{-0.35}$ & $0.557^{+0.053}_{-0.031}$ & & $1100^{+150}_{-100}$ & $5.53^{+0.33}_{-0.47}$ & $0.585^{+0.099}_{-0.041}$ & \\
\text{\citet{Oreshenko2020AJ} \texttt{AMES-cond}} \tablenotemark{\ssmall b}& - & - & - & & $1300^{+100}_{-100}$ & $5.5^{+0.434}_{-0.866}$ & - & & $930^{+70}_{-30}$ & $5.73^{+0.224}_{-0.399}$ & - & \\
\text{\citet{Oreshenko2020AJ} \texttt{HELIOS}}  & $800^{+14}_{-100}$ & $5.08^{+0.62}_{-0.68}$ & $0.686^{+0.592}_{-0.109}$ & & $1530^{+145}_{-127}$ & $5.54^{+0.22}_{-1.56}$ & $0.593^{+0.119}_{-0.030}$ & & $1180^{+239}_{-181}$ & $5.32^{+0.47}_{-1.28}$ & $0.610^{+0.234}_{-0.041}$ &\\
\text{\citet{Oreshenko2020AJ} \texttt{HELIOS}} \tablenotemark{\ssmall b}& - & - & - & & $1300^{+102}_{-97}$ & $5.62^{+0.269}_{-1.12}$ & - & & $900^{+100}_{-100}$ & $5.86^{+0.0482}_{-0.321}$ & - &\\
\text{\citet{Kitzmann2020ApJ} equil.} & $730^{+18}_{-17}$  & $4.61^{+0.08}_{-0.08}$ & $1.000^{+0.010}_{-0.008}$ & & $1339^{+19}_{-19}$  & $5.49^{+0.06}_{-0.10}$ & $0.533^{+0.001}_{-0.001}$ & & $768^{+26}_{-25}$  & $5.11^{+0.05}_{-0.05}$ &$0.533^{+0.001}_{-0.001}$ & \\
\text{\citet{Kitzmann2020ApJ} free} & $703^{+17}_{-20}$  & $5.01^{+0.13}_{-0.19}$ & $1.277^{+0.003}_{-0.004}$ & & $1420^{+16}_{-16}$  & $5.62^{+0.07}_{-0.07}$ & $0.303^{+0.001}_{-0.001}$ & & $992^{+22}_{-21}$  & $4.85^{+0.17}_{-0.19}$ &$0.504^{+0.002}_{-0.001}$ & \\
\hline
\enddata
\tablenotetext{a}{\scriptsize This value was derived and is based on the assumption that $\sqrt{f}=R$}
\tablenotetext{b}{\scriptsize The values are based on a two-parameter retrieval model but otherwise analogous to \citet{Oreshenko2020AJ} three-parameter retrievals}
\tablenotetext{c}{\scriptsize This value was derived and is based on the measured dynamical mass and assuming $R = R$ ($1 \pm 0.1$) R$_\mathrm{J}$}
\end{deluxetable*}

The median retrieved spectra show large deviations up to one order of magnitude compared with the observational data that varies among the different grids. Indeed, some of the model spectra do not seem to fit the data at all, with their flux values either strongly over- or underestimating the actual brown dwarf spectra. Such poor fits of grid spectra are not uncommon for brown dwarfs, particularly in comparison to traditional retrieval methods (cf.\ \citealt{Line2015ApJ,Kitzmann2020ApJ}), suggesting that the self-consistent models are missing important ingredients to properly represent brown dwarf atmospheres. Alternately, approximations made in the models, such as chemical equilibrium, may not be entirely valid for these objects. On the other hand, classical retrieval methods that fit spectral data more accurately typically neglect the actual physical and chemical processes that govern brown dwarf atmospheres.

Despite the overall low quality of spectral fits (Figure~\ref{fig:DataApplication}), the suite of random forest retrievals obtain similar posterior parameter values, albeit with large error bars. For Gl 570 D, the retrieved effective temperatures are all between 800~K and 1000~K, while retrieved surface gravities are mostly between $4.8 \lesssim \log{g} \lesssim 5.2$. Three notable exceptions for the latter are the A+12c, P+20 \& K+21 grids that yield much smaller $\log{g}$ values, closer to 4. The scaling factor $f$ is retrieved to be around $\approx 0.8$ for most grids, with the three grids B+97, P+20 \& M+21 grids favoring values closer to unity. These values correspond to radii of 0.8--1.0~R$_\mathrm{Jup}$. In comparing to literature values, we note that most prior retrieval or grid-fitting studies of Gl 570 D obtain lower effective temperatures, below 800 K. The surface gravities obtained by these studies, on the other hand, mostly fall into the same range as the ones we obtain with the random forest retrieval. 

For $\epsilon$ Indi Ba and Bb, random forest retrieval results agree better with previous studies. Notably, both previous studies and model grids in this paper favor scaling factors well below unity, from 0.3 to 0.7 for both components (corresponding to radii of only 0.5--0.8~R$_\mathrm{Jup}$) with only B+97 and S+12 yielding values closer to unity. We emphasize that the scale factor can signify mismatches between the morphologies of grid spectra and observations and not the radius, which explains the large uncertainty intervals for this parameter for all three sources.

\subsection{Retrieval results for the L-T sequence}
\label{sect:retrieval_LT_sequence}

We applied the same retrieval procedure to our L dwarf and T dwarf sequence. Since clouds are expected to be present in some of the former spectral class, this comparison allows to better evaluate the impact of clouds on model grids such as the B+06/c, H+07/c, and M+11/c grids. 

\begin{figure*}[ht!]
\begin{center}
\includegraphics[width=0.98\textwidth]{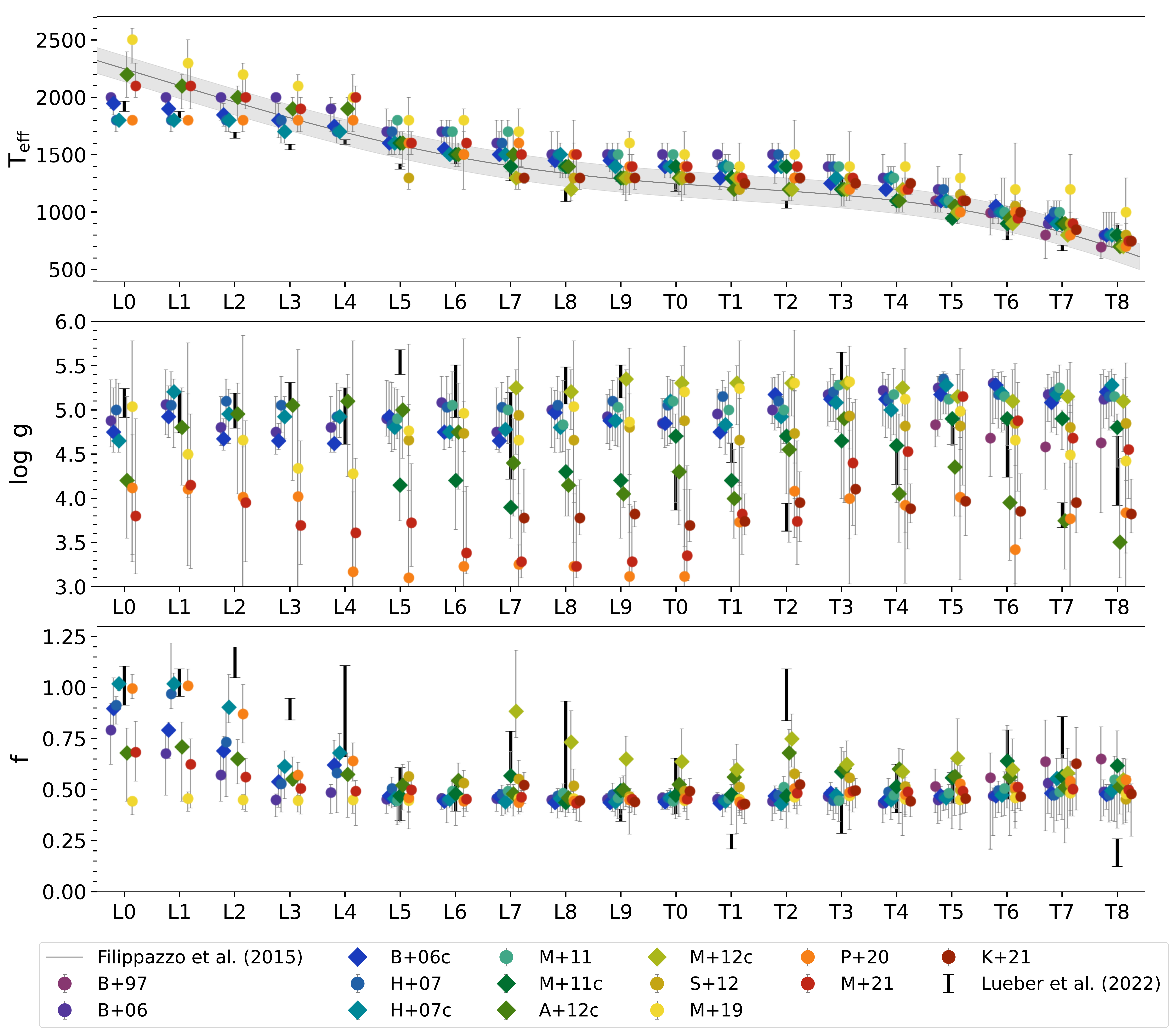}
\end{center}
\vspace{-0.1in}
\caption{Comparing the retrieved posterior values for effective temperature (top), surface gravity (middle), and scale factor (bottom) from our suite of brown dwarf retrievals across the L-T sequence. Shown are all permutations of the theoretical grid models for each spectral standard, as well as the posterior parameters retrieved by Helios-r2 \citep{Lueber2022ApJ...930..136L}. 1$\sigma$ uncertainties are shown. The sixth-order field age polynomial spectral type/effective temperature relation of \cite{Filippazzo2015ApJ}, and its 1$\sigma$ RMS are shown in the uppermost panel as a gray bar. It should be noted that object-specific uncertainties have been considered according to Section~\ref{sect:RF_grid_preparation}, and are, therefore, data-driven.}
\label{fig:DataApplicationSequence}
\end{figure*}

Figure~\ref{fig:DataApplicationSequence} displays the posterior values for effective temperature, surface gravity, and scaling factor as a function of spectral type and for each of the model grids. We also compare to results from our previous retrievals using a Bayesian nested sampling method, restricted spectra, all molecules, and gray clouds with \texttt{Helios-r2} as described in \citet{Lueber2022ApJ...930..136L}. The effective temperatures are generally predicted consistently and tend to follow the field age spectral type/effective temperature trend of \citet{Filippazzo2015ApJ}. This is particularly the case for the A+12c and M+21 models that span the full temperature range. Indeed, the model posteriors for $T_\mathrm{eff}$ for several models are effectively limited to the effective temperature range of the respective grid.

Retrieved surface gravities are consistently around $\log{g}~\approx~5$, appropriate for field low mass stars and brown dwarfs. The notable exceptions are M+11c, A+12c, M+21, and K+21 model grids that produce consistently low surface gravities likely due to the greater range in surface gravity covered by these models. Indeed, the fact that most of the grids are restricted to values between $4.5 \leq \log{g} \leq 5.5$, the tight trend in surface gravity is potentially spurious. This outcome is consistent with the grid inter-comparison discussed in Section~\ref{sect:Grid_comparison}, and further emphasizes that near-infrared spectral retrievals with grid models tend to yield poor predictions for surface gravities. \citet{Zhang2021bApJ...921...95Z} and \citet{Zalesky2022ApJ...936...44Z} specifically discuss discrepancies between the $\log{g}$ values based on grid model fits and brown dwarf evolution models, finding that the former 
typically underestimate the surface gravity. 

A remarkable trend in these fits is the consistently small values inferred in the scaling factor $f$, roughly 0.4--0.5 for spectral types later than L3 with small error bars (typically 0.1). There are a few exceptions to this behavior; for example, the M+12c grid occasionally yields larger $f$ values. The Bayesian retrieval method applied in \citet{Lueber2022ApJ...930..136L} also found small scale factors for mid L dwarfs and T dwarfs, albeit with more source-to-source variation and large posterior uncertainties.

It is clear that these scale factors should not be interpreted as a simple measure of brown dwarf radius, as it leads to un-physically small radii (0.6--0.7~R$_\mathrm{Jup}$) inconsistent with evolutionary models \citep{Burrows1993RvMP}. The exact reasons for this behavior are not yet fully understood, but may be related to temperature overestimation, unaccounted-for errors in the flux calibration of the spectra or missing ingredients in the models \citep{Kitzmann2020ApJ}.

\begin{figure*}[ht!]
\gridline{\fig{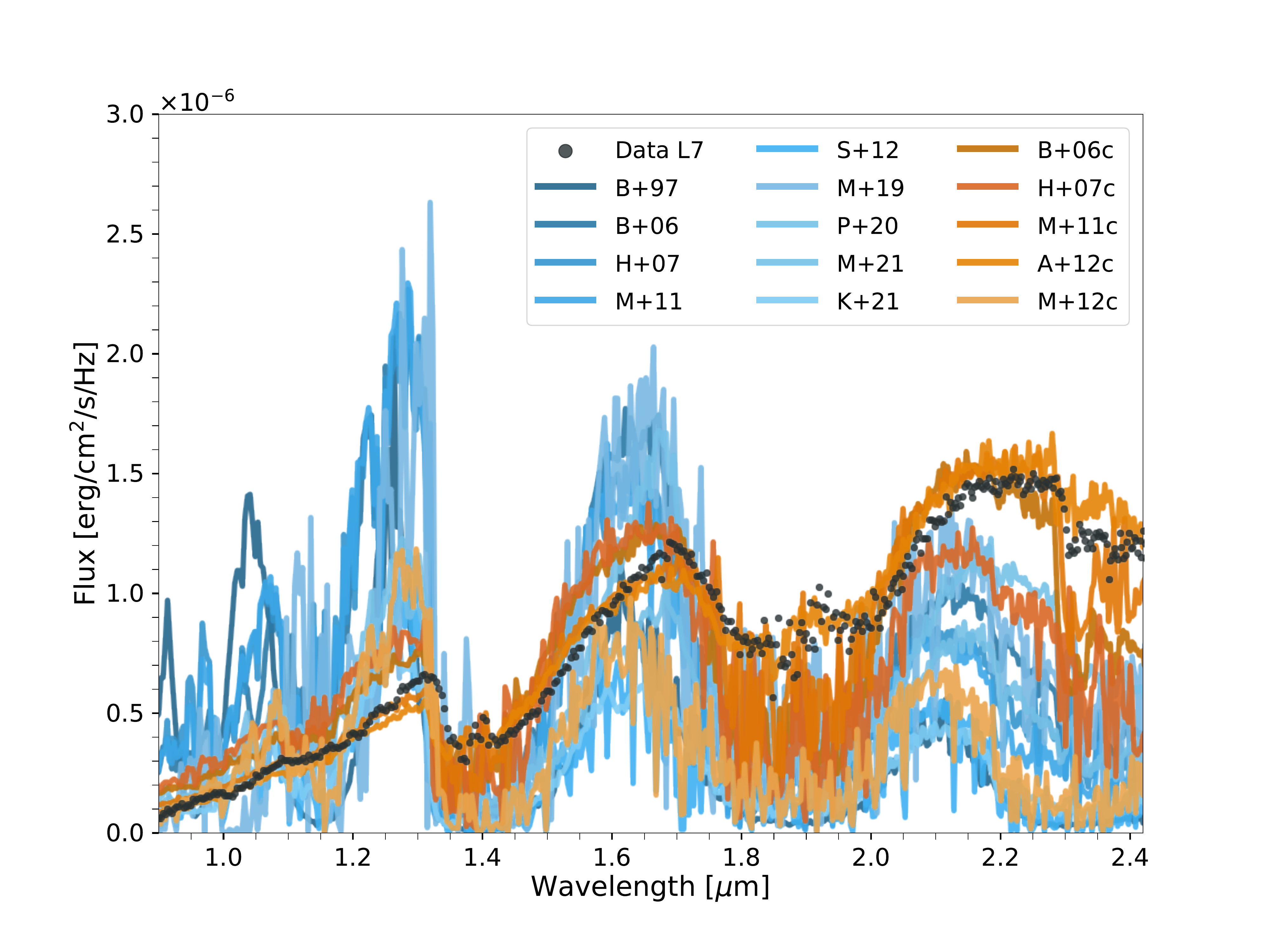}{0.5\textwidth}{L7 dwarf: 2MASS J01033203+1935361}
\fig{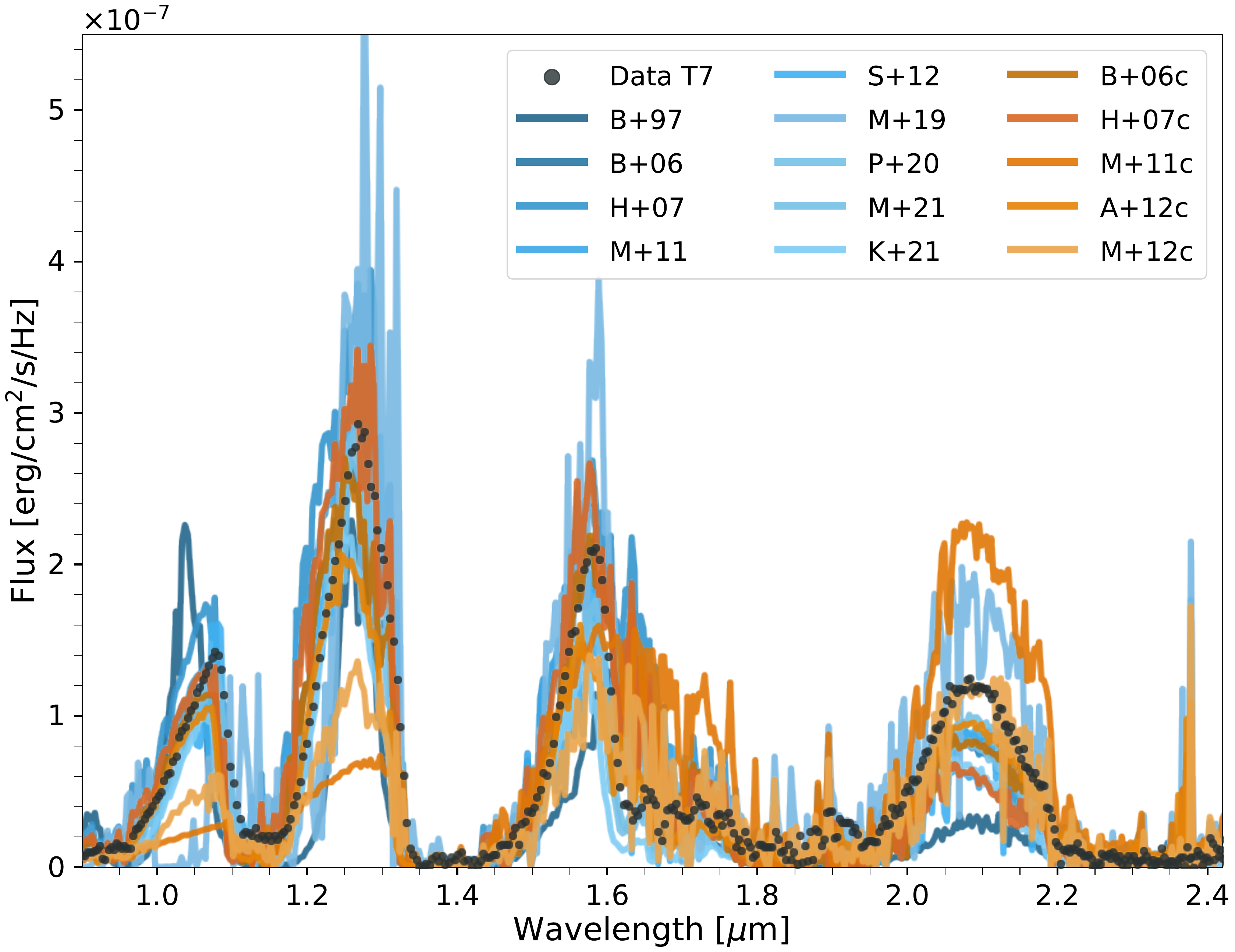}{0.49\textwidth}{T7 dwarf: 2MASS J07271824+1710012}}
\caption{Near-infrared spectra L7 (left) and T7 (right) standards (black dots), compared to median retrieved spectra from the 14 model grids. The latter are shown in shades of blue for cloud free models and shades of orange for cloud models. For clarity, we have omitted data uncertainties in the plot, but it should be noted that object-specific uncertainties have been considered according to Section~\ref{sect:RF_grid_preparation}, and are, therefore, data-driven.}
\label{fig:L7_T7_Comparison}          
\end{figure*}

Figure~\ref{fig:L7_T7_Comparison} shows comparisons of the median spectra for each grid with the observed data for two spectral types: an L7 dwarf and a T7 dwarf. Model fits to the former are expected to require the presence of clouds, while the photosphere of the latter, like Gl 570 D, is assumed to be largely free of clouds (with the possible exception of salt or sulfide clouds; \citealt{Morley2012ApJ...756..172M}). A well-known problem of existing model grid spectra is their inability to reproduce the general slope of near-infrared spectra of L-dwarfs, in particular the shape of the H-band peak. For example, studies by \cite{Manjavacas2014A&A...564A..55M, Manjavacas2016MNRAS.455.1341M} discuss this phenomenon for the \texttt{BT-Settl} model grid (A+12c, \citealt{Allard2012EAS....57....3A}). On the other hand, this model grid is able to reproduce the spectral energy distributions of late T dwarfs quite well \citep{Burningham2011MNRAS.414.3590B}. If one focuses on mid and late T dwarfs, which are expected to be cloud-free, generally both cloudy and cloud-free model grids fit the data equally well. In contrast, for L-dwarfs only cloudy model grids fit the data adequately and produce plausible posterior values. The cloud-free models for the L7 dwarf in Figure~\ref{fig:L7_T7_Comparison} tend to result in reduced $\chi_r^2$-values at the $\sim 10$ level, whereas the best fitting cloudy models have a reduced $\chi_r^2$-value at the $\sim 1$ level.

\section{Discussion}
\label{sect:Discusion}

\begin{figure*}[ht!]
\begin{center}
\includegraphics[width=0.7\textwidth]{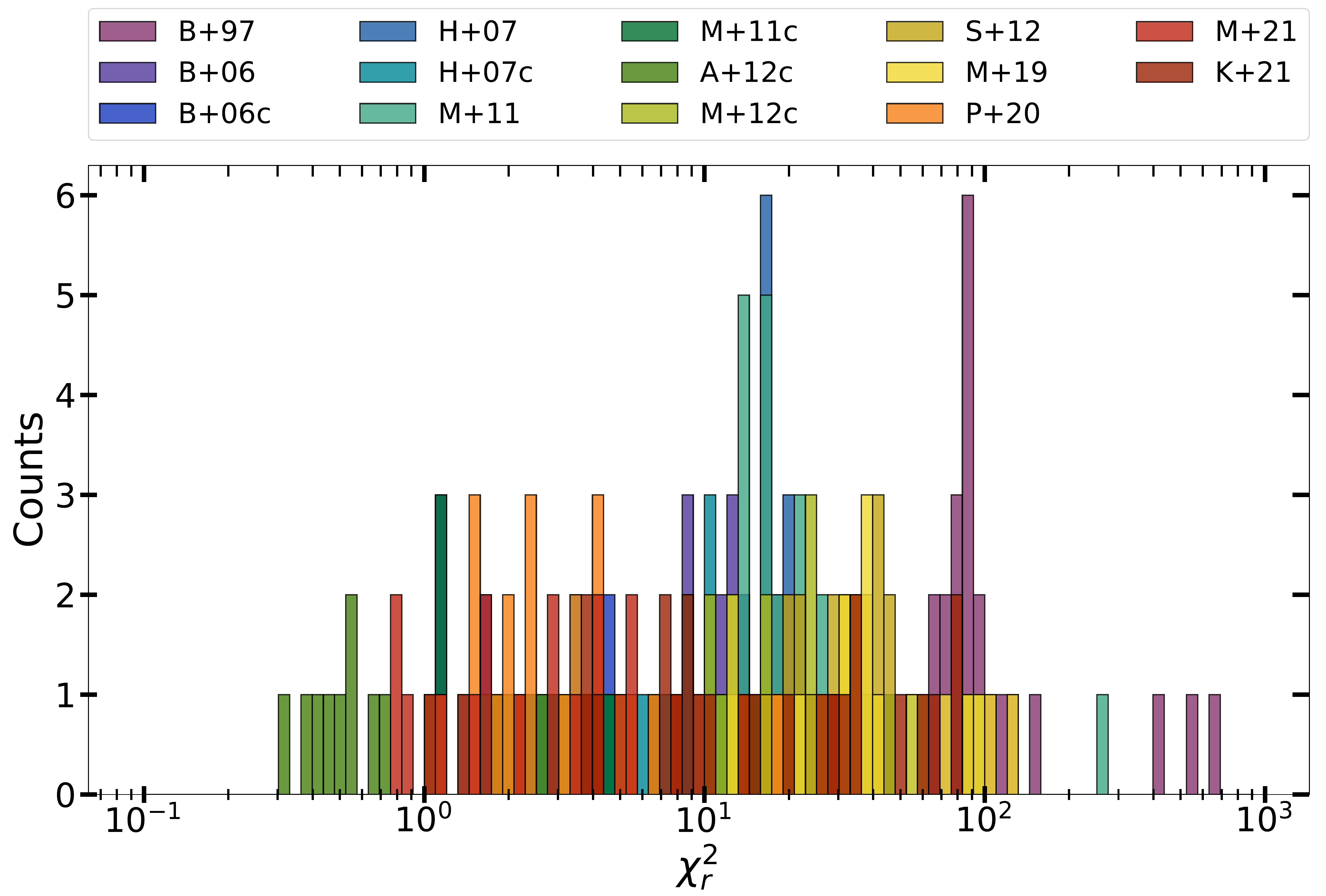}
\end{center}
\vspace{-0.1in}
\caption{Reduced chi-square values $\chi_r^2$ associated with models fits to our curated dataset, which includes 3 benchmark brown dwarfs (Gl 570D, $\epsilon$ Indi Ba and Bb) as well as a sample of 19 L and T dwarfs across the L-T sequence. Each color represents one of the 14 legacy model grids.}
\label{fig:Chi2_reduced}
\end{figure*}

In the current study, we have performed a systematic study of the information content of 14 previously published model grids of brown dwarf spectra.  One of our main conclusions is that inferring $\log{g}$ from spectra is a model-dependent exercise.  In particular, the A+12, M+21, and K+21 model grids (\texttt{BT-Settl}, \texttt{Sonora Bobcat} \& \texttt{Sonora Cholla}) tend to predict $\log{g} \sim 3$--4 (cgs units).  Since we were not involved in the construction of these model grids, we are unable to identify the underlying reasons for this systematic prediction of somewhat low surface gravity values (compared to $\log{g} \sim 5$ by other model grids).  Our ability to reach such an outcome is both a strength and a weakness of our supervised machine learning method, which allows us to  use model grids as training sets without necessarily being privy to all of the details involved in constructing these models.

Figure \ref{fig:Chi2_reduced} shows the distribution of $\chi^2_r$ values for all of the model fits to data that were performed in the current study.  The presence of model fits with $\chi^2_r \approx 0.4$--1 indicates that there are no obvious mistakes in the implementation of our random forest fitting procedure.  The presence of bad fits ($\chi^2_r \gg 1$) is either due to the mismatch between model assumptions and data (e.g. fitting a cloudy object with a cloud-free model) or hints at missing physics or chemistry in a model grid--despite it being self-consistent.  Clearly, self-consistency alone does not guarantee accuracy.

The current study highlights a longstanding puzzle in the research of brown dwarfs, which is the formation of clouds and their influence on the emergent spectra (see reviews by \citealt{Marley2013cctp.book..367M}, \citealt{Helling2014A&ARv..22...80H} or \citealt{Madhusudhan2018haex.bookE.104M}).  Ideally, one wishes to model the formation and evolution of clouds from first principles.  In practice, studies published in the literature by different groups have adopted a heterogeneous set of assumptions and approximations in their calculations. While the Tsuji model (e.g., \citealt{Tsuji2002ApJ}) uses an ad-hoc cloud particle size of 10$^2 \mu$m, \cite{Allard2001ApJ} and \cite{Ackerman2001ApJ} assumed power-law and broad log-normal size distribution, respectively.  \cite{Burrows2011ApJ} and \cite{Madhusudhan2011ApJ...737...34M} assumed a size distribution informed by cumulus clouds on Earth, the so-called Deirmendjian distribution, while describing the vertical spatial structure using several free parameters. More recently, the tendency is to utilize the cloud parameterization proposed by \cite{Ackerman2001ApJ}, e.g., \texttt{TauREx} \citep{AlRefaie2021ApJ...917...37A} or \texttt{Sonora Diamondback} (Morley et al., in prep.). 

It is worth emphasizing that, as soon as clouds are parameterized within a model grid, self-consistency is, strictly speaking, lost.  In other words, a truly self-consistent model should be able to predict the size distribution, chemical composition and spatial structure of clouds without having to specify these properties as parameters. Pragmatically, practitioners will continue to parameterize cloud models and iterate them until radiative-convective equilibrium is reached.  The strict definition of self-consistency remains a long-term, and possibly unattainable, goal for the brown dwarf community.  

A related issue is that these cloud models are sophisticated enough that the measured spectra typically do not encode enough information to test or falsify some of these assumptions.  In other words, the puzzle of understanding clouds in brown dwarfs is two-fold: the formidable task of formulating a first-principles cloud model free of ad-hoc parameters, and the subsequent validation of this model using data.

\section*{acknowledgments}
We acknowledge partial financial support from the Swiss National Science Foundation and the European Research Council (via a Consolidator Grant to KH; grant number 771620), as well as administrative support from the Center for Space and Habitability (CSH). Additionally, we gratefully acknowledge the open-source libraries in the Python programming language that made this work possible: \textbf{scikit.learn} \citep{scikit-learn2011}, \textbf{numpy} \citep{Harris2020Natur.585..357H}, \textbf{matplotlib} \citep{Hunter2007CSE.....9...90H}, and \textbf{astropy} \citep{astropy:2013, astropy:2018, astropy:2022}.

\bibliographystyle{aasjournal}
\bibliography{reference}

\appendix
\label{sect: Appendix}

Figure~\ref{fig:Effect of interpolation} represents the real versus predicted (RvP) comparisons of the random forest approach. Training and predicting was done using a single grid. RvP comparisons were done for effective temperature T$_\mathrm{eff}$, surface gravity $\log{g}$, and scaling factor $f$. The size of the grid in the two dimensions $\log{g}$ and temperature T$_\mathrm{eff}$ are represented by specification in `\#'.\\

\begin{figure}[h!]
\gridline{\fig{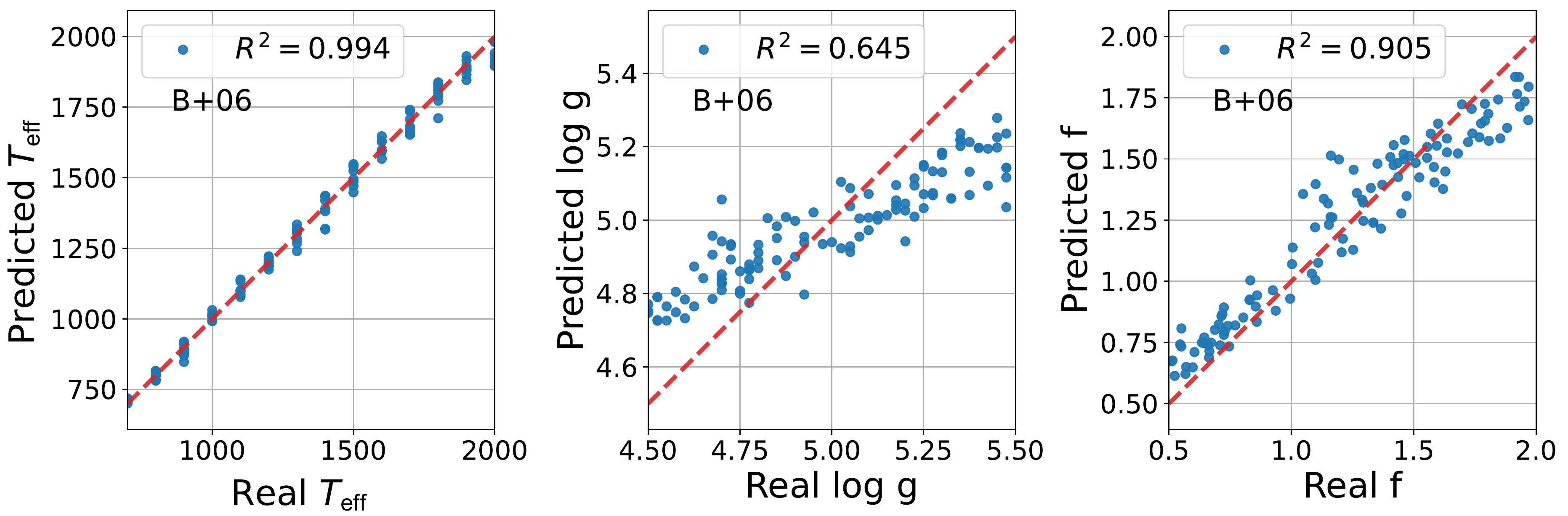}{0.49\textwidth}{B+06: \# T$_\mathrm{eff}$ = 14, \# $\log{g}$ = 40}
          \fig{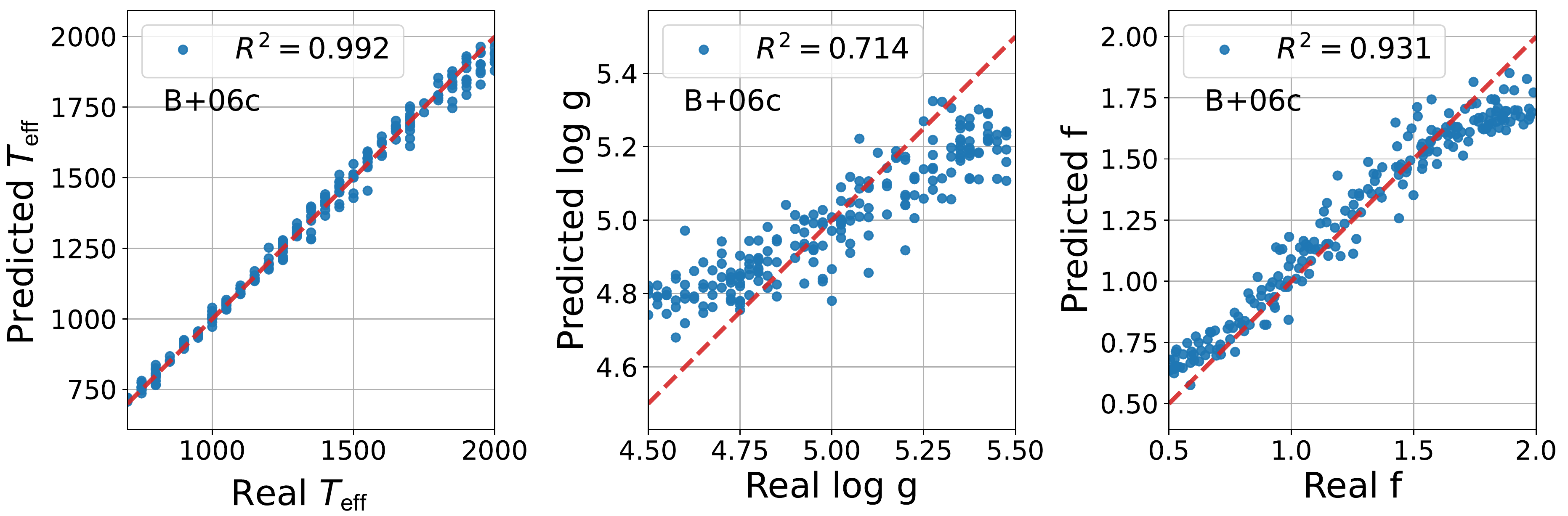}{0.49\textwidth}{B+06c: \# T$_\mathrm{eff}$ = 27, \# $\log{g}$ = 40}}
\gridline{\fig{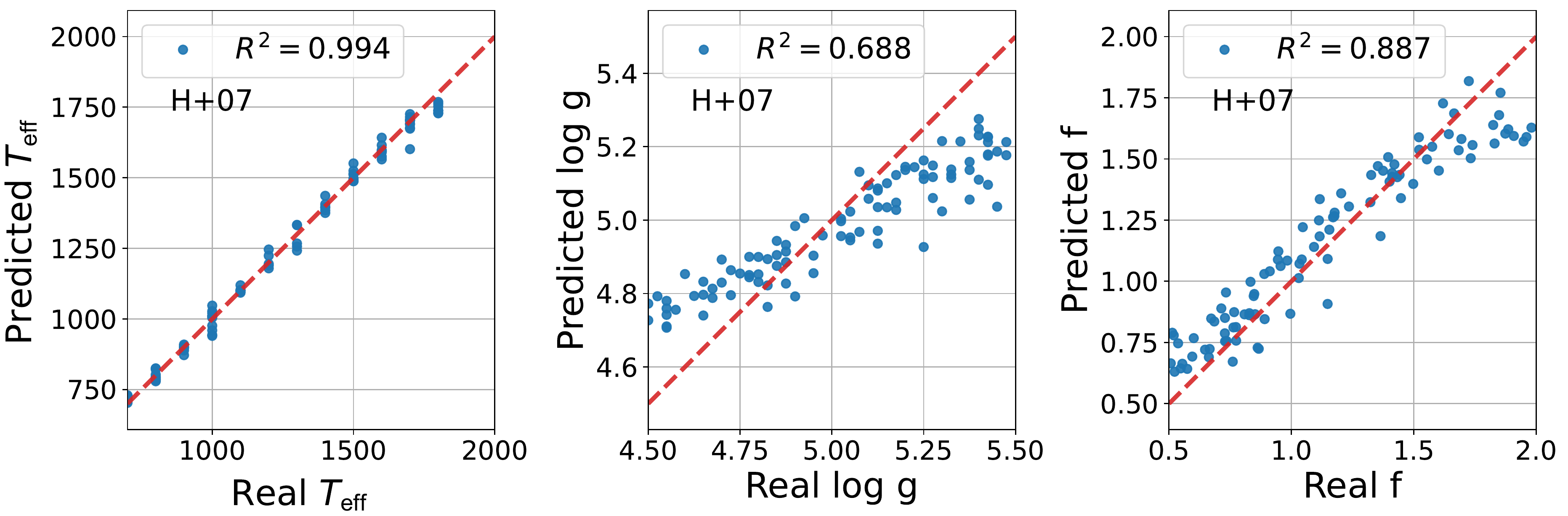}{0.49\textwidth}{H+07: \# T$_\mathrm{eff}$ = 12, \# $\log{g}$ = 40}
          \fig{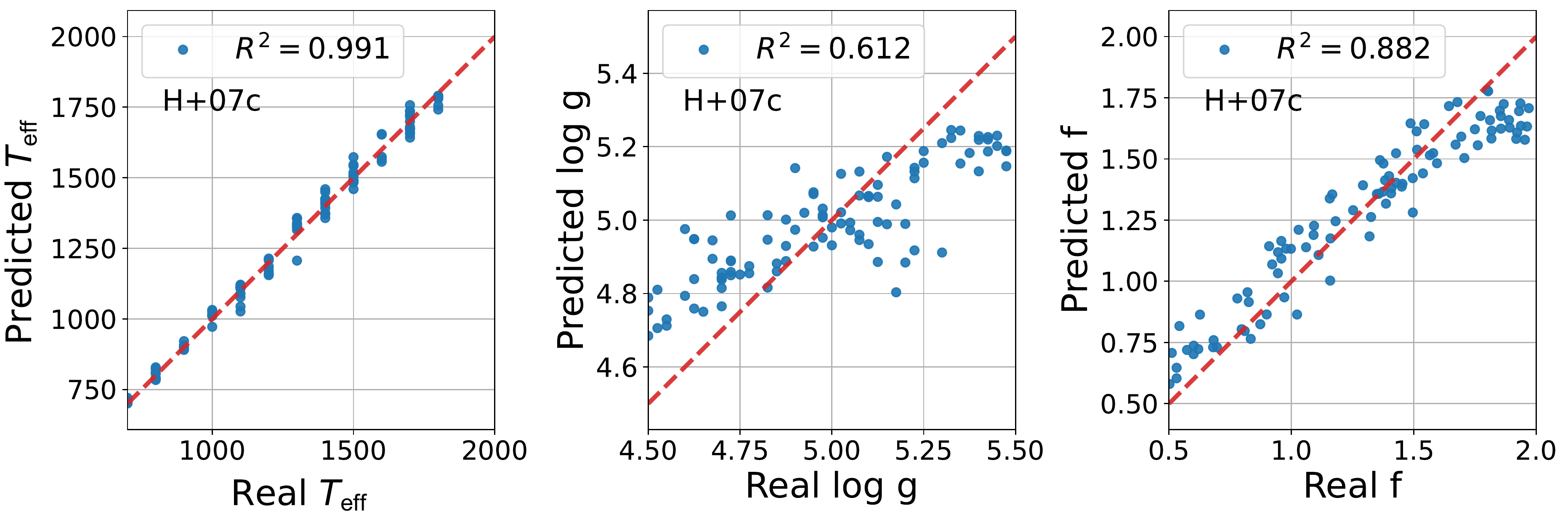}{0.49\textwidth}{H+07c: \# T$_\mathrm{eff}$ = 12, \# $\log{g}$ = 40}}
\gridline{\fig{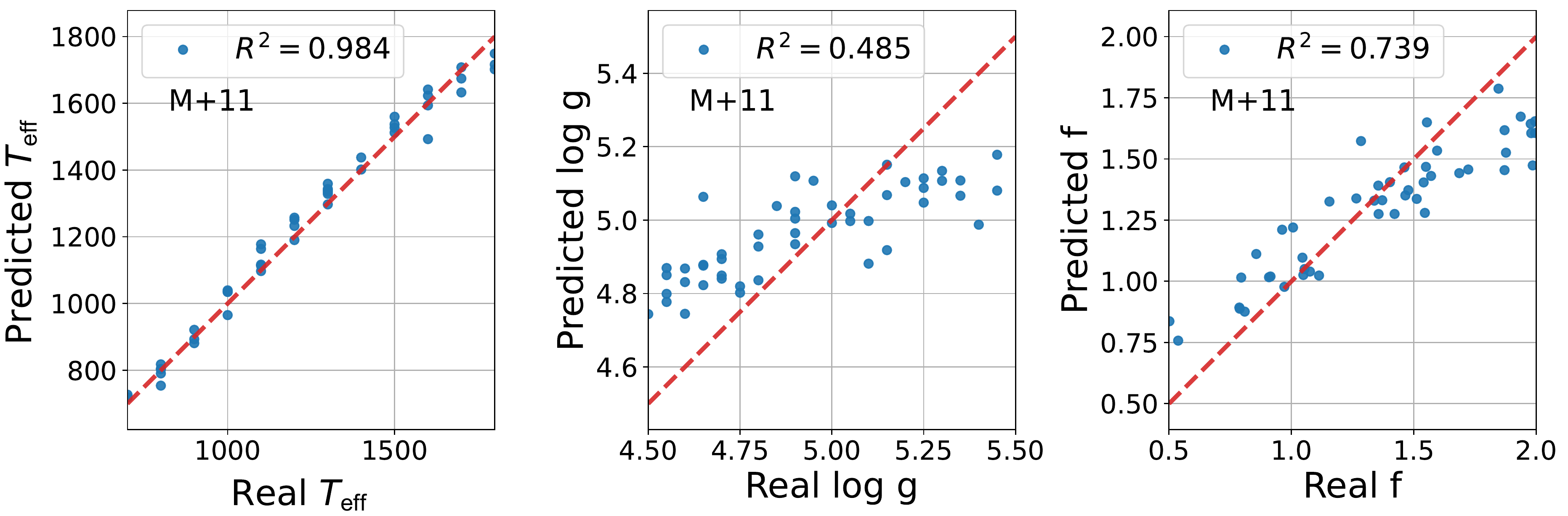}{0.49\textwidth}{M+11: \# T$_\mathrm{eff}$ = 12, \# $\log{g}$ = 20}
          \fig{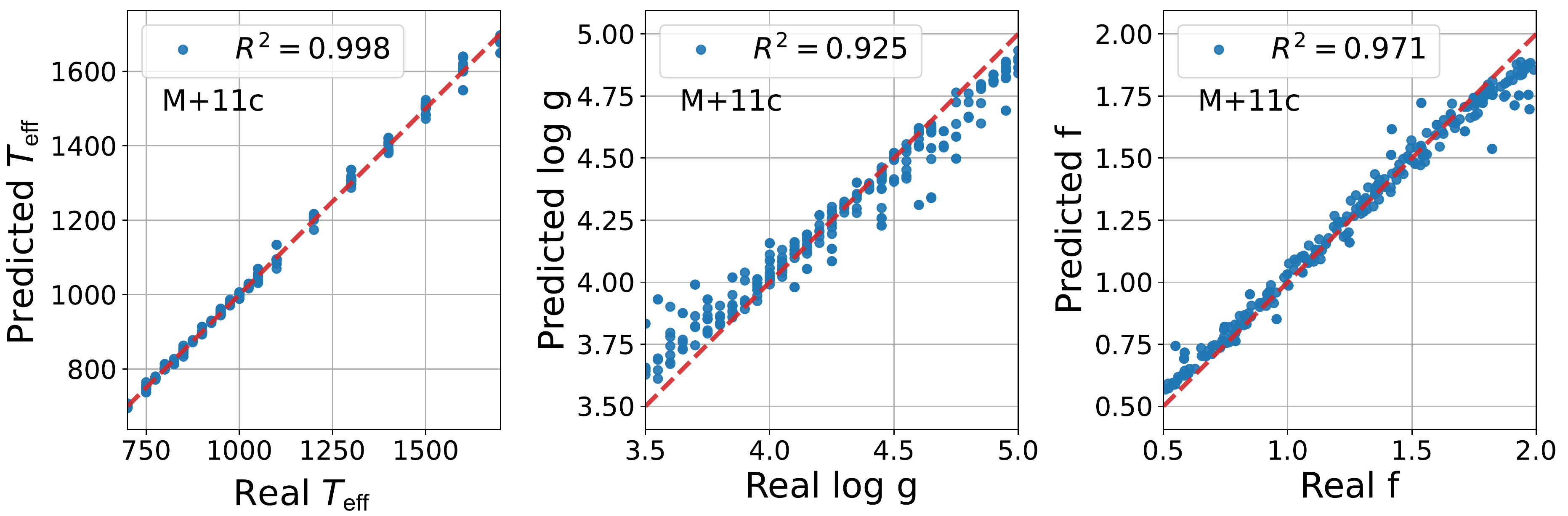}{0.49\textwidth}{M+11c: \# T$_\mathrm{eff}$ = 23, \# $\log{g}$ = 30}}
\gridline{\fig{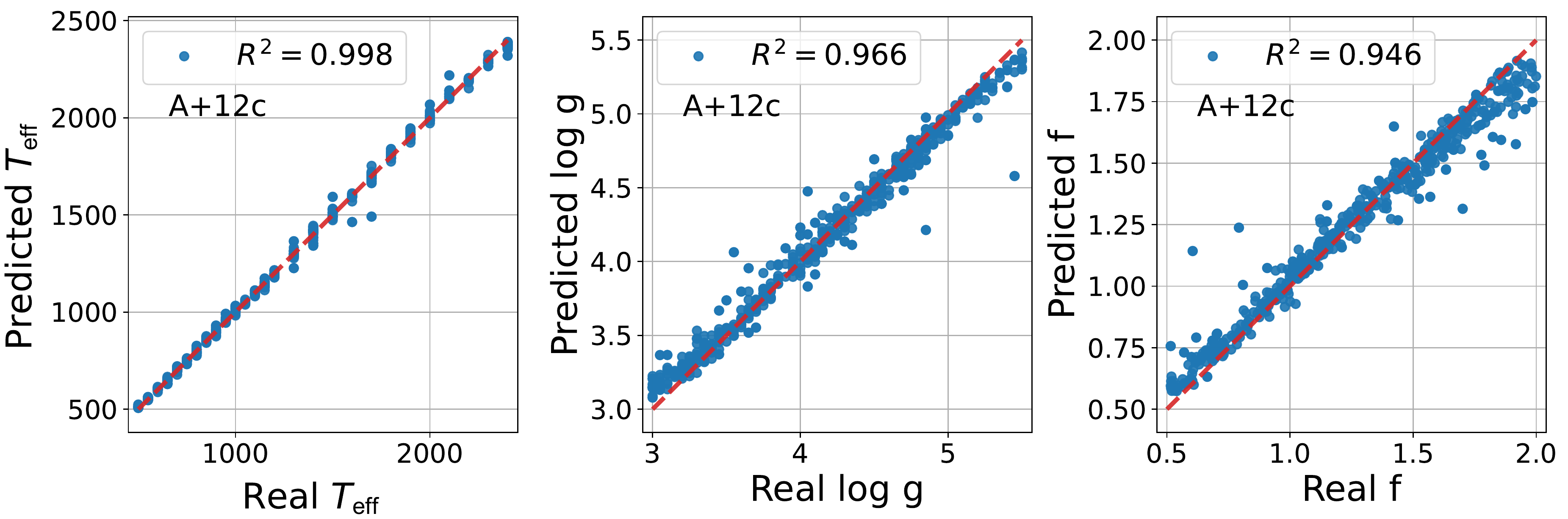}{0.49\textwidth}{A+12c: \# T$_\mathrm{eff}$ = 27, \# $\log{g}$ = 50}
          \fig{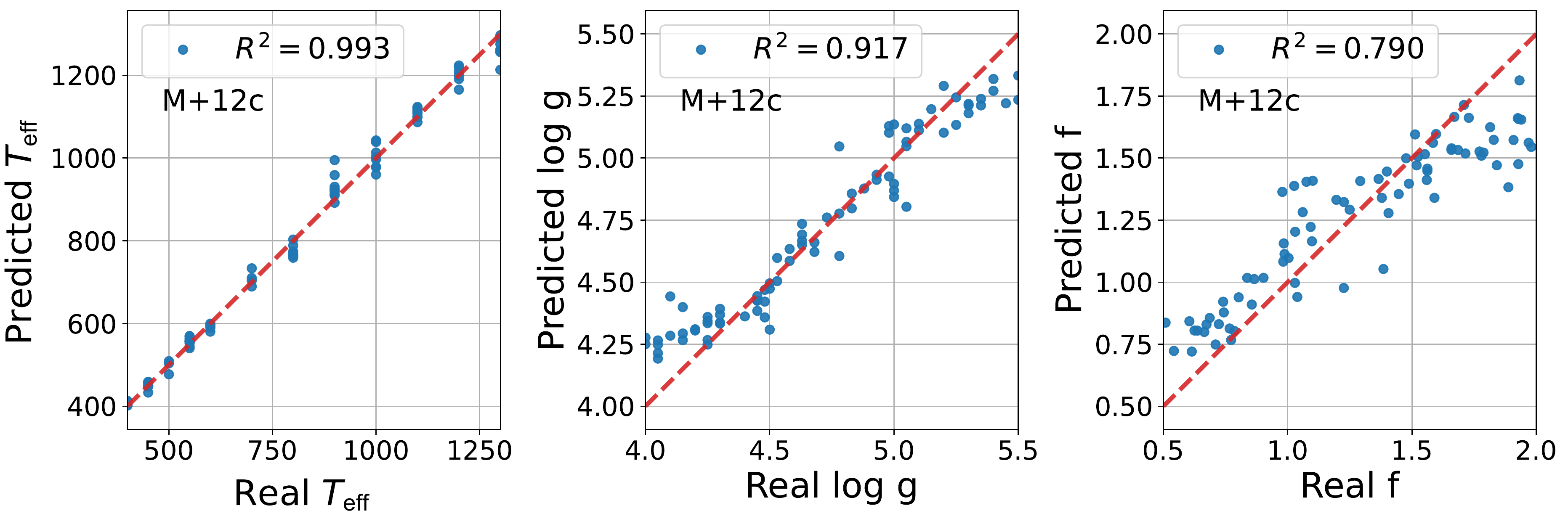}{0.49\textwidth}{M+12c: \#  T$_\mathrm{eff}$ = 12, \# $\log{g}$ = 32}}
\caption{RvP comparison of the random forest training and testing approach when using a single grid. Each panel represents the RvP comparisons for effective temperature T$_\mathrm{eff}$, surface gravity $\log{g}$, and scaling factor $f$. Each grid was interpolated individually for $\log{g}$ (see text for details) and binned on the spectral resolution and wavelength coverage of the $\epsilon$ Indi Ba spectrum. Model uncertainties have been considered according to Section~\ref{sect:RF_grid_preparation}.}
\label{fig:Effect of interpolation}          
\end{figure}

\begin{figure}[ht!]
\gridline{\fig{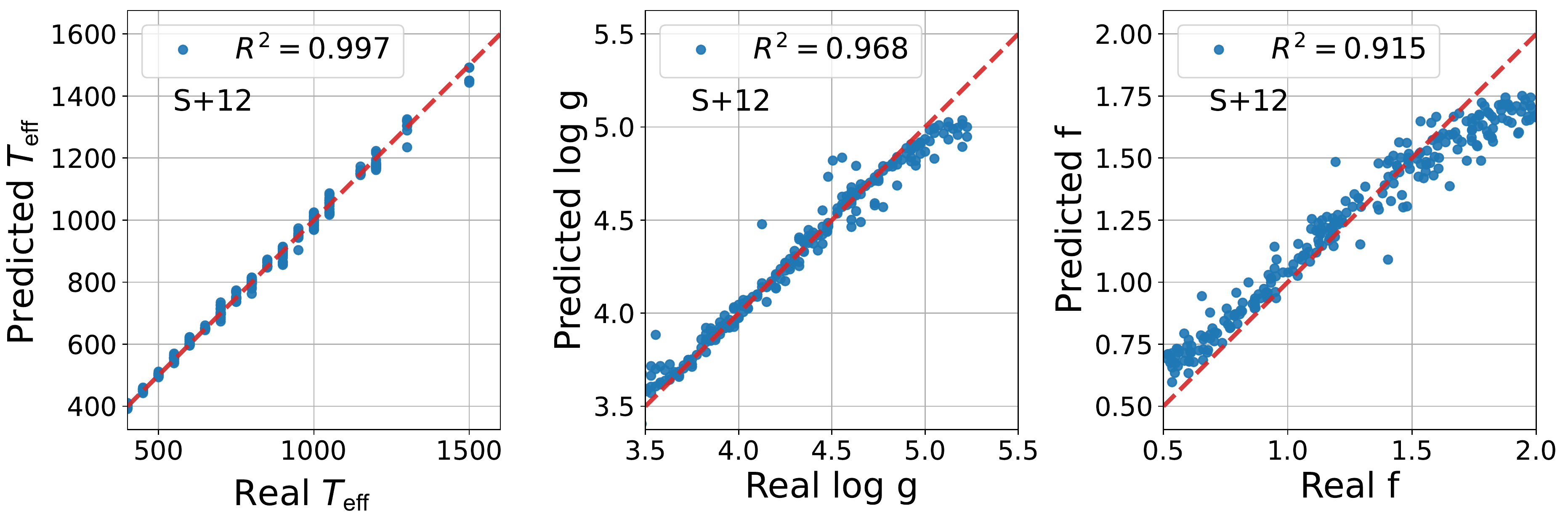}{0.49\textwidth}{S+12: \# T$_\mathrm{eff}$ = 20, \# $\log{g}$ = 62}
          \fig{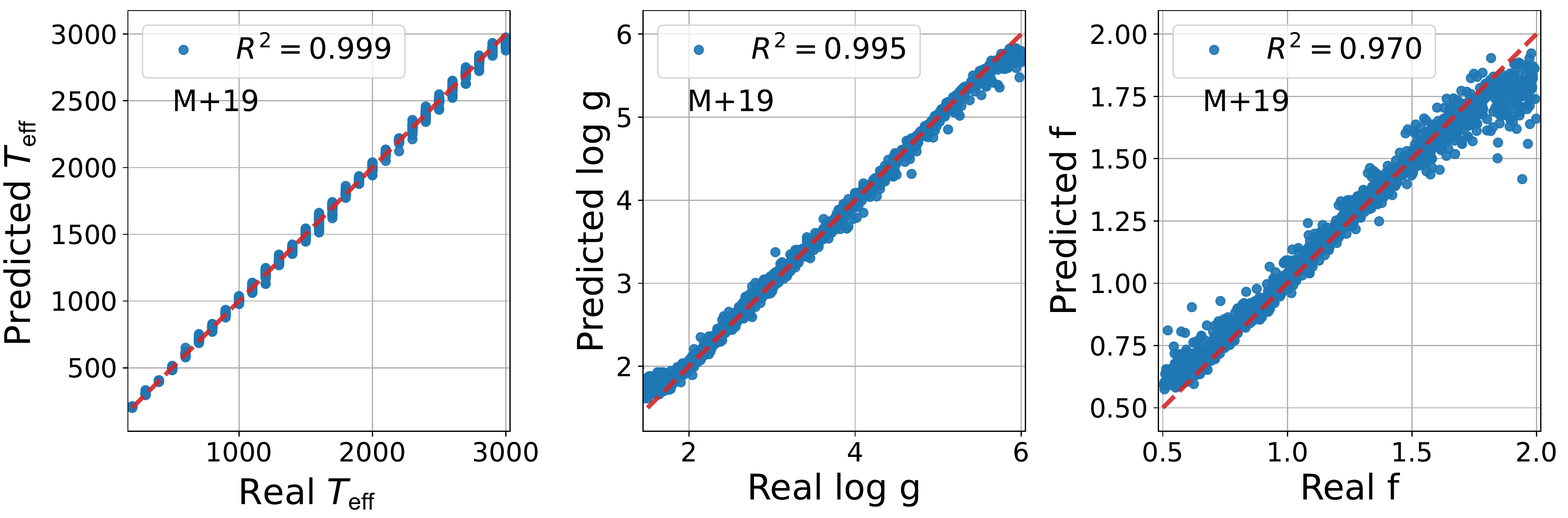}{0.49\textwidth}{M+19: \# T$_\mathrm{eff}$ = 29, \# $\log{g}$ = 230}}
\gridline{\fig{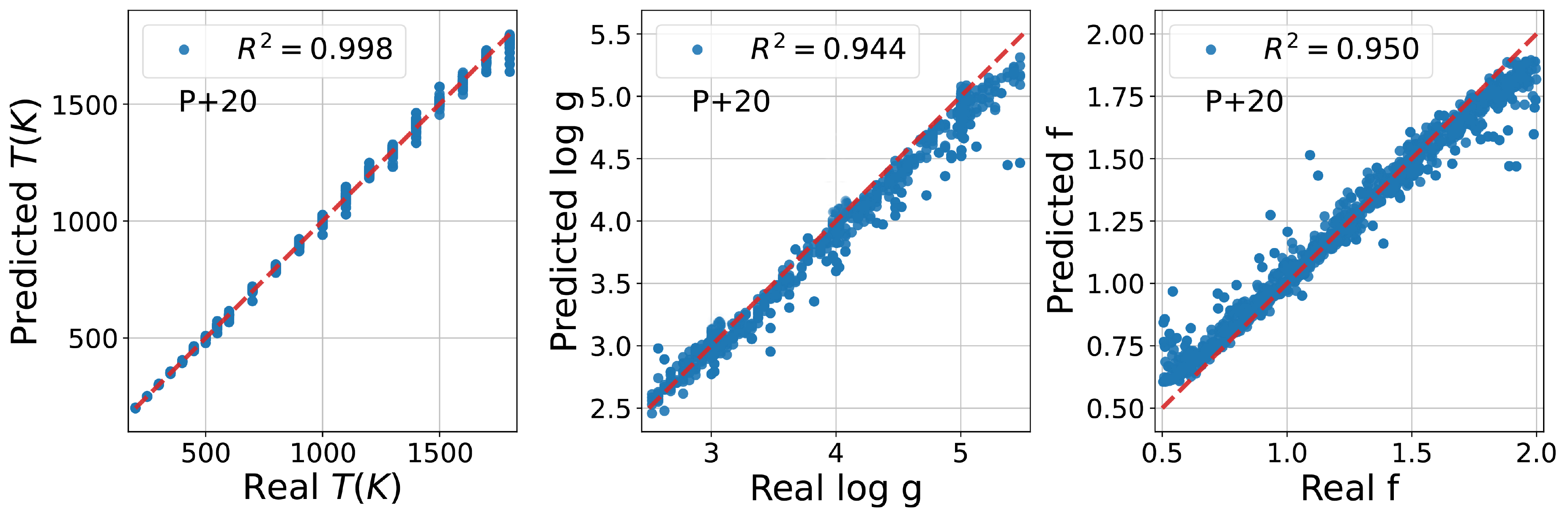}{0.49\textwidth}{P+20: \# T$_\mathrm{eff}$ = 20, \# $\log{g}$ = 100}
          \fig{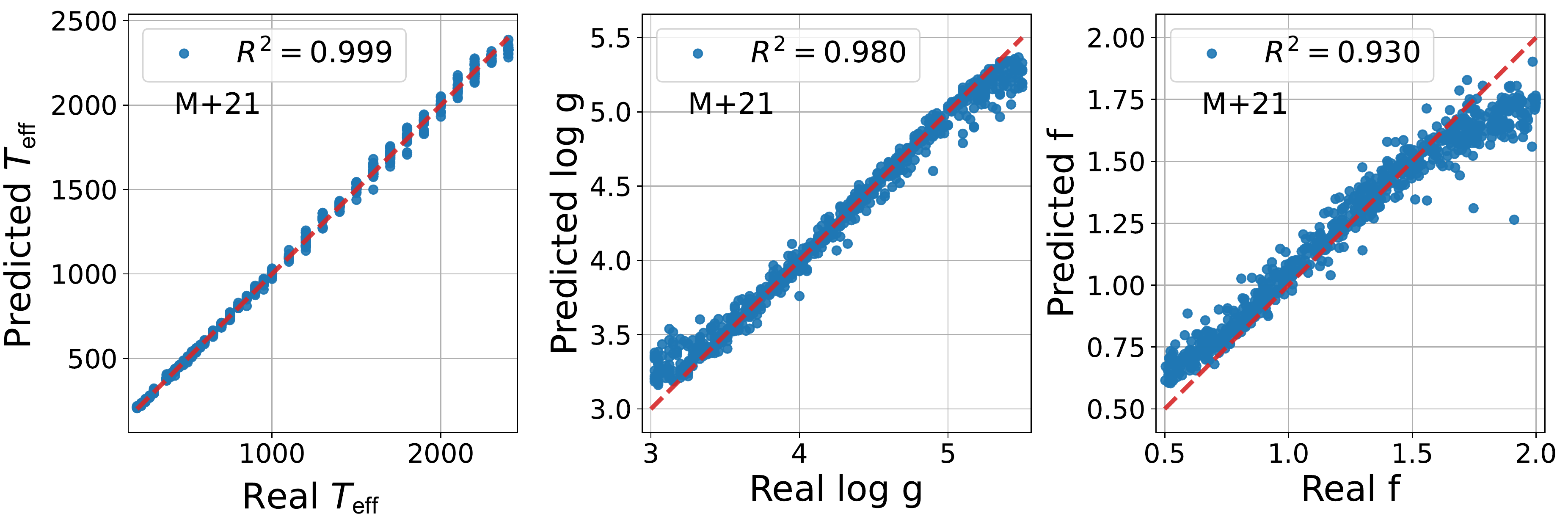}{0.49\textwidth}{M+21: \# T$_\mathrm{eff}$ = 36, \# $\log{g}$ = 100}}
\gridline{\fig{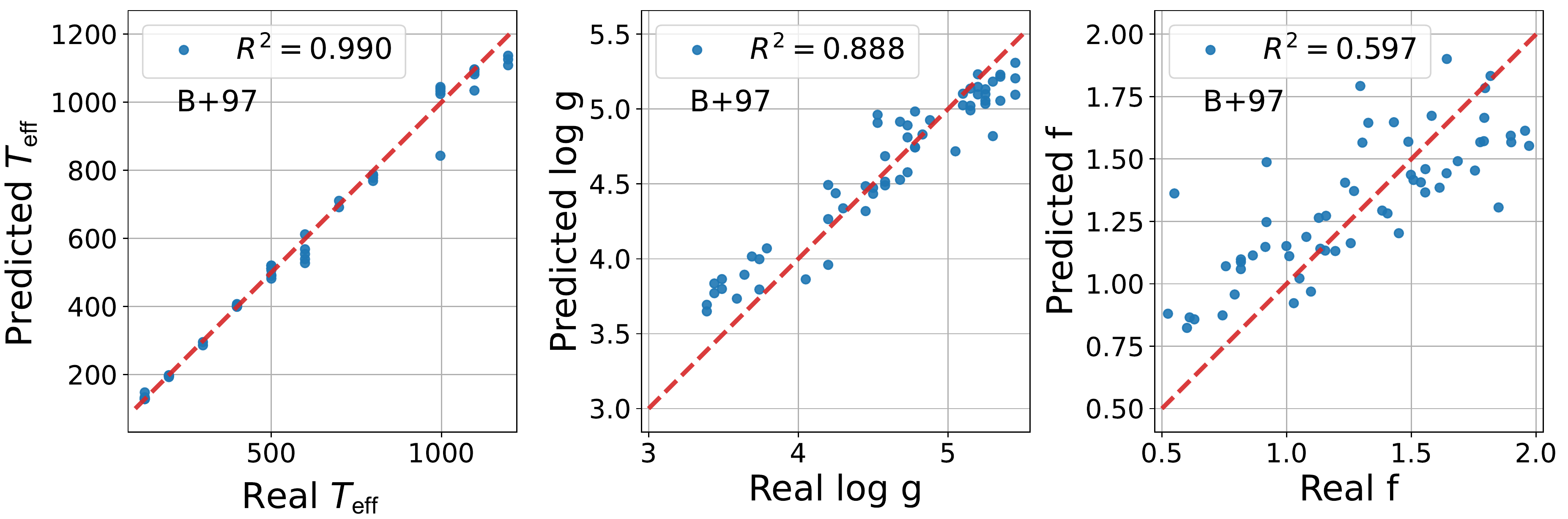}{0.49\textwidth}{B+97: \# T$_\mathrm{eff}$ = 11, \# $\log{g}$ = 26}
          \fig{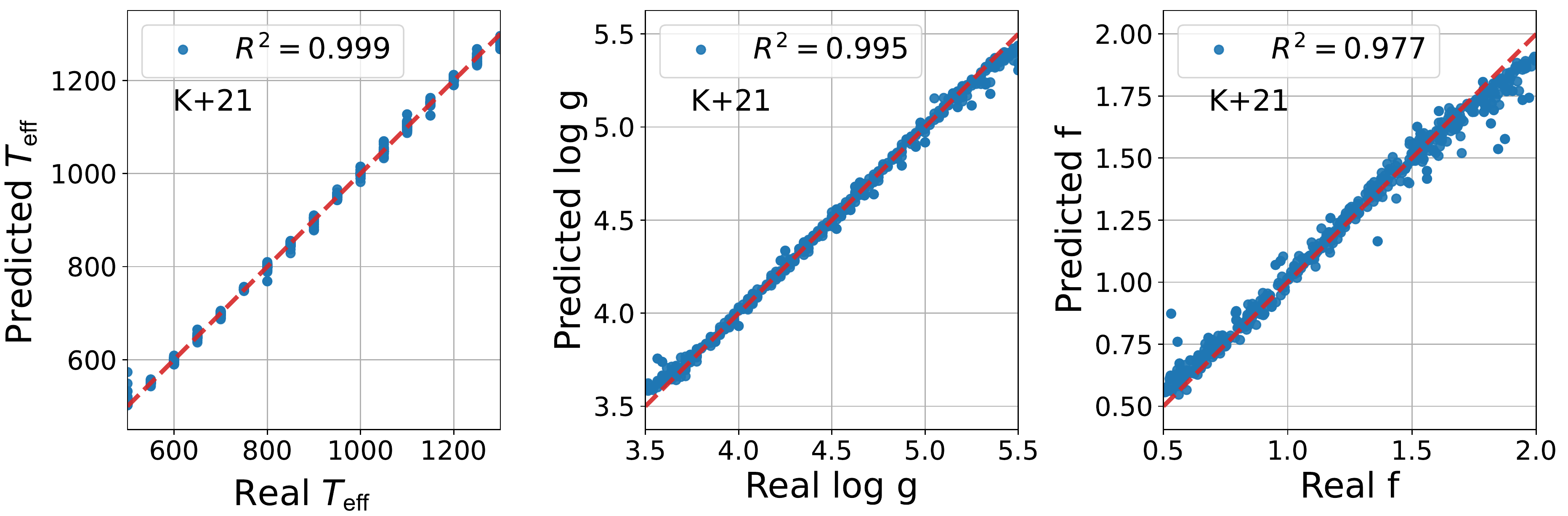}{0.49\textwidth}{K+21: \# T$_\mathrm{eff}$ = 17, \# $\log{g}$ = 80}}   
\figurenum{\ref{fig:Effect of interpolation}}
\caption{continued. Model uncertainties have been considered according to section~\ref{sect:RF_grid_preparation}.}     
\end{figure}

\begin{figure}[ht!]
    \centering
    \includegraphics[width=6.8cm]{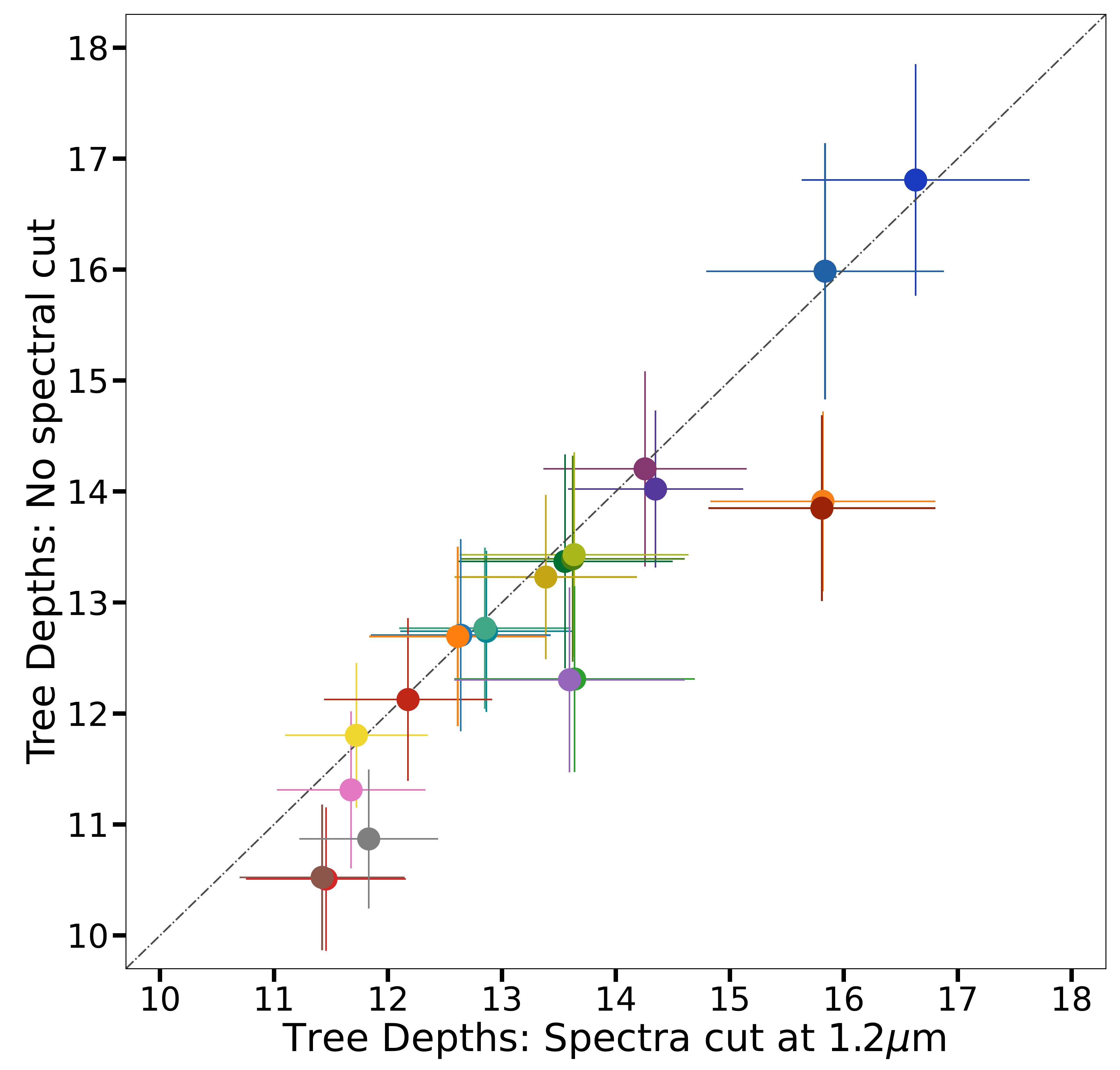}
    \caption{Mean tree depths for random forest calculations with full versus restricted wavelength ranges, corresponding to Figure~\ref{fig:Effect_of_wavelength_coverage}.}
    \label{fig:Treedepths}
\end{figure}

\newpage
In Figure~\ref{fig:Treedepths}, we present an example of the mean tree depths obtained from random forest calculations using full and restricted wavelength ranges, corresponding to Figure~\ref{fig:Effect_of_wavelength_coverage}. We observe that the runs associated with spectra with restricted wavelength ranges (cut off blueward of 1.2 microns) have slightly larger tree depths, although the underlying reason for this outcome remains elusive. For completeness, Tables~\ref{tab:data_posteriors_L_dwarfs} and \ref{tab:data_posteriors_T_dwarfs} record the outcomes of the random forest retrieval algorithm for the SpeX prism standard L0-L9 \& T0-T8 brown dwarfs, respectively.

\begin{deluxetable*}{cccccccccccc}[ht!]
\tabletypesize{\scriptsize}
\tablecolumns{12}
\tablewidth{\columnwidth}
\tablecaption{Results of the random forest retrieval algorithm for the SpeX prism standard brown dwarfs and L0 - L9 spectral classes.}
\label{tab:data_posteriors_L_dwarfs}
\tablehead{
 \colhead{Model} & \colhead{Parameter} & \colhead{L0} & \colhead{L1} & \colhead{L2} & \colhead{L3} & \colhead{L4} & \colhead{L5} & \colhead{L6} & \colhead{L7} & \colhead{L8} & \colhead{L9}
}
\startdata
\text{B+97} & T$_\mathrm{eff}$ \text{(K)} & $1200^{+0}_{-0}$ & $1200^{+0}_{-99}$ & $1200^{+0}_{-99}$ & $1200^{+0}_{-99}$ & $1200^{+0}_{-99}$ & $1200^{+0}_{-99}$ & $1100^{+99}_{-100}$ & $1200^{+0}_{-99}$ & $1100^{+99}_{-100}$ & $1100^{+99}_{-100}$	\\
 & $\log{g}$ (\text{cm s}$^{−2}$) & $5.10^{+0.40}_{-0.05}$ & $5.20^{+0.25}_{-0.42}$ & $5.25^{+0.05}_{-0.20}$ & $5.15^{+0.20}_{-0.52}$ &$5.15^{+0.30}_{-0.52}$ & $5.15^{+0.15}_{-0.47}$ & $4.88^{+0.42}_{-0.30}$ & $5.15^{+0.20}_{-0.52}$ & $4.78^{+0.57}_{-0.25}$ & $4.78^{+0.47}_{-0.20}$	\\
 & $f$ & $1.870^{+0.111}_{-0.659}$ & $1.490^{+0.152}_{-0.471}$ & $1.520^{+0.463}_{-0.519}$ & $1.640^{+0.221}_{-0.476}$ & $1.700^{+0.212}_{-0.345}$ & $1.780^{+0.187}_{-0.561}$ & $1.460^{+0.354}_{-0.488}$ & $1.340^{+0.519}_{-0.340}$ & $1.600^{+0.296}_{-0.636}$ & $1.200^{+0.429}_{-0.459}$	\\
 & $\chi_r^2$ & 65.31 & 66.22 & 70.39 & 77.86 & 73.87 & 85.84 & 89.57 & 95.36 & 104.46 & 119.01 \\ \hline
\text{B+06} & T$_\mathrm{eff}$ \text{(K)}  &$2000^{+0}_{-100}$ & $2000^{+0}_{-100}$ & $2000^{+0}_{-200}$ & $2000^{+0}_{-200}$ & $1900^{+100}_{-200}$ & $1700^{+200}_{-100}$ & $1700^{+100}_{-100}$ & $1600^{+200}_{-100}$ & $1500^{+200}_{-100}$ & $1500^{+100}_{-100}$	\\
 & $\log{g}$ (\text{cm s}$^{−2}$) & $4.88^{+0.46}_{-0.30}$ & $5.06^{+0.39}_{-0.34}$ & $4.80^{+0.35}_{-0.20}$ & $4.75^{+0.40}_{-0.25}$ &$4.80^{+0.35}_{-0.28}$ & $4.90^{+0.43}_{-0.20}$ &
 $5.08^{+0.30}_{-0.35}$ & $4.75^{+0.50}_{-0.20}$ & $5.00^{+0.25}_{-0.43}$ & $4.92^{+0.25}_{-0.30}$	\\
 & $f$ & $1.380^{+0.391}_{-0.304}$ & $1.110^{+0.464}_{-0.291}$ & $0.867^{+0.296}_{-0.233}$ & $0.593^{+0.193}_{-0.065}$ & $0.670^{+0.233}_{-0.097}$ & $0.597^{+0.162}_{-0.051}$ & $0.604^{+0.122}_{-0.065}$ & $0.604^{+0.140}_{-0.068}$ & $0.593^{+0.144}_{-0.050}$ & $0.595^{+0.098}_{-0.067}$	\\ 
 & $\chi_r^2$ & 8.46 &  8.02 & 10.33 & 8.65 & 11.82 & 10.65 & 12.86 & 16.05 & 16.23 & 13.90\\ \hline
 \text{B+06c} & T$_\mathrm{eff}$ \text{(K)} &$1950^{+50}_{-100}$ & $1900^{+100}_{-100}$ & $1850^{+100}_{-100}$ & $1800^{+150}_{-150}$ & $1750^{+200}_{-50}$ & $1600^{+200}_{-100}$ & $1550^{+100}_{-50}$ & $1500^{+200}_{-50}$ &  $1450^{+50}_{-100}$ & $1450^{+50}_{-100}$	\\
 & $\log{g}$ (\text{cm s}$^{−2}$) & $4.75^{+0.50}_{-0.23}$ & $4.92^{+0.35}_{-0.23}$ & $4.67^{+0.23}_{-0.13}$ & $4.65^{+0.20}_{-0.10}$ & $4.62^{+0.25}_{-0.10}$ & $4.92^{+0.40}_{-0.30}$ & $4.75^{+0.30}_{-0.20}$ & $4.65^{+0.40}_{-0.13}$ & $4.97^{+0.25}_{-0.45}$ & $4.88^{+0.35}_{-0.33}$	\\
 & $f$ & $1.620^{+0.227}_{-0.350}$ & $1.380^{+0.321}_{-0.092}$ & $1.140^{+0.314}_{-0.173}$ & $0.795^{+0.288}_{-0.191}$ & $0.984^{+0.125}_{-0.281}$ & $0.634^{+0.115}_{-0.107}$ & $0.571^{+0.068}_{-0.057}$ & $0.626^{+0.023}_{-0.086}$ & $0.559^{+0.041}_{-0.035}$ & $0.560^{+0.067}_{-0.034}$	\\ 
 & $\chi_r^2$ & 1.14 & 1.10 & 1.15 & 1.66 & 1.69 & 2.48 & 2.91 & 4.58 & 6.47 & 5.70 \\ \hline
\text{H+07} & T$_\mathrm{eff}$ \text{(K)} &$1800^{+0}_{-100}$ & $1800^{+0}_{-100}$ & $1800^{+0}_{-100}$ & $1800^{+0}_{-200}$ & $1700^{+100}_{-100}$ & $1700^{+100}_{-200}$ & $1700^{+100}_{-200}$ & $1600^{+200}_{-100}$ & $1500^{+200}_{-100}$ & $1500^{+100}_{-200}$	\\
 & $\log{g}$ (\text{cm s}$^{−2}$) & $5.00^{+0.35}_{-0.25}$ & $5.05^{+0.38}_{-0.10}$ & $5.10^{+0.25}_{-0.28}$ & $5.05^{+0.33}_{-0.30}$ & $4.92^{+0.38}_{-0.30}$ & $4.83^{+0.48}_{-0.30}$ & $5.03^{+0.35}_{-0.35}$ & $5.03^{+0.28}_{-0.38}$ & $5.05^{+0.33}_{-0.38}$ & $5.10^{+0.23}_{-0.45}$	\\
 & $f$ & $1.660^{+0.215}_{-0.096}$ & $1.790^{+0.166}_{-0.578}$ & $1.240^{+0.615}_{-0.420}$ & $0.774^{+0.321}_{-0.187}$ & $0.891^{+0.447}_{-0.205}$ & $0.721^{+0.221}_{-0.129}$ & $0.579^{+0.245}_{-0.066}$ & $0.652^{+0.236}_{-0.102}$ & $0.637^{+0.254}_{-0.082}$ & $0.648^{+0.191}_{-0.133}$	\\ 
  & $\chi_r^2$ & 18.05 & 14.01 & 16.79 & 19.78 & 22.52 & 15.89 & 16.04 & 20.52 & 20.73 & 17.64\\  \hline
 \text{H+07c} & T$_\mathrm{eff}$ \text{(K)} &$1800^{+0}_{-0}$ & $1800^{+0}_{-0}$ & $1800^{+0}_{-100}$ & $1700^{+100}_{-100}$ & $1700^{+100}_{-0}$ & $1600^{+100}_{-100}$ & $1500^{+300}_{-0}$ & $1500^{+100}_{-100}$ & $1500^{+100}_{-200}$ & $1400^{+100}_{-100}$	\\
 & $\log{g}$ (\text{cm s}$^{−2}$) & $4.65^{+0.65}_{-0.13}$ & $5.20^{+0.15}_{-0.63}$ & $4.95^{+0.28}_{-0.33}$ & $4.92^{+0.23}_{-0.25}$ & $4.92^{+0.30}_{-0.23}$ & $4.80^{+0.45}_{-0.25}$ & $4.75^{+0.43}_{-0.18}$ & $4.78^{+0.35}_{-0.13}$ & $4.80^{+0.40}_{-0.28}$ & $4.88^{+0.25}_{-0.28}$	\\
 & $f$ & $1.900^{+0.077}_{-0.056}$ & $1.900^{+0.099}_{-0.126}$ & $1.640^{+0.180}_{-0.372}$ & $0.968^{+0.365}_{-0.179}$ & $1.120^{+0.201}_{-0.219}$ & $0.607^{+0.095}_{-0.036}$ & $0.596^{+0.098}_{-0.070}$ & $0.567^{+0.137}_{-0.012}$ & $0.601^{+0.141}_{-0.068}$ & $0.569^{+0.164}_{-0.064}$	\\ 
& $\chi_r^2$ & 3.48 & 2.53 & 2.98 & 4.27 & 3.51 & 4.44 & 6.11 & 10.48 & 8.41 & 9.42\\ \hline
\text{M+11} & T$_\mathrm{eff}$ \text{(K)} &$1800^{+0}_{-100}$ & $1800^{+0}_{-100}$ & $1800^{+0}_{-100}$ & $1700^{+100}_{-100}$ & $1700^{+100}_{-0}$ & $1800^{+0}_{-200}$ & $1700^{+0}_{-200}$ &   $1700^{+100}_{-300}$ & $1400^{+100}_{-100}$ & $1500^{+100}_{-200}$	\\
 & $\log{g}$ (\text{cm s}$^{−2}$) & $5.22^{+0.20}_{-0.38}$ & $5.12^{+0.30}_{-0.33}$ & $5.05^{+0.30}_{-0.25}$ & $5.00^{+0.38}_{-0.33}$ & $5.05^{+0.25}_{-0.38}$ & $4.90^{+0.33}_{-0.23}$ & $5.05^{+0.38}_{-0.40}$ & $5.00^{+0.38}_{-0.38}$ & $4.83^{+0.50}_{-0.30}$ & $5.00^{+0.35}_{-0.30}$ 	\\
 & $f$ & $1.820^{+0.184}_{-0.255}$ & $1.730^{+0.177}_{-0.324}$ & $1.390^{+0.427}_{-0.589}$ & $0.854^{+0.338}_{-0.223}$ & $0.873^{+0.496}_{-0.218}$ & $0.585^{+0.274}_{-0.044}$ & $0.686^{+0.223}_{-0.111}$ & $0.685^{+0.231}_{-0.081}$ & $0.662^{+0.208}_{-0.110}$ & $0.616^{+0.240}_{-0.070}$	\\ 
  & $\chi_r^2$ & 15.68 & 13.98 & 16.25 & 18.01 & 21.06 & 14.31 & 16.70 & 21.00 & 27.99 & 18.27 \\ \hline
 \text{M+11c} & T$_\mathrm{eff}$ \text{(K)} &$1700^{+0}_{-0}$ & $1700^{+0}_{-0}$ & $1700^{+0}_{-100}$ & $1600^{+100}_{-100}$ & $1600^{+100}_{-100}$ & $1600^{+100}_{-100}$ & $1500^{+200}_{-100}$ & $1400^{+100}_{-100}$ & $1400^{+0}_{-100}$ & $1300^{+200}_{-100}$	\\
 & $\log{g}$ (\text{cm s}$^{−2}$) & $4.55^{+0.45}_{-0.00}$ & $4.55^{+0.30}_{-0.85}$ & $4.55^{+0.20}_{-0.35}$ & $4.20^{+0.50}_{-0.50}$ & $4.25^{+0.60}_{-0.60}$ & $4.15^{+0.65}_{-0.40}$ & $4.20^{+0.55}_{-0.55}$ & $3.90^{+0.55}_{-0.35}$ & $4.30^{+0.50}_{-0.50}$ & $4.20^{+0.75}_{-0.65}$	\\
 & $f$ & $1.950^{+0.009}_{-0.000}$ & $1.960^{+0.040}_{-0.000}$ & $1.750^{+0.201}_{-0.060}$ & $1.380^{+0.406}_{-0.276}$ & $1.440^{+0.448}_{-0.167}$ & $0.634^{+0.295}_{-0.069}$ & $0.658^{+0.358}_{-0.106}$ & $0.866^{+0.364}_{-0.268}$ & $0.560^{+0.157}_{-0.018}$ & $0.698^{+0.266}_{-0.144}$	\\ 
  & $\chi_r^2$ & 4.50 & 2.43 & 1.05 & 1.19 & 1.19 & 1.16 & 1.42 & 1.78 & 1.71 & 2.94\\ \hline
\text{A+12c} & T$_\mathrm{eff}$ \text{(K)} &$2200^{+200}_{-200}$ & $2100^{+100}_{-200}$ & $2000^{+100}_{-200}$ & $1900^{+100}_{-100}$ & $1900^{+100}_{-200}$ & $1600^{+100}_{-200}$ & $1500^{+100}_{-100}$ & $1500^{+200}_{-100}$ & $1400^{+100}_{-100}$ & $1300^{+100}_{-150}$	\\
 & $\log{g}$ (\text{cm s}$^{−2}$) & $4.20^{+0.95}_{-0.65}$ & $4.80^{+0.45}_{-0.65}$ & $4.95^{+0.35}_{-0.90}$ & $5.05^{+0.05}_{-0.85}$ & $5.10^{+0.30}_{-0.85}$ & $5.00^{+0.15}_{-0.55}$ &
 $4.75^{+0.55}_{-0.65}$ & $4.40^{+0.50}_{-0.60}$ & $4.15^{+0.40}_{-0.35}$ & $4.05^{+0.95}_{-0.15}$	\\
 & $f$ & $1.120^{+0.544}_{-0.285}$ & $1.190^{+0.619}_{-0.277}$ & $1.050^{+0.276}_{-0.226}$ & $0.824^{+0.074}_{-0.254}$ & $0.879^{+0.487}_{-0.242}$ & $0.758^{+0.279}_{-0.231}$ & $0.809^{+0.175}_{-0.199}$ & $0.657^{+0.250}_{-0.106}$ & $0.614^{+0.166}_{-0.023}$ & $0.707^{+0.232}_{-0.154}$	\\ 
 & $\chi_r^2$ & 0.55 & 0.47 & 0.70 & 0.41 & 0.37 & 0.55 & 0.51 & 0.33 & 0.75 & 1.98 \\ \hline
\text{M+12c} & T$_\mathrm{eff}$ \text{(K)} &$1300^{+0}_{-0}$ & $1300^{+0}_{-0}$ & $1300^{+0}_{-0}$ &      $1300^{+0}_{-0}$ & $1300^{+0}_{-0}$ &  $1300^{+0}_{-100}$ & $1300^{+0}_{-100}$ &  $1300^{+0}_{-100}$ &  $1200^{+100}_{-100}$ & $1300^{+0}_{-200}$	\\
 & $\log{g}$ (\text{cm s}$^{−2}$) & $5.10^{+0.30}_{-0.32}$ & $5.00^{+0.30}_{-0.50}$ & $5.20^{+0.00}_{-0.22}$ & $5.25^{+0.00}_{-0.94}$ & $5.45^{+0.00}_{-0.30}$ & $5.20^{+0.15}_{-0.42}$ & $5.25^{+0.20}_{-0.42}$ & $5.25^{+0.20}_{-0.75}$ & $5.20^{+0.25}_{-0.42}$ & $5.35^{+0.10}_{-0.47}$	\\
 & $f$ & $1.940^{+0.025}_{-0.100}$ & $1.850^{+0.129}_{-0.000}$ & $1.910^{+0.079}_{-0.311}$ & $1.900^{+0.000}_{-0.221}$ & $1.970^{+0.000}_{-0.283}$ & $1.650^{+0.286}_{-0.358}$ & $1.440^{+0.303}_{-0.311}$ & $1.590^{+0.298}_{-0.694}$ & $1.240^{+0.558}_{-0.356}$ & $1.050^{+0.540}_{-0.258}$ \\	
& $\chi_r^2$ & 94.89 & 88.48 & 62.79 & 42.37 & 37.39 & 26.42 & 24.70 & 29.74 & 24.55 & 20.38 \\ \hline
\text{S+12} & T$_\mathrm{eff}$ \text{(K)} &$1500^{+0}_{-0}$ & $1500^{+0}_{-0}$ & $1500^{+0}_{-0}$ &  $1500^{+0}_{-0}$ & $1500^{+0}_{-0}$ & $1300^{+200}_{-100}$ & $1500^{+0}_{-300}$ &       $1300^{+200}_{-100}$ & $1300^{+200}_{-150}$ & $1300^{+0}_{-150}$	\\
 & $\log{g}$ (\text{cm s}$^{−2}$) &  $4.73^{+0.40}_{-0.18}$ & $4.58^{+0.67}_{-0.05}$ & $5.12^{+0.13}_{-0.47}$ & $4.68^{+0.45}_{-0.15}$ & $4.73^{+0.35}_{-0.13}$ & $4.66^{+0.40}_{-0.18}$ & $4.73^{+0.37}_{-0.20}$ & $4.94^{+0.26}_{-0.89}$ & $4.66^{+0.35}_{-0.71}$ &
  $4.80^{+0.38}_{-0.53}$	\\
 & $f$ & $1.920^{+0.073}_{-0.061}$ & $1.950^{+0.035}_{-0.099}$ & $1.810^{+0.085}_{-0.187}$ & $1.530^{+0.352}_{-0.623}$ & $1.610^{+0.331}_{-0.326}$ & $0.852^{+0.587}_{-0.173}$ & $0.780^{+0.380}_{-0.147}$ & $0.826^{+0.367}_{-0.215}$ & $0.751^{+0.306}_{-0.185}$ & $0.626^{+0.425}_{-0.089}$	\\ 
  & $\chi_r^2$ & 46.04 & 38.96 & 27.58 & 19.81 & 21.19 & 32.47 & 19.66 & 44.78 & 38.70 & 35.54 \\ \hline
\text{M+19} & T$_\mathrm{eff}$ \text{(K)} &$2500^{+100}_{-200}$ & $2300^{+200}_{-200}$ & $2200^{+100}_{-200}$ &  $2100^{+100}_{-300}$ & $2000^{+200}_{-200}$ & $1800^{+200}_{-100}$ & $1800^{+100}_{-200}$ &     $1700^{+200}_{-200}$ & $1500^{+300}_{-100}$ & $1600^{+100}_{-200}$	\\
 & $\log{g}$ (\text{cm s}$^{−2}$) &  $5.04^{+0.74}_{-1.76}$ & $4.50^{+1.26}_{-1.26}$ & $4.66^{+1.18}_{-1.86}$ & $4.34^{+1.34}_{-1.98}$ & $4.28^{+1.50}_{-1.68}$ & $4.76^{+0.98}_{-2.03}$ & $4.96^{+0.84}_{-2.08}$ & $4.66^{+1.16}_{-1.70}$ & $5.04^{+0.74}_{-1.54}$ &
  $4.86^{+0.86}_{-0.84}$	\\
 & $f$ & $0.573^{+0.152}_{-0.048}$ & $0.604^{+0.114}_{-0.073}$ & $0.588^{+0.103}_{-0.064}$ & $0.583^{+0.132}_{-0.058}$ & $0.587^{+0.150}_{-0.071}$ & $0.584^{+0.139}_{-0.061}$ & $0.593^{+0.101}_{-0.064}$ & $0.607^{+0.149}_{-0.074}$ & $0.604^{+0.107}_{-0.081}$ & $0.594^{+0.089}_{-0.065}$	\\ 
 & $\chi_r^2$ & 4.95 &  5.72 & 7.93 & 12.75 & 12.42 & 22.34 & 25.34 & 28.11 & 31.72 & 34.55 \\ \hline
\text{P+20} & T$_\mathrm{eff}$ \text{(K)} &$1800^{+0}_{-0}$ & $1800^{+0}_{-0}$ & $1800^{+0}_{-100}$ &  $1800^{+0}_{-100}$ & $1800^{+0}_{-100}$ & $1600^{+100}_{-100}$ & $1500^{+200}_{-0}$ &     $1600^{+0}_{-200}$ & $1500^{+100}_{-100}$ & $1400^{+100}_{-100}$	\\
 & $\log{g}$ (\text{cm s}$^{−2}$) &  $4.12^{+0.60}_{-0.75}$ & $4.10^{+1.08}_{-0.09}$ & $4.01^{+0.57}_{-1.01}$ & $4.02^{+1.07}_{-1.02}$ & $3.17^{+0.90}_{-0.40}$ & $3.10^{+0.99}_{-0.33}$ & $3.23^{+0.85}_{-0.40}$ & $3.25^{+0.22}_{-0.38}$ & $3.23^{+0.82}_{-1.15}$ &
  $3.12^{+0.81}_{-0.14}$	\\
 & $f$ & $1.850^{+0.111}_{-0.162}$ & $1.880^{+0.056}_{-0.193}$ & $1.560^{+0.329}_{-0.333}$ & $0.867^{+0.273}_{-0.151}$ & $1.030^{+0.230}_{-0.201}$ & $0.609^{+0.152}_{-0.081}$ & $0.577^{+0.101}_{-0.044}$ & $0.563^{+0.132}_{-0.027}$ & $0.571^{+0.085}_{-0.062}$ & $0.586^{+0.139}_{-0.065}$	\\ 
 & $\chi_r^2$ & 3.17 & 1.96 & 1.87 & 2.31 & 2.01 & 2.40 & 3.58 & 4.17 & 5.45 & 4.03 \\ \hline
\text{M+21} & T$_\mathrm{eff}$ \text{(K)} &$2100^{+200}_{-100}$ & $2100^{+200}_{-200}$ & $2000^{+200}_{-100}$ & $1900^{+100}_{-200}$ & $2000^{+100}_{-200}$ & $1600^{+100}_{-100}$ & $1600^{+0}_{-100}$ & $1500^{+100}_{-100}$ & $1500^{+0}_{-200}$ & $1400^{+100}_{-100}$	\\
 & $\log{g}$ (\text{cm s}$^{−2}$) & $3.80^{+1.10}_{-0.65}$ & $4.15^{+0.80}_{-0.94}$ & $3.95^{+0.93}_{-0.67}$ & $3.69^{+0.96}_{-0.44}$ & $3.61^{+0.84}_{-0.47}$ & $3.72^{+0.67}_{-0.49}$ & $3.38^{+0.75}_{-0.23}$ & $3.28^{+0.59}_{-0.18}$ & $3.23^{+0.92}_{-0.13}$ & $3.28^{+0.63}_{-0.13}$	\\
 & $f$ & $1.030^{+0.326}_{-0.348}$ & $0.990^{+0.492}_{-0.291}$ & $0.850^{+0.386}_{-0.210}$ & $0.715^{+0.286}_{-0.148}$ & $0.685^{+0.382}_{-0.125}$ & $0.706^{+0.131}_{-0.128}$ & $0.595^{+0.138}_{-0.060}$ & $0.623^{+0.171}_{-0.070}$ & $0.551^{+0.194}_{-0.029}$ & $0.600^{+0.150}_{-0.083}$	\\ 
 & $\chi_r^2$ & 0.84 & 0.76 & 1.15 & 1.61 & 1.58 & 2.90 & 2.89 & 5.43 & 4.05 & 4.19 \\ \hline
\text{K+21} & T$_\mathrm{eff}$ \text{(K)} & $1300^{+0}_{-0}$ & $1300^{+0}_{-0}$ & $1300^{+0}_{-0}$ & $1300^{+0}_{-0}$ & $1300^{+0}_{-0}$ & $1300^{+0}_{-50}$ & $1300^{+0}_{-50}$ & $1300^{+0}_{-50}$ & $1300^{+0}_{-100}$ & $1300^{+0}_{-100}$	\\
 & $\log{g}$ (\text{cm s}$^{−2}$) & $4.17^{+0.53}_{-0.18}$ & $4.68^{+0.30}_{-0.78}$ & $4.57^{+0.83}_{-1.03}$ & $3.90^{+0.65}_{-0.39}$ & $4.13^{+0.33}_{-0.46}$ & $3.88^{+0.48}_{-0.34}$ & $3.90^{+0.50}_{-0.29}$ & $3.78^{+0.55}_{-0.16}$ & $3.78^{+0.43}_{-0.19}$ & $3.82^{+0.15}_{-0.14}$	\\
 & $f$ &  $1.910^{+0.049}_{-0.054}$ & $1.910^{+0.061}_{-0.055}$ & $1.970^{+0.023}_{-0.124}$ & $1.780^{+0.165}_{-0.174}$ & $1.910^{+0.063}_{-0.022}$ & $0.943^{+0.248}_{-0.199}$ & $0.852^{+0.148}_{-0.235}$ & $0.760^{+0.208}_{-0.137}$ & $0.584^{+0.241}_{-0.049}$ & $0.567^{+0.141}_{-0.034}$\\
 & $\chi_r^2$ & 81.99 & 76.91 & 58.18 & 31.46 & 34.48 & 28.71 & 27.26 & 35.02 & 19.66 & 14.84 \\ \hline
\enddata
\end{deluxetable*}

\begin{deluxetable*}{cccccccccccc}[ht!]
\tabletypesize{\scriptsize}
\tablecolumns{11}
\tablewidth{\columnwidth}
\tablecaption{Results of the random forest retrieval algorithm for the SpeX prism standard brown dwarfs and T0 - T8 spectral classes.}
\label{tab:data_posteriors_T_dwarfs}
\tablehead{ \colhead{Model} & \colhead{Parameter} & \colhead{T0} & \colhead{T1} & \colhead{T2} & \colhead{T3} & \colhead{T4} & \colhead{T5} & \colhead{T6} & \colhead{T7} & \colhead{T8}
}
\startdata 
B+97 & T$_\mathrm{eff}$ \text{(K)} &$1100^{+99}_{-0}$ & $1100^{+99}_{-100}$ & $1100^{+99}_{-100}$ &  $1100^{+99}_{-100}$ &  $1100^{+99}_{-100}$ & $1100^{+99}_{-100}$ & $997^{+100}_{-198}$ & $799^{+198}_{-200}$ & $699^{+100}_{-100}$	\\
 & $\log{g}$ (\text{cm s}$^{−2}$) & $5.05^{+0.25}_{-0.37}$ & $4.93^{+0.32}_{-0.35}$ & $4.83^{+0.57}_{-0.25}$ & $4.98^{+0.42}_{-0.40}$ &$4.88^{+0.42}_{-0.30}$ & $4.83^{+0.37}_{-0.25}$ & $4.68^{+0.62}_{-0.43}$ & $4.58^{+0.67}_{-0.48}$ & $4.63^{+0.77}_{-0.79}$\\
 & $f$ & $1.210^{+0.475}_{-0.279}$ & $1.230^{+0.507}_{-0.485}$ & $1.330^{+0.509}_{-0.579}$ & $1.460^{+0.401}_{-0.498}$ & $1.280^{+0.378}_{-0.384}$ & $0.745^{+0.297}_{-0.195}$ & $0.843^{+0.809}_{-0.276}$ & $1.020^{+0.777}_{-0.474}$ & $1.050^{+0.697}_{-0.367}$\\ 
 & $\chi_r^2$ & 88.02 & 86.69 & 92.38 & 76.89 & 84.90 & 78.19 & 121.27 & 145.13 & 410.34 \\ \hline
B+06 & T$_\mathrm{eff}$ \text{(K)} &$1500^{+100}_{-100}$ & $1500^{+0}_{-100}$ & $1500^{+100}_{-100}$ & $1400^{+100}_{-100}$ & $1300^{+200}_{-100}$ & $1200^{+200}_{-200}$ & $1000^{+100}_{-100}$ & $900^{+200}_{-100}$ & $800^{+200}_{-100}$	\\
 & $\log{g}$ (\text{cm s}$^{−2}$) & $4.85^{+0.28}_{-0.18}$ & $4.95^{+0.28}_{-0.25}$ & $5.00^{+0.36}_{-0.15}$ & $5.17^{+0.15}_{-0.35}$ &$5.22^{+0.20}_{-0.40}$ & $5.25^{+0.13}_{-0.55}$ &
 $5.30^{+0.15}_{-0.45}$ & $5.17^{+0.23}_{-0.30}$ & $5.12^{+0.33}_{-0.33}$	\\
 & $f$ & $0.609^{+0.129}_{-0.061}$ & $0.600^{+0.053}_{-0.076}$ & $0.574^{+0.216}_{-0.071}$ & $0.625^{+0.204}_{-0.087}$ & $0.551^{+0.221}_{-0.027}$ & $0.587^{+0.264}_{-0.075}$ & $0.633^{+0.446}_{-0.093}$ & $0.781^{+0.341}_{-0.241}$ & $0.679^{+0.269}_{-0.139}$	\\ 
& $\chi_r^2$ & 12.21 & 13.72 & 8.84 & 9.15 & 12.71 & 11.73 & 21.15 & 27.33 & 35.82 \\ \hline
B+06c & T$_\mathrm{eff}$ \text{(K)} &$1400^{+100}_{-100}$ & $1300^{+150}_{-100}$ & $1400^{+100}_{-150}$ & $1250^{+250}_{-100}$ & $1200^{+100}_{-50}$ & $1100^{+100}_{-100}$ & $1050^{+100}_{-100}$ & $950^{+100}_{-100}$ & $800^{+200}_{-100}$	\\
 & $\log{g}$ (\text{cm s}$^{−2}$) & $4.85^{+0.28}_{-0.28}$ & $4.75^{+0.45}_{-0.18}$ & $5.17^{+0.20}_{-0.40}$ & $5.12^{+0.35}_{-0.25}$ & $5.12^{+0.23}_{-0.43}$ & $5.17^{+0.23}_{-0.28}$ & $5.28^{+0.10}_{-0.38}$ & $5.08^{+0.29}_{-0.33}$ & $5.20^{+0.18}_{-0.30}$	\\
 & $f$ & $0.556^{+0.068}_{-0.053}$ & $0.553^{+0.071}_{-0.020}$ & $0.632^{+0.197}_{-0.117}$ & $0.662^{+0.193}_{-0.127}$ &  $0.585^{+0.157}_{-0.054}$ & $0.635^{+0.283}_{-0.103}$ & $0.634^{+0.238}_{-0.113}$ & $0.667^{+0.273}_{-0.121}$ & $0.652^{+0.172}_{-0.109}$	\\ 
 & $\chi_r^2$ & 4.45 & 8.06 & 5.01 & 8.48 & 14.38 & 15.16 & 20.57 & 26.45 & 36.69 \\ \hline
H+07 & T$_\mathrm{eff}$ \text{(K)} &$1400^{+200}_{-0}$ & $1400^{+200}_{-100}$ & $1500^{+100}_{-100}$ & $1400^{+100}_{-100}$ & $1300^{+100}_{-100}$ & $1200^{+100}_{-100}$ & $1000^{+100}_{-100}$ & $1000^{+100}_{-100}$ & $800^{+200}_{-100}$	\\
 & $\log{g}$ (\text{cm s}$^{−2}$) & $5.05^{+0.33}_{-0.45}$ & $5.15^{+0.18}_{-0.33}$ & $5.10^{+0.30}_{-0.30}$ & $5.20^{+0.20}_{-0.33}$ & $5.12^{+0.28}_{-0.33}$ & $5.35^{+0.08}_{-0.40}$ &
 $5.17^{+0.25}_{-0.38}$ & $5.20^{+0.23}_{-0.40}$ & $5.15^{+0.28}_{-0.38}$	\\
 & $f$ & $0.612^{+0.127}_{-0.085}$ & $0.592^{+0.192}_{-0.069}$ & $0.644^{+0.174}_{-0.072}$ & $0.583^{+0.212}_{-0.035}$ & $0.683^{+0.102}_{-0.151}$ & $0.619^{+0.154}_{-0.076}$ & $0.693^{+0.246}_{-0.111}$ & $0.639^{+0.410}_{-0.110}$ & $0.644^{+0.306}_{-0.121}$	\\ 
 & $\chi_r^2$ & 15.90 & 16.05 & 11.81 & 10.95 & 14.15 & 15.98 & 24.16 & 26.39 & 41.24\\ \hline
 H+07c & T$_\mathrm{eff}$ \text{(K)} &$1400^{+0}_{-100}$ & $1400^{+100}_{-100}$ & $1400^{+100}_{-12}$ & $1300^{+100}_{-100}$ & $1300^{+100}_{-100}$ & $1100^{+200}_{-0}$ &
 $1100^{+300}_{-0}$ & $900^{+200}_{-100}$ & $800^{+200}_{-100}$	\\
 & $\log{g}$ (\text{cm s}$^{−2}$) & $5.10^{+0.25}_{-0.48}$ & $4.83^{+0.43}_{-0.33}$ & $4.92^{+0.23}_{-0.40}$ & $5.08^{+0.25}_{-0.30}$ & $5.00^{+0.20}_{-0.35}$ & $5.28^{+0.13}_{-0.33}$ &
 $5.19^{+0.24}_{-0.44}$ & $5.17^{+0.25}_{-0.45}$ & $5.28^{+0.18}_{-0.35}$	\\
 & $f$ & $0.615^{+0.111}_{-0.101}$ & $0.571^{+0.121}_{-0.060}$ & $0.547^{+0.179}_{-0.013}$ & $0.646^{+0.208}_{-0.090}$ & $0.591^{+0.223}_{-0.045}$ &  $0.613^{+0.132}_{-0.069}$ & $0.643^{+0.245}_{-0.074}$ & $0.829^{+0.475}_{-0.230}$ & $0.699^{+0.343}_{-0.151}$	\\ 
 & $\chi_r^2$ &  10.68 & 8.68 & 7.41 & 10.15 & 11.20 & 15.45 & 23.10 & 28.83 & 42.73 \\ \hline
M+11 & T$_\mathrm{eff}$ \text{(K)} &$1500^{+0}_{-100}$ & $1400^{+100}_{-100}$ & $1400^{+100}_{-100}$ & $1400^{+0}_{-200}$ & $1300^{+100}_{-100}$ & $1100^{+200}_{-0}$ &
 $1000^{+300}_{-100}$ & $1000^{+100}_{-100}$ & $800^{+200}_{-100}$	\\
 & $\log{g}$ (\text{cm s}$^{−2}$) & $5.10^{+0.23}_{-0.30}$ & $5.00^{+0.43}_{-0.38}$ & $5.00^{+0.43}_{-0.38}$ & $5.28^{+0.05}_{-0.53}$ & $5.17^{+0.30}_{-0.45}$ & $5.12^{+0.23}_{-0.23}$ &
 $5.15^{+0.23}_{-0.38}$ & $5.25^{+0.26}_{-0.15}$ & $5.15^{+0.25}_{-0.28}$	\\
 & $f$ & $0.574^{+0.155}_{-0.042}$ & $0.606^{+0.102}_{-0.062}$ & $0.734^{+0.276}_{-0.122}$ & $0.585^{+0.236}_{-0.076}$ & $0.646^{+0.145}_{-0.070}$ &  $0.665^{+0.279}_{-0.099}$ & $0.721^{+0.303}_{-0.186}$ & $0.683^{+0.264}_{-0.154}$ & $0.819^{+0.470}_{-0.266}$	\\
 & $\chi_r^2$ & 14.22 & 16.02 & 13.79 & 10.13 & 13.51 & 16.36 & 22.63 & 25.30 & 255.51 \\ \hline
M+11c & T$_\mathrm{eff}$ \text{(K)} &$1400^{+0}_{-200}$ & $1300^{+100}_{-100}$ & $1400^{+100}_{-200}$ & $1200^{+100}_{-150}$ & $1100^{+200}_{-100}$ & $950^{+150}_{-50}$ &
 $900^{+150}_{-100}$ & $900^{+75}_{-100}$ & $800^{+100}_{-100}$	\\
 & $\log{g}$ (\text{cm s}$^{−2}$) & $4.70^{+0.25}_{-0.75}$ & $4.20^{+0.60}_{-0.35}$ & $4.70^{+0.25}_{-0.65}$ & $4.65^{+0.30}_{-0.65}$ & $4.60^{+0.38}_{-0.65}$ & $4.90^{+0.05}_{-0.30}$ &
 $4.90^{+0.05}_{-0.30}$ & $4.90^{+0.10}_{-0.35}$ & $4.80^{+0.15}_{-0.45}$	\\
 & $f$ & $0.642^{+0.265}_{-0.098}$ & $0.653^{+0.343}_{-0.114}$ & $0.637^{+0.363}_{-0.097}$ & $0.908^{+0.294}_{-0.276}$ & $0.744^{+0.253}_{-0.169}$ &  $0.843^{+0.582}_{-0.187}$ & $1.030^{+0.528}_{-0.403}$ & $0.858^{+0.487}_{-0.244}$ & $0.974^{+0.702}_{-0.399}$	\\ 
& $\chi_r^2$ & 2.25 & 4.90 & 3.92 & 12.80 & 13.00 & 18.80 &22.01 & 33.70 & 47.30\\ \hline
A+12c & T$_\mathrm{eff}$ \text{(K)} &$1300^{+100}_{-150}$ & $1200^{+100}_{-100}$ & $1200^{+100}_{-100}$ & $1200^{+100}_{-150}$ & $1100^{+100}_{-100}$ & $1050^{+50}_{-100}$ & $950^{+100}_{-100}$ & $900^{+50}_{-100}$ & $700^{+150}_{-50}$	\\
 & $\log{g}$ (\text{cm s}$^{−2}$) & $4.30^{+1.00}_{-0.60}$ & $4.00^{+0.55}_{-0.45}$ & $4.55^{+0.75}_{-1.05}$ & $4.90^{+0.40}_{-0.95}$ & $4.05^{+0.90}_{-0.55}$ & $4.35^{+0.45}_{-0.55}$ &
 $3.95^{+0.60}_{-0.65}$ & $3.75^{+0.65}_{-0.55}$ & $3.50^{+0.95}_{-0.40}$	\\
 & $f$ & $0.765^{+0.363}_{-0.209}$ & $0.844^{+0.167}_{-0.203}$ & $1.120^{+0.669}_{-0.269}$ & $0.948^{+0.401}_{-0.188}$ & $0.941^{+0.615}_{-0.230}$ &  $0.858^{+0.440}_{-0.167}$ & $0.846^{+0.658}_{-0.255}$ & $0.737^{+0.630}_{-0.159}$ & $0.743^{+0.315}_{-0.171}$	\\
  & $\chi_r^2$ & 1.06 & 2.56 & 1.81 & 3.07 & 3.46 & 4.32 & 9.50 & 15.22 & 16.46 \\ \hline
M+12c & T$_\mathrm{eff}$ \text{(K)} &$1300^{+0}_{-200}$ & $1300^{+0}_{-200}$ & $1200^{+100}_{-100}$ & $1200^{+100}_{-100}$ & $1200^{+100}_{-200}$ & $1000^{+100}_{-100}$ & $900^{+200}_{-100}$ &  $800^{+100}_{-0}$ & $700^{+100}_{-0}$	\\
 & $\log{g}$ (\text{cm s}$^{−2}$) & $5.30^{+0.20}_{-0.47}$ & $5.30^{+0.20}_{-0.47}$ & $5.30^{+0.10}_{-0.37}$ & $5.30^{+0.20}_{-0.42}$ & $5.25^{+0.20}_{-0.32}$ & $5.15^{+0.25}_{-0.27}$ &
 $5.10^{+0.35}_{-0.47}$ & $5.15^{+0.30}_{-0.62}$ & $5.10^{+0.25}_{-0.57}$	\\
 & $f$ & $1.020^{+0.331}_{-0.377}$ & $0.928^{+0.716}_{-0.294}$ & $1.280^{+0.321}_{-0.279}$ & $0.989^{+0.461}_{-0.267}$ & $0.908^{+0.723}_{-0.247}$ &  $1.060^{+0.464}_{-0.444}$ & $0.932^{+0.465}_{-0.346}$ & $0.890^{+0.620}_{-0.284}$ & $0.814^{+0.495}_{-0.244}$\\	
 & $\chi_r^2$ & 16.85 & 13.95 & 12.24 & 10.29 & 10.07 & 11.52 & 17.13 & 22.91 & 35.76\\ \hline
S+12 & T$_\mathrm{eff}$ \text{(K)} &$1300^{+200}_{-150}$ & $1200^{+100}_{-50}$ & $1300^{+200}_{-100}$ & $1300^{+0}_{-150}$ & $1200^{+100}_{-50}$ & $1150^{+50}_{-150}$ &
 $1050^{+100}_{-150}$ & $900^{+100}_{-100}$ & $800^{+100}_{-100}$	\\
 & $\log{g}$ (\text{cm s}$^{−2}$) & $4.88^{+0.30}_{-0.55}$ & $4.66^{+0.47}_{-0.51}$ & $4.73^{+0.47}_{-0.33}$ & $4.93^{+0.20}_{-0.37}$ & $4.82^{+0.25}_{-0.50}$ & $4.82^{+0.18}_{-0.27}$ &
 $4.85^{+0.15}_{-0.48}$ & $4.80^{+0.25}_{-0.45}$ & $4.85^{+0.30}_{-0.65}$	\\
 & $f$ & $0.686^{+0.193}_{-0.139}$ & $0.735^{+0.291}_{-0.195}$ & $0.887^{+0.200}_{-0.247}$ & $0.840^{+0.191}_{-0.191}$ & $0.745^{+0.294}_{-0.105}$ &  $0.703^{+0.446}_{-0.161}$ & $0.700^{+0.427}_{-0.108}$ & $0.765^{+0.337}_{-0.137}$ & $0.595^{+0.234}_{-0.073}$	\\ 
& $\chi_r^2$ & 28.47 & 33.30 & 18.09 & 14.04 & 22.26 & 16.92 & 23.69 & 32.09 & 40.01 \\ \hline
\text{M+19} & T$_\mathrm{eff}$ \text{(K)} &$1500^{+200}_{-100}$ & $1400^{+200}_{-100}$ & $1500^{+300}_{-100}$ &  $1400^{+300}_{-200}$ & $1400^{+200}_{-200}$ & $1300^{+200}_{-100}$ & $1200^{+400}_{-200}$ &     $1200^{+300}_{-300}$ & $1000^{+300}_{-200}$	\\
 & $\log{g}$ (\text{cm s}$^{−2}$) &  $5.20^{+0.52}_{-1.74}$ & $5.24^{+0.54}_{-1.42}$ & $5.30^{+0.60}_{-1.74}$ & $5.32^{+0.40}_{-1.78}$ & $5.12^{+0.58}_{-1.50}$ & $4.98^{+0.72}_{-1.18}$ & $4.66^{+0.86}_{-1.62}$ & $4.49^{+1.05}_{-1.87}$ & $4.42^{+0.94}_{-1.58}$ \\
 & $f$ & $0.583^{+0.122}_{-0.060}$ & $0.595^{+0.129}_{-0.054}$ & $0.623^{+0.130}_{-0.094}$ & $0.635^{+0.158}_{-0.095}$ & $0.600^{+0.199}_{-0.069}$ & $0.588^{+0.168}_{-0.057}$ & $0.616^{+0.267}_{-0.078}$ & $0.664^{+0.257}_{-0.121}$ & $0.656^{+0.323}_{-0.114}$ \\ 
& $\chi_r^2$ &  34.66 & 37.87 & 30.41 & 42.21 & 38.52 & 53.97 & 71.25 & 94.88 & 122.69 \\ \hline
\text{P+20} & T$_\mathrm{eff}$ \text{(K)} &$1400^{+0}_{-100}$ & $1300^{+100}_{-100}$ & $1300^{+100}_{-100}$ &  $1200^{+100}_{-100}$ & $1200^{+0}_{-100}$ & $1000^{+100}_{-0}$ & $1000^{+0}_{-100}$ &     $800^{+100}_{-0}$ & $700^{+100}_{-0}$ 	\\
 & $\log{g}$ (\text{cm s}$^{−2}$) &  $3.12^{+0.98}_{-1.02}$ & $3.73^{+0.45}_{-0.73}$ & $4.08^{+1.14}_{-0.08}$ & $4.00^{+0.77}_{-0.97}$ & $3.92^{+0.15}_{-0.88}$ & $4.01^{+0.68}_{-0.93}$ & $3.42^{+1.20}_{-0.42}$ & $3.77^{+1.20}_{-0.78}$ & $3.84^{+1.69}_{-0.46}$ 	\\
 & $f$ & $0.604^{+0.111}_{-0.088}$ & $0.567^{+0.152}_{-0.064}$ & $0.716^{+0.187}_{-0.138}$ & $0.675^{+0.421}_{-0.152}$ & $0.650^{+0.230}_{-0.108}$ & $0.773^{+0.145}_{-0.207}$ & $0.736^{+0.292}_{-0.200}$ & $0.801^{+0.334}_{-0.229}$ & $0.816^{+0.272}_{-0.236}$ 	\\ 
& $\chi_r^2$ & 2.12 & 2.49 & 1.53 & 1.55 & 1.45 & 1.14 & 4.05 & 6.46 & 8.05 \\ \hline
M+21 & T$_\mathrm{eff}$ \text{(K)} &$1400^{+100}_{-100}$ & $1300^{+100}_{-0}$ & $1400^{+0}_{-100}$ & $1300^{+0}_{-100}$ & $1200^{+100}_{-0}$ & $1100^{+0}_{-100}$ & $950^{+50}_{-50}$ &     $900^{+50}_{-50}$ & $750^{+50}_{-50}$	\\
 & $\log{g}$ (\text{cm s}$^{−2}$) & $3.35^{+1.02}_{-0.28}$ & $3.82^{+0.75}_{-0.45}$ & $3.74^{+0.91}_{-0.49}$ & $4.40^{+0.68}_{-0.71}$ & $4.53^{+0.58}_{-1.10}$ & $5.15^{+0.30}_{-1.00}$ & $4.88^{+0.43}_{-0.80}$ & $4.68^{+0.43}_{-0.73}$ & $4.55^{+0.33}_{-0.55}$	\\
 & $f$ &  $0.595^{+0.223}_{-0.053}$ & $0.537^{+0.161}_{-0.023}$ &$0.663^{+0.229}_{-0.107}$ & $0.692^{+0.241}_{-0.120}$ & $0.677^{+0.143}_{-0.112}$ & $0.686^{+0.246}_{-0.097}$ & $0.735^{+0.246}_{-0.195}$ & $0.714^{+0.306}_{-0.162}$ & $0.706^{+0.256}_{-0.152}$ \\ 
& $\chi_r^2$ & 2.16 & 3.55 & 1.39 & 1.74 & 1.04 & 0.79 & 3.83 & 4.92 & 5.69 \\ \hline
K+21 & T$_\mathrm{eff}$ \text{(K)} &$1300^{+0}_{-50}$ & $1250^{+50}_{-50}$ & $1300^{+0}_{-50}$ & $1250^{+50}_{-50}$ & $1250^{+50}_{-100}$ & $1100^{+50}_{-100}$ & $1000^{+100}_{-50}$ & $850^{+100}_{-50}$ & $750^{+50}_{-50}$	\\
 & $\log{g}$ (\text{cm s}$^{−2}$) & $3.69^{+0.41}_{-0.18}$ & $3.74^{+0.31}_{-0.15}$ & $3.95^{+0.35}_{-0.44}$ & $4.10^{+1.02}_{-0.51}$ & $3.88^{+0.28}_{-0.16}$ & $3.97^{+1.03}_{-0.39}$ & $3.85^{+0.43}_{-0.31}$ & $3.95^{+0.45}_{-0.34}$ & $3.82^{+0.40}_{-0.21}$	\\
 & $f$ &  $0.686^{+0.129}_{-0.149}$ & $0.545^{+0.224}_{-0.038}$ & $0.761^{+0.235}_{-0.138}$ & $0.698^{+0.196}_{-0.135}$ & $0.573^{+0.172}_{-0.063}$ & $0.608^{+0.161}_{-0.069}$ & $0.631^{+0.160}_{-0.089}$ & $0.999^{+0.523}_{-0.413}$ & $0.659^{+0.476}_{-0.110}$ \\  
 & $\chi_r^2$ & 9.06 & 7.96 & 8.47 & 3.68 & 3.87 & 4.14 & 7.10 & 10.78 & 13.84 \\ \hline
\enddata
\end{deluxetable*}

\end{document}